\documentclass[aps,pra,longbibliography,showpacs,amsmath,amssymb,amsfonts,superscriptaddress,twocolumn]{revtex4-1}

\usepackage{graphics,bbm}
\usepackage[colorlinks=true,citecolor=blue,linkcolor=blue]{hyperref}
\usepackage[usenames]{color}
\usepackage{graphicx}
\usepackage{subfigure}
\usepackage{epsfig,times}
\usepackage{epsf}
\usepackage{dcolumn}
\usepackage{bm}
\usepackage{color}
\usepackage{verbatim}
\usepackage{epstopdf}
\usepackage{latexsym}
\usepackage{psfrag}
\usepackage{xcolor}
\usepackage{amsthm}

\def\ln{{\rm ln}}

\def\p{{\rm\bf p}}
\def\q{{\rm\bf q}}

\def\f{{\rm\bf f}}
\def\F{{\rm\bf F}}
\def\U{{\rm U}}

\def\om{\omega}

\newcommand{\beqa}{\begin{eqnarray}}
\newcommand{\eeqa}{\end{eqnarray}}

\newcommand{\bra}[1]{\left\langle #1\right|}
\newcommand{\ket}[1]{\left|#1\right\rangle}
\newcommand{\braket}[2]{\left\langle #1|#2\right\rangle}

\newcommand{\la}{\left\langle}
\newcommand{\ra}{\right\rangle}
\newcommand{\pd}{\partial}
\newcommand{\de}[1]{\delta\left(#1\right)}
\newcommand{\td}{\mathrm{d}}

\newcommand{\etal}{\textit{et al. }}
\newcommand{\e}[1]{\exp{\left(#1\right)}}
\newcommand{\lo}[1]{\ln{\left(#1\right)}}

\newcommand{\com}[2]{\left[#1,\,#2\right]}

\newcommand{\bla}{bla\\bla\\bla\\bla\\bla}

\newcommand{\PRA}{Phys. Rev. A}

\newcommand{\PRE}{Phys. Rev. E}
\newcommand{\PRL}{Phys. Rev. Lett.}

\newcommand{\EPL}{EPL (Europhys. Lett.)}
\newcommand{\RMP}{Rev. Mod. Phys.}
\newcommand{\NJP}{New. J. Phys.}

\newcommand{\mb}[1]{\mbox{\boldmath$#1$}}
\newcommand{\mc}[1]{\mathcal{#1}}

\newcommand{\mrm}[1]{\mathrm{#1}}

\begin{document}

\title{Classical and quantum shortcuts to adiabaticity for scale-invariant driving}
\author{Sebastian Deffner}
\affiliation{Department of Chemistry and Biochemistry and Institute for Physical Science and Technology, University of Maryland, College Park, Maryland 20742, USA}
\author{Christopher Jarzynski}
\affiliation{Department of Chemistry and Biochemistry and Institute for Physical Science and Technology, University of Maryland, College Park, Maryland 20742, USA}
\author{Adolfo del Campo}
\affiliation{Theoretical Division,  Los Alamos National Laboratory, Los Alamos, NM 87545, USA}
\affiliation{Center for Nonlinear Studies,  Los Alamos National Laboratory, Los Alamos, NM 87545, USA}

\begin{abstract}
A shortcut to adiabaticity is a driving protocol that reproduces in a short time the same final state that would result from an adiabatic, infinitely slow process. A  powerful technique to engineer such shortcuts relies on the use of auxiliary counterdiabatic fields. Determining the explicit form of the required fields has generally proven to be complicated. We present explicit counterdiabatic driving protocols  for scale-invariant dynamical processes, which describe for instance expansion and transport.  To this end, we use the formalism of generating functions, and unify previous approaches independently developed in classical and quantum studies.  The resulting framework is applied to the design of shortcuts to adiabaticity for a large class of classical and quantum, single-particle,  non-linear, and many-body systems.
\end{abstract}


\maketitle

\section{Introduction\label{sec:intro}}

Modern research in nanoengineering develops increasingly small devices, which operate in a regime described by effective classical dynamics \cite{collins_1997,huang_2003} or  quantum mechanics \cite{kane_1998,kinoshita_2006}. Achieving a fast coherent control with high-fidelity \cite{avron_1988,kral_2007} is a  ubiquitous goal shared by a variety of fields and technologies, including quantum sensing and metrology \cite{gio06}, finite-time thermodynamics \cite{andresen_1984}, quantum simulation \cite{trabesinger_2012}, and adiabatic quantum computation \cite{nielsen_00}. The quantum adiabatic theorem \cite{messiah_1966}, however, appears as a no-go theorem for excitation-free ultrafast processes.  As a result, an increasing amount of theoretical and experimental research is targeting the design of shortcuts to adiabaticity (STA), i.e., non-adiabatic processes that reproduce in a finite-time the same final state that would result from an adiabatic, infinitely slow protocol \cite{torrontegui_2013}. 

A variety of techniques have been developed to engineer STA: the use of dynamical invariants \cite{chen_2010,chen_2011}, the inversion of scaling laws \cite{campo_2011,campo_boshier_2012}, the fast-forward technique \cite{masuda_2010,masuda_2011,torrontegui_2012}, and  counterdiabatic driving, also known as transitionless quantum driving \cite{demirplak_rice_2003,demirplak_rice_2005,berry_2009}. Among these  techniques, counterdiabatic  driving (CD)  is unique in that it drives the dynamics precisely through the adiabatic manifold of the system Hamiltonian. In addition, it enjoys a wide applicability. In its original formulation \cite{demirplak_rice_2003,demirplak_rice_2005,berry_2009}, one considers a time-dependent Hamiltonian $\hat{H}_0(t)$ with instantaneous eigenvalues $\{\varepsilon_n(t)\}$ and eigenstates $\{\ket{n(t)}\}$. In the limit of infinitely slow variation of $\hat{H}_0(t)$ a solution of the dynamics is given by
\beqa
\label{eq:adiabatic_approximation}
\ket{\psi_n(t)}=e^{-\frac{i}{\hbar}\int_0^t\td s\,\varepsilon_n(s)-\int_0^t\td s\la n|\partial_{s}n\ra }\ket{n(t)}. 
\eeqa
In this adiabatic limit no transitions between eigenstates occur \cite{messiah_1966}, and each eigenstate acquires a time-dependent phase that can be separated into a dynamical and a geometric contribution \cite{berry_1984}, represented by the two terms inside the exponential in the above expression.

Now consider a non-adiabatic Hamiltonian $\hat{H}_0(t)$.
In the CD paradigm, a corresponding Hamiltonian $\hat{H}(t)$ is constructed, such that the adiabatic approximation associated with $\hat{H}_0(t)$ \eqref{eq:adiabatic_approximation} is an exact solution of the dynamics generated by $\hat{H}(t)$ under the time-dependent Schr\" odinger equation.
Writing the time-evolution operator as $\hat{U}(t)=\sum_n \ket{\psi_n(t)}\bra{n(0)}$, one arrives at an explicit expression for $\hat{H}(t)$ \cite{demirplak_rice_2003,demirplak_rice_2005,berry_2009}:
\begin{equation}
\label{q01}
\hat{H}= \hat{H}_0+\hat{H}_1=\hat{H}_0+ i\hbar\sum_n\left(\ket{\partial_tn}\bra{n}-\braket{n}{\partial_t n}\ket{n}\bra{ n}\right)\,.
\end{equation}
Here, the {\it auxiliary CD Hamiltonian} $\hat{H}_1(t)$ enforces evolution along the adiabatic manifold of $\hat{H}_0(t)$: if a system is prepared in an eigenstate $\ket{n(0)}$ of $\hat{H}_0(0)$ and subsequently evolves under $\hat{H}(t)$, then the term $\hat{H}_1(t)$ effectively suppresses the non-adiabatic transitions out of $\ket{n(t)}$ that would arise in the absence of this term. Note that the evolution is non-adiabatic with respect to the full {\it CD Hamiltonian} $\hat{H}= \hat{H}_0+\hat{H}_1$. 

The CD Hamiltonian $\hat{H}(t)$ has been the object of intense study. It was found that the higher the speed of evolution, the larger is the intensity of the required auxiliary CD field \cite{demirplak_rice_2008,campo_rams_zurek_2012}. Experimental demonstration of driving protocols inspired by the CD technique have recently been reported in single two-level systems \cite{bason_2011,zhang_2013}. In the many-body case,  $\hat{H}_1(t)$ generally includes non-local and multi-body  interactions \cite{campo_rams_zurek_2012,campo_2013}. Local driving protocols can be derived for unitarily equivalent Hamiltonians \cite{campo_2013}, an approach which has proven useful as well in single-particle systems \cite{bason_2011,zhang_2013,mostafazadeh_2001,torrontegui_2013}.

However, the computation of the auxiliary term $\hat{H}_1(t)$ requires knowledge of the spectral properties of the instantaneous system Hamiltonian $\hat{H}_0(t)$ at all times. This constraint has limited the range of applicability of the method to the control of few-level systems \cite{demirplak_rice_2003,demirplak_rice_2005,berry_2009} and non-interacting matter-waves in time-dependent harmonic traps \cite{muga_2010,torrontegui_2011,deng_2013}.

Recently a classical analogue of CD was proposed, namely \textit{dissipationless classical driving} \cite{jarzynski_2013}.  Here, for a time-dependent classical Hamiltonian $H_0(q,p,t)$, one seeks an auxiliary term $H_1(q,p,t)$ such that under the Hamiltonian dynamics generated by $H = H_0 + H_1$, the classical adiabatic invariant of $H_0$ is conserved exactly. For systems with one degree of freedom an explicit solution of this problem, analogous to \eqref{eq:adiabatic_approximation} above, was obtained (see Eq.~(32) of Ref.~\cite{jarzynski_2013}).  Moreover it was argued that this classical solution can be useful in constructing the quantal CD Hamiltonian $\hat H(t)$, bypassing the spectral decomposition of $\hat H_0(t)$.  This was illustrated for arbitrary power-law traps (including the particle-in-a-box as a limiting case), for which simple expressions for $\hat H_1(t)$ in terms of position and momentum operators were obtained and quantized. Further progress was achieved 
using scaling laws in expansions and compressions for a wide-variety of single-particle, nonlinear, and many-body quantum systems \cite{campo_2013}.

Our aim in this paper is to find an experimentally realizable CD Hamiltonian \eqref{q01} for scale invariant processes, without using the explicit spectral decomposition of $\hat{H}_0(t)$. Scale-invariant driving is generated by transformations of $\hat{H}_0(t)$ for which the  density profile (and all local correlations in real space) is preserved up to scaling and translation. Using this property, we start with a single quantum particle in a one-dimensional potential, from which we will develop a general framework to find local CD protocols for multi-particle quantum systems, obeying both linear and non-linear dynamics. We will use methods from classical Hamiltonian dynamics, namely the formalism of generating functions, to treat dissipationless classical driving by the same means. Our approach also allows one to treat arbitrary external potentials, beyond the validity of perturbation theory.

The paper is organized as follows: we will begin in Sec.~\ref{sec:xp} by deriving an expression for the CD Hamiltonian \eqref{q01} for the scale-invariant driving of a quantum system with one degree of freedom. Section~\ref{sec:gen} is dedicated to classical Hamiltonian dynamics, in which the classical version of  $\hat{H}_1$ can be rewritten in a local form using linear canonical transformations and the formalism of generating functions. These findings will be generalized and applied in Sec.~\ref{sec:multi} to a broad family of many-body quantum systems.  Specific protocols for arbitrary trapping potentials will be discussed in Sec.~\ref{sec:trap}. Section~\ref{sec:nonlin} is dedicated to nonlinear systems, with emphasis in mean-field theories. In Sec.~\ref{sec:scale},  we will discuss the relation of CD to more general scaling laws, before we explicitly engineer STA in Sec.~\ref{sec:eng}. We close with a summary and discussion in Sec.~\ref{sec:con}.

\section{Counterdiabatic Hamiltonian for scale-invariant driving\label{sec:xp}}

Generally, it appears to be hardly feasible to find closed form expressions, i.e., expressions that do not depend on the full spectral decomposition of $\hat{H}_0(t)$, for the auxiliary term in the CD Hamiltonian $\hat{H}(t)$  \eqref{q01}. Recently, it has been shown that scale-invariance greatly facilitates this task for processes that describe self-similar expansions and compressions in a time-dependent trap \cite{campo_2013}, including the family of power-law potentials as a special case \cite{jarzynski_2013}. More generally, scale-invariant driving refers to transformations of the system Hamiltonian associated with  a set of external control parameters $\mb{\lambda}(t)=(\lambda_1(t),\dots,\lambda_n(t))$ which can be absorbed by scaling of coordinates, time, energy, and possibly other variables to rewrite the transformed Hamiltonian in its original form up to a multiplicative factor. If only the potential term $\U(q,\mb{\lambda}(t))$ is modulated, its overall shape does not change under $\mb{\lambda}(0)\rightarrow\mb{\lambda}(t)$. For the time being, we focus on a quantum system with a single degree of freedom,
\begin{equation}
\label{q02}
\hat{H}_0(t)=\frac{p^2}{2m}+\U(q,\mb{\lambda}(t))=\frac{p^2}{2m}+\frac{1}{\gamma^2}\,\U_0\left(\frac{q-f}{\gamma}\right)\,,
\end{equation}
where $\mb{\lambda}=(\gamma,f)$ and $\U_0(q)=\U(q,\mb{\lambda}(0))$. Note that generally $\gamma=\gamma(t)$ and $f=f(t)$ are both allowed to be time-dependent, but we assume that they are independent of each other. This time-dependence encompasses transport processes ($\gamma(t)=1$), dilations (such as an expansion or compression, with $f(t)=0$) and combined dynamics, which are the focus of our attention and elements of the dynamical group of the system Hamiltonian, the universal covering group of $SU(1,1)$, $\overline{SU(1,1)}$ \cite{gambardella_1975}.

Our goal is to rewrite the auxiliary term $\hat{H}_1(t)$ \eqref{q01} into a form that does not rely on the spectral decomposition of $\hat{H}_0(t)$. Let $\psi_n^0(q)=\braket{n}{q}$ be an eigenfunction of the Hamiltonian $\hat{H}_0(\gamma=1,f=0)$, then $\psi_n(q,\gamma,f)=\alpha(\gamma)\,\psi_n^0\left((q-f)/\gamma\right)$ is an eigenfunction of $\hat{H}_0(\gamma,f)$, where $\alpha(\gamma)=1/\sqrt{\gamma}$ is a normalization constant. The proof of this statement can be found in appendix \ref{sec:scalef}.

Now, we want to use this symmetry to simplify $\hat{H}_1(t)$ in Eq.~\eqref{q01}. We have,
\begin{equation}
\label{q04}
\hat{H}_1(t)=i\hbar \dot{\mb{\lambda}}\cdot \,\sum\limits_m\,\left(\ket{\nabla_{\mb{\lambda}}\,m}\bra{m}-\braket{m}{\nabla_{\mb{\lambda}}\,m}\ket{m}\bra{m}\right)\,,
\end{equation}
which reads in space representation
\begin{equation}
\label{q05}
\begin{split}
\hat{H}_1(t)&=i\hbar\dot{\mb{\lambda}}\cdot\,\sum\limits_m\,\int\td q\,\ket{q}\,\nabla_{\mb\lambda}\,\psi_m(q,\mb{\lambda})\,\bra{m}\\
&\quad-i\hbar\dot{\mb{\lambda}}\cdot\,\sum\limits_m\,\int\td q\,\braket{m}{q}\,\nabla_{\mb\lambda} \psi_m(q,\mb{\lambda}) \ket{m}\bra{m}\,. 
\end{split}
\end{equation}
To simplify this expression,  we note that
\begin{equation}
\label{q07}
\begin{split}
&\nabla_{\mb{\lambda}}\psi_n(q,\mb{\lambda})=\\
&\frac{\alpha'(\gamma)}{\alpha(\gamma)}\,\psi_n(q,\mb{\gamma})-\frac{q-f}{\gamma}\,\pd_q\psi_n(q,\gamma),\,\,-\pd_q\psi_n(q,\mb{\gamma})\,.
\end{split}
\end{equation}
For the sake of clarity, let us treat both terms of $\hat{H}_1(t)$ in \eqref{q05} separately. We obtain for the first term
\begin{equation}
\label{q08}
\begin{split}
&i\hbar\dot{\mb{\lambda}}\cdot\,\sum\limits_m\,\int\td q\,\ket{q}\,\nabla_{\mb\lambda}\,\psi_m(q,\mb{\lambda})\,\bra{m}\\
&\quad=\frac{\dot{\gamma}}{\gamma}\,\left(q-f\right)p+i\hbar\dot{\gamma}\,\frac{\alpha'(\gamma)}{\alpha(\gamma)}+\dot{f} \,p\,,
\end{split}
\end{equation}
while the second term reduces to
\begin{equation}
\label{q09}
\begin{split}
&-i\hbar\dot{\mb{\lambda}}\cdot\,\sum\limits_m\,\int\td q\,\braket{m}{q}\,\nabla_{\mb\lambda} \psi_m(q,\mb{\lambda}) \ket{m}\bra{m}\\
&\quad\quad\quad=-\frac{i\hbar\dot{\gamma}}{2\gamma}-i\hbar\dot{\gamma}\,\frac{\alpha'(\gamma)}{\alpha(\gamma)}\,.
\end{split}
\end{equation}
Note that the second component of $\nabla_{\mb{\lambda}}\psi_n(q,\mb{\lambda})$ does not contribute, since the wavefunction vanishes at infinity due to normalizability. In conclusion, we obtain the explicit expression of  the auxiliary CD Hamiltonian,
\begin{equation}
\label{q10}
\hat{H}_1(t)=\frac{\dot{\gamma}}{2\gamma}\left[\left(q-f\right)\,p+p\,\left(q-f\right)\right]+\dot{f}\,p\,,
\end{equation}
where we used $\com{q-f}{p}=i\hbar$. Notice that $\hat{H}_1(t)$ in Eq.~\eqref{q10} is of the general form $\hat{H}_1\propto (qp+pq)$, 
which was found for a time-dependent harmonic-trap \cite{muga_2010}  and more generally in Refs.~\cite{jarzynski_2013,campo_2013} for the class of potentials
\begin{equation}
\label{q11}
\U(q,\gamma(t))=\frac{A}{\gamma^2}\,\left(\frac{q}{\gamma}\right)^b\,,
\end{equation}
where $b\in\{2,4,6,\dots\}$, and $A>0$. See as well \cite{jarzynski_2013,dimartino_2013} for a discussion of the limiting case $b\rightarrow \infty$, that of a box-like confinement. Obviously this class \eqref{q11} belongs to the more general scale-invariant potentials introduced above in Eq. \eqref{q02}.

Equation~\eqref{q10} is our first main result. For all driving protocols under which the original Hamiltonian $\hat{H}_0(t)$ is scale-invariant, i.e., where the time-dependent potential is of the form \eqref{q02}, the auxiliary term $\hat{H}_1(t)$ takes the closed form \eqref{q10}. In particular, $\hat{H}_1(t)$ is independent of the explicit energy eigenfunctions, and only depends on the anticommutator, $\hat{H}_1\propto \{q,p\}=qp+pq$, the generator of dilations. 
As a result, CD applies not only to single eigenstates, but also to non-stationary quantum superpositions and mixed states.
However, the expression \eqref{q10} is still not particularly practical as non-local  Hamiltonians \footnote{Hamiltonians that include products of space and momentum operator, $q$ and $p$, are \textit{non-local}, whereas \textit{local} Hamiltonians contain only terms that depend on at most sums of $q$ and $p$.} are hard to realize in the laboratory. We continue our analysis by explicitly constructing coordinate transformations, which allow us to write $\hat{H}_1(t)$ in local form, i.e., where $\hat{H}_1(t)$ depends only on position. 
In order to do so we will use the classical version of CD as a guide.

\section{Scale-invariant driving -- a case for generating functions\label{sec:gen}}

We now turn to \textit{dissipationless classical driving} \cite{jarzynski_2013,deng_2013}, the classical analogue of quantum counterdiabatic driving. For scale-invariant Hamiltonians the connection between the quantum and classical cases is particularly close, and the corresponding auxiliary CD terms $\hat{H}_1(t)$ and $H_1(t)$ are essentially identical, up to quantization.

In complete analogy with the quantum case we consider a classical Hamiltonian with one degree of freedom,
\begin{equation}
\label{classicalH0}
H_0(z,t) = H_0(z;\mb{\lambda}(t)) = \frac{p^2}{2m}+\U(q,\mb{\lambda}(t)) \, ,
\end{equation}
where $z=(q,p)$ is a point in phase space. The classical adiabatic invariant is given by
\begin{equation}
\label{eq:adiabaticInvariant_omega}
\omega(z,\mb{\lambda}) = \Omega\left( H_0(z,\mb{\lambda}),\mb{\lambda} \right) \, ,
\end{equation}
where
\begin{equation}
\label{eq:adiabaticInvariant_E}
\Omega(E,\mb{\lambda}) = \int\td z\,\Theta(E-H_0(z,\mb{\lambda}))
\end{equation}
is the volume of phase space enclosed by the energy shell $E$ of $H_0(z,\mb{\lambda})$.
In the adiabatic limit, $\omega(z(t),\mb{\lambda}(t))$ remains constant along a Hamiltonian trajectory $z(t)$ evolving under $H_0(z,\mb{\lambda}(t))$, just as the quantum number $n$ remains constant in the quantum case.
We now consider non-adiabatic driving of the parameters $\mb{\lambda}(t)$, and we seek an auxiliary CD term
\begin{equation}
\label{eq:class_aux_CD}
H_1(z,t)=\dot{\mb{\lambda}} \cdot \mb{\xi}(z,\mb{\lambda}(t))\,,
\end{equation}
resembling Eq.~\eqref{q04}, such that $\omega$ remains constant at arbitrary driving speed, for any trajectory evolving under the Hamiltonian $H(z,t)=H_0(z,\mb{\lambda}(t)) + H_1(z,t)$.

It is useful to picture dissipationless driving in terms of an ensemble of trajectories evolving under $H(z,t)$, with initial conditions sampled from an energy shell $E(0)$ of $H_0(z;\mb{\lambda}(0))$.
Since the value of $\omega$ is preserved for every trajectory in this ensemble, at any later time $t>0$ these trajectories populate a single energy shell $E(t)$ of $H_0(z;\mb{\lambda}(t))$, determined by the condition $\Omega(E(t),\mb{\lambda}(t)) = \Omega(E(0),\mb{\lambda}(0))$, which defines the \textit{adiabatic energy shell}.

As discussed in \cite{jarzynski_2013}, it is useful to view $\mb{\xi}(z,\mb{\lambda})$ as a generator of infinitesimal transformations $z \rightarrow z + {\rm d}z$, with
\begin{equation}
\label{eq:dz_generator}
{\rm d}z = {\rm d}\mb{\lambda} \cdot \{z,\mb{\xi}\} \, ,
\end{equation}
where $\{A,B\} = \partial_qA\cdot\partial_pB - \partial_pA\cdot\partial_qB$ is the Poisson bracket.
Equation~\eqref{eq:dz_generator} provides a rule for converting a small change of parameters, ${\rm d}\mb{\lambda}$, into a small displacement in phase space, ${\rm d}z$.
In order to achieve dissipationless classical driving, the energy shells of $H_0(\mb{\lambda})$ must be mapped, under Eq.~\eqref{eq:dz_generator}, onto those of $H_0(\mb{\lambda} + {\rm d}\mb{\lambda})$ with
\begin{equation}
\label{eq:condition}
\omega(z + {\rm d}z,\mb{\lambda} + {\rm d}\mb{\lambda}) = \omega(z,\mb{\lambda}) \, .
\end{equation}
When this condition is satisfied, the term $H_1 = \dot{\mb{\lambda}}\cdot\mb{\xi}$ provides precisely the counterdiabatic driving required to preserve the value of $\omega$.
Thus, to construct the CD Hamiltonian, we must find the function $\mb{\xi}(z,\mb{\lambda})$ that generates infinitesimal deformations of the adiabatic energy shell, as per Eqs.~\eqref{eq:dz_generator} and \eqref{eq:condition}.

Our scale-invariant Hamiltonian
\begin{equation}
H_0(z;\gamma,f) = \frac{p^2}{2m}+\frac{1}{\gamma^2}\,\U_0\left(\frac{q-f}{\gamma}\right)
\end{equation}
satisfies
\begin{equation}
\begin{split}
H_0(q+a,p;\gamma,f+a) &= H_0(q,p;\gamma,f)\,, \\
H_0\Bigl(rq , \frac{p}{r} ; r\gamma , rf\Bigr) &= \frac{1}{r^2} H_0(q,p;\gamma,f)\,, \\
\Omega(E,\gamma,f) &= \Omega(\gamma^2E,1,0)\,,
\end{split}
\end{equation}
for any real $a$ and positive $r$.
Using these properties we can verify by direct substitution that the canonical mapping
\begin{equation}
(q,p) \rightarrow \Bigl( q + {\rm d}f + \frac{{\rm d}\gamma}{\gamma} (q-f) , p - \frac{{\rm d}\gamma}{\gamma} p \Bigr)
\label{dqp}
\end{equation}
satisfies Eq.~\eqref{eq:condition}.
The change $f \rightarrow f+{\rm d}f$ produces a coordinate translation, while under the change $\gamma \rightarrow \gamma + {\rm d}\gamma$, the adiabatic energy shell is stretched along the coordinate $q-f$ and compressed along the momentum $p$.
The infinitesimal transformation  \eqref{dqp} is generated by
\begin{equation}
\label{eq:xi_scaling}
\xi_\gamma = \frac{(q-f)p}{\gamma}\,, 
\quad
\xi_f = p  \, ,
\end{equation}
as verified by substitution into Eq.~\eqref{eq:dz_generator}, with $\mb{\lambda}=(\gamma,f)$ and $\mb{\xi} = (\xi_\gamma,\xi_f)$.
Combining Eqs.~\eqref{eq:class_aux_CD} and \eqref{eq:xi_scaling} we arrive at
\begin{equation}
\label{q13}
H_1(z,t)=\frac{\dot{\gamma}}{\gamma}\left(q-f\right)\,p+\dot{f}\,p\,,
\end{equation}
the classical counterpart of (\ref{q10}).
With this auxiliary  term, the value of $\omega(z(t),\mb{\lambda}(t))$ remains constant along a trajectory evolving under the Hamiltonian $H=H_0 + H_1$, for any protocol $\mb{\lambda}(t)$.  To illustrate this general result, we derive $H_1(t)$ for an analytically solvable example, namely the parametric Morse oscillator in appendix \ref{sec:morse}.

Equation~\eqref{q13} gives us a \textit{nonlocal} CD Hamiltonian that accomplishes dissipationless classical driving.
Our goal now is to find a coordinate transformation mapping $(q,p)$ to a set of new variables $(\bar{q},\bar{p})$, and a corresponding Hamiltonian $\bar{H}(\bar{q},\bar{p},t)$ whose dynamics 
(in $q$-space) is identical to that under $ H(q,p,t)$, and for which $\bar{H}(\bar{q},\bar{p},t)$ is {\it local}, i.e. it is the sum of a kinetic energy term $\sim\bar{p}^2$ and a function $\bar{\U}(\bar{q},t)$. In classical mechanics such problems can be elegantly solved by the formalism of generating functions \cite{goldstein_1959}.

We briefly recall the main idea. Let $h_1(q_1,p_1,t)$ be a time-dependent Hamiltonian, written in terms of coordinates $(q_1,p_1)$ in a two-dimensional phase space.
Now consider new coordinates $(q_2,p_2)$ that are related to $(q_1,p_1)$ by a time-dependent canonical transformation:
\begin{equation}
q_2 = q_2(q_1,p_1,t) \quad,\quad p_2 = p_2(q_1,p_1,t) \, .
\end{equation}
Since canonical transformations are invertible, we can alternatively express the ``old'' coordinates $(q_1,p_1)$ as functions of the ``new'' ones, $(q_2,p_2)$.
If a function $F(q_1,p_2,t)$ can be constructed such that the relationship between the two coordinate sets is given by
\begin{equation}
\label{q14}
p_1=\frac{\pd F}{\pd q_1}, \hspace{1em} q_2=\frac{\pd F}{\pd p_2},\hspace{1em} 
\end{equation}
then $F(q_1,p_2,t)$ is called a generating function of a type-2 canonical transformation \cite{goldstein_1959}.
We can then define a Hamiltonian
\begin{equation}
\label{eq:h2h1}
h_2(q_2,p_2,t) = h_1 + \frac{\pd F}{\pd t}
\end{equation}
that generates trajectories \textit{equivalent} to those of $h_1(q_1,p_1,t)$.
By this we mean that solutions to the equations
\begin{equation}
\label{q15}
\dot{q}_1=\frac{\pd h_1}{\pd p_1},\hspace{1em} \dot{p}_1=-\frac{\pd h_1}{\pd q_1}\,,
\end{equation}
when rewritten in the new coordinates become solutions to
\begin{equation}
\label{q16}
\dot{q}_2=\frac{\pd h_2}{\pd p_2},\hspace{1em} \dot{p}_2=-\frac{\pd h_2}{\pd q_2}\,.
\end{equation}
The function $F(q_1,p_2,t)$ thus encodes the transformation of both the variables \eqref{q14} and the Hamiltonian \eqref{eq:h2h1}.

In what follows we shall apply this approach to three different sets of coordinates related by canonical transformations.
The CD Hamiltonian for scale-invariant dynamics with the non-local term \eqref{q13} reads
\begin{equation}
\label{q17}
H(q,p,t)=\frac{p^2}{2 m}+\frac{1}{\gamma^2}\,\U_0\left(\frac{q-f}{\gamma}\right)+\frac{\dot{\gamma}}{\gamma}\left(q-f\right)\,p+\dot{f}\,p\,.
\end{equation}
Now we define a type-2 generating function
\begin{equation}
\label{q20}
F(q,\bar{p},t)=q(\bar{p}-m\dot{f})-\frac{m}{2}\frac{\dot{\gamma}}{\gamma}(q-f)^2+\frac{m}{2}\int^t_0\td s \dot{f}^2\,,
\end{equation}
and we use it with \eqref{q14} and \eqref{eq:h2h1} to construct a canonical transformation to coordinates $(\bar{q},\bar{p})$, obtaining
\begin{equation}
\label{q21}
\bar{q}=q\quad , \quad \bar{p}=(p+m\dot{f})+m\frac{\dot{\gamma}}{\gamma}(q-f)\,,
\end{equation}
and
\begin{equation}
\label{q18}
\bar{H}(\bar{q},\bar{p},t)=\frac{\bar{p}^2}{2 m}+\frac{1}{\gamma^2}\,\U_0\left(\frac{\bar{q}-f}{\gamma}\right)-\frac{m}{2}\frac{\ddot{\gamma}}{\gamma}\,(\bar{q}-f)^2-m\ddot{f}\bar{q}\,.
\end{equation}
The Hamiltonians $H(q,p,t)$ and $\bar{H}(\bar{q},\bar{p},t)$ generate equivalent trajectories, in the sense of Eqs.~\eqref{q15} and \eqref{q16}.
Moreover, since $q=\bar{q}$ these trajectories are identical in \textit{configuration space}.
This can be verified independently by considering the second-order differential equation for the coordinates  $q$ and $\bar{q}$. In either case we have,
\begin{equation}
\label{q22}
m\ddot{q}=-\frac{1}{\gamma^3}\,\U_0'[(q-f)/\gamma] +m \frac{\ddot{\gamma}}{\gamma}\left(q-f\right)+m \ddot{f}\,.
\end{equation}

Comparing $H$ \eqref{q17} and $\bar{H}$ \eqref{q18}, we see that the non-local terms in the former are replaced by local terms in the latter, which we conceptually identify as a local formulation of $H_1(t)$ \eqref{q13}. The first of these new terms in $\bar{H}$ is an inverted harmonic oscillator whose stiffness is proportional to the acceleration of the scaling factor, cf. also Refs.~\cite{campo_boshier_2012,campo_2013}. The second term is the classical analog of a Duru transformation in transport processes \cite{duru_1989}. 

\begin{figure}
\includegraphics[width=0.98\linewidth]{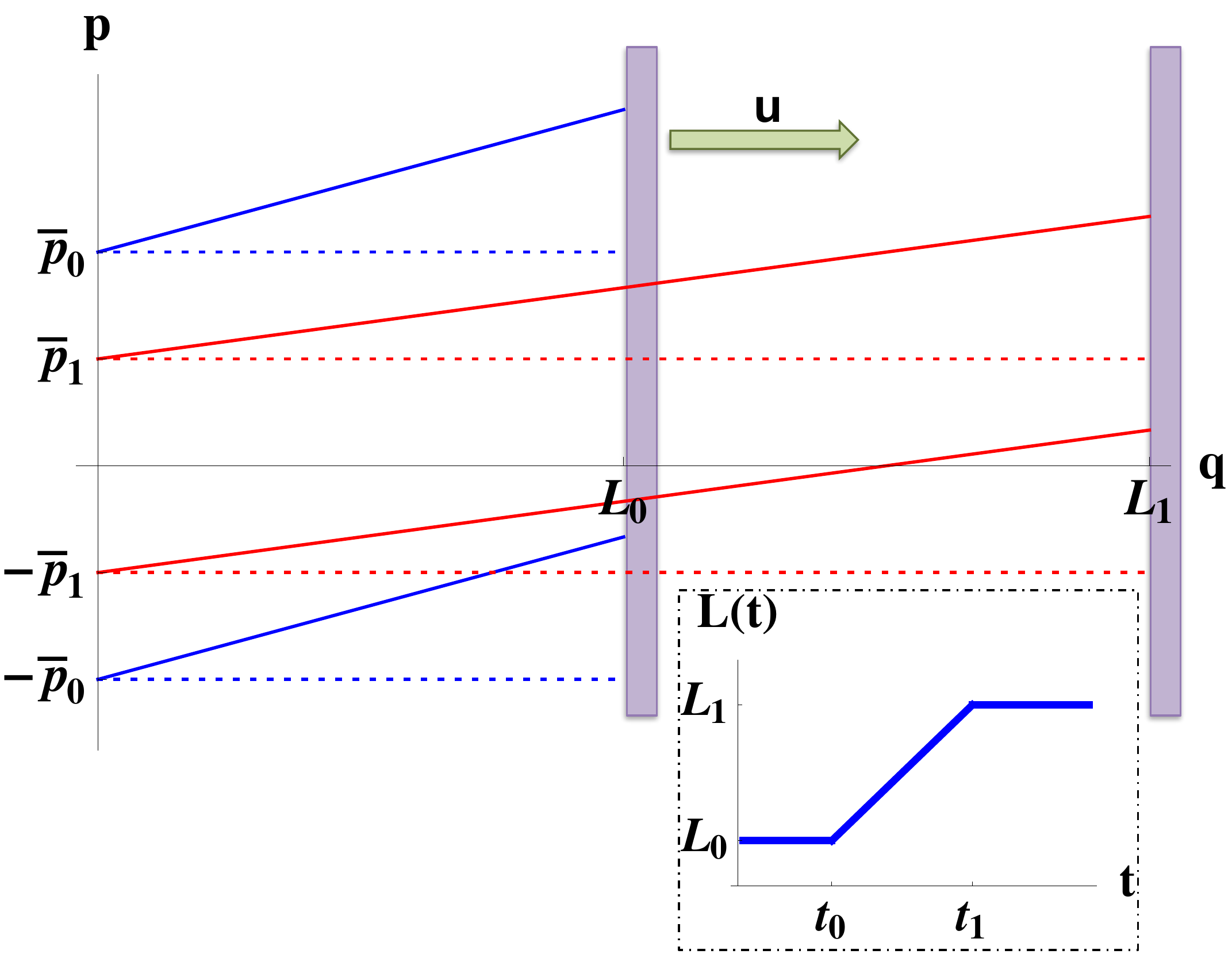}
\caption{\label{fig_box} {\bf Shortcut to adiabaticity based on dissipationless classical driving.} 
 Energy shells for a particle in a one-dimensional box \eqref{q32}, in a time-dependent piston of width $L(t)$ which  changes at constant rate $u$ for $t_0\leq t\leq t_1$ (inset). Energy shells corresponding to $H_0(\bar{z};L)$ are shown as a pair of parallel, dotted line segments of length $L$, at momenta $\pm\bar{p}$. The solid lines represent a level surface of the adiabatic invariant $I$ corresponding to the full, counter-adiabatic Hamiltonian $H(\bar{z};L)$ \eqref{q33}. At $t=t_0$ the force $f(q,t)$ \eqref{q34} induces a ``jump'' of trajectories $\bar{z}$ from dashed to solid line, for $t_0\leq t\leq t_1$ the adiabatic energy shell is deformed invariantly, and finally at $t=t_1$ force $f(q,t)$ \eqref{q34} induces jumps back to $\bar{p}$.}
\end{figure}

Now consider a protocol in which the parameters $f$ and $\gamma$ are fixed outside some interval $t_0 \le t \le t_1$ (as in the inset of Fig.~\ref{fig_box}), and imagine  trajectories $z(t)$ and $\bar{z}(t)$ that evolve under $H$ and $\bar{H}$, respectively, from identical initial conditions at $t<t_0$.
These equivalent trajectories are related by \eqref{q21} at every instant in time.
This immediately implies that $z(t)$ and $\bar{z}(t)$ are identical in \textit{phase space} for $t<t_0$, then their momenta diverge during the interval $t_0 \le t \le t_1$, and finally they meet again at $t=t_1$ and remain identical thereafter.
Since the adiabatic invariant $\omega$ is preserved exactly along the trajectory $z(t)$, it follows that along the trajectory $\bar{z}(t)$ the initial value of $\omega$ (at $t<t_0$) is identical to the final values of $\omega$ (at $t>t_1$), even if it varies at intermediate times.
Thus the local Hamiltonian \eqref{q18} provides a shortcut to adiabaticity, provided the parameters $f$ and $\gamma$ are fixed initially and finally.

To gain further insight, let us construct a new canonical transformation,  to variables $(\tilde{q},\tilde{p})$, using
\begin{equation}
\label{q23}
F(q,\tilde{p},t)=\frac{1}{\gamma}\left(q-f\right)\tilde{p} \,.
\end{equation}
Applying \eqref{q14} and \eqref{eq:h2h1} we get
\begin{equation}
\label{q24}
\tilde{q}=\frac{q-f}{\gamma}\quad ,\quad \tilde{p}=\gamma p\, ,
\end{equation}
and 
\begin{equation}
\label{q19}
\tilde{H}(\tilde{q},\tilde{p},t)=\frac{1}{\gamma^2}\left[\frac{\tilde{p}^2}{2m}+\U_0(\tilde{q})\right]\,.
\end{equation}
The transformation \eqref{q24} is a linear dilation of the coordinate and the reciprocal contraction in momentum space.

The fact that $\tilde{H}(\tilde{q},\tilde{p},t)$ is time-\textit{in}dependent, apart from the factor $1/\gamma^2$, has two interesting consequences. First, the quantity $I(\tilde{q},\tilde{p})=\gamma^2 \tilde{H}$ is a dynamical invariant, as follows from direct inspection of Hamilton's equations. 
If we picture a level surface of $I$ as a closed loop in $\tilde{z}$-space, then under $\tilde{H}$ a trajectory $\tilde{z}(t)$ simply evolves round and round this loop, at a speed proportional to $1/\gamma(t)^2$.
The function $I$ can be expressed in any of the three sets of phase space coordinates considered above. The resulting functions
\begin{equation}
\label{q27}
\begin{split}
I(q,p,t)&=\gamma^2 \frac{p^2}{2 m}+\U_0\left(\frac{q-f}{\gamma}\right)\,,\\
I(\bar{q},\bar{p},t)&=\frac{\gamma^2}{2m}\left[\bar{p}-m\frac{\dot{\gamma}}{\gamma}(\bar{q}-f)-m\dot{f}\right]^2+\U_0\left(\frac{\bar{q}-f}{\gamma}\right)\,,\\
I(\tilde{q},\tilde{p})&=\frac{\tilde{p}^2}{2 m}+\U_0(\tilde{q})\,,
\end{split}
\end{equation}
are all dynamical invariants, along Hamiltonian trajectories generated by $H(z,t)$,  $\bar{H}(\bar{z},t)$ and $\tilde{H}(\tilde{z},t)$,  respectively.
This follows from the equivalence of the trajectories $z(t)$, $\bar{z}(t)$, and $\tilde{z}(t)$, but it can also be verified by inspection of Hamilton's equations.

The invariance of $I$ allows us to visualize the evolution of these trajectories, as each one clings to a level surface of $I$ expressed in the given phase space coordinates.
If $f$ is not varied with time, then a level surface of $I(q,p,t)$ gets stretched along $q$ and contracted along $p$ as $\gamma$ increases with time (or the other way around if $\gamma$ decreases); and a level surface of $I(\bar{q},\bar{p},t)$ additionally acquires a {\it shear} along the momentum direction, proportional to $\dot\gamma$, as illustrated by the pairs of diagonal lines in Fig.~\ref{fig_box}.
If $f$ is varied with time, then a level surface of $I(q,p,t)$ undergoes translation along the coordinate $q$, and level surface of $I(\bar{q},\bar{p},t)$ additionally undergoes a displacement along $p$ by an amount $m\dot f$.

Second, if we introduce the new time-like variable \cite{berry_klein_84}
\begin{equation}
\label{q28}
\tau(t)=\int_0^t \td s\,\gamma^{-2}(s)\,,
\end{equation}
we obtain
\begin{equation}
\label{q29}
\frac{\td \tilde{q}}{\td \tau}=\frac{\tilde{p}}{m}\quad \mathrm{and}\quad\frac{\td \tilde{p}}{\td \tau}=-\U_0'(\tilde{q})\,,
\end{equation}
which describe motion under a time-independent Hamiltonian, whose energy shells are the level surfaces of $I(\tilde{q},\tilde{p})$.
Let $(\tilde{q}(\tau),\tilde{p}(\tau))$ denote a particular solution to these equations of motion. Inverting the canonical transformations in Eqs.~\eqref{q24} and \eqref{q25}, we can immediately use this solution to construct trajectories generated by the Hamiltonians $H$ and $\bar{H}$, namely:
\begin{equation}
\label{q30}
q(t)=\gamma \tilde{q}(\tau),\quad p(t)=\frac{1}{\gamma}\,\tilde{p}(\tau)
\end{equation}
and
\begin{equation}
\label{q31}
\bar{q}(t)=\gamma \tilde{q}(\tau)+f,\quad \bar{p}(t)=\frac{1}{\gamma}\,\tilde{p}(\tau)+m\dot{\gamma}\tilde{q}(\tau)+m\dot{f}\,.
\end{equation}
Hence, trajectories generated by the time-dependent Hamiltonians $H$ and $\bar{H}$ can be constructed directly from trajectories evolving under a time-independent Hamiltonian $\tilde{H}$.
This further emphasizes the equivalence between these trajectories.
We will exploit these observations in the following discussion of shortcuts for time-dependent multi-particle quantum systems.

Energy-like dynamical invariants such as $I$ were intensely studied in the mathematical literature for classical and quantum dynamics. In particular, it can be shown that if (and only if) an energy-like invariant exists, then one can find a coordinate transformation as discussed in the present analysis  \cite{lewis_1968,lewis_1969a,gunther_1977,leach_1977,leach_1977b,leach_1977a,mostafazadeh_2001,lohe_2009}.

For completeness, we note that the transformation from $(\bar{q},\bar{p})$ to $(\tilde{q},\tilde{p})$ is generated by the function
\begin{equation}
\label{q25}
\begin{split}
F(\bar{q},\tilde{p},t)=&\,\,\frac{1}{\gamma}(\bar{q}-f)(\tilde{p}+m\gamma f)+\frac{m}{2}\frac{\dot{\gamma}}{\gamma}(\bar{q}-f)^2\\
&+\frac{m}{2}\int^t_0\td s\,(\dot{\gamma}^2+2\ddot{\gamma}\gamma)\,,
\end{split}
\end{equation}
for which we have
\begin{equation}
\label{q26}
\tilde{q}=\frac{1}{\gamma}(\bar{q}-f)\quad \mathrm{and}\quad \tilde{p}=\gamma(\bar{p}-m\dot{f})-m\dot{\gamma}(\bar{q}-f)\,.
\end{equation}

\paragraph*{An illustrative example -- particle in time-dependent box}
For a particle in a time-dependent box the form of the new Hamiltonian $\bar{H}(\bar{q},\bar{p},t)$ \eqref{q18} can be understood intuitively. Consider a particle of mass $m$ inside a one-dimensional box with hard walls at $q=0$ and $q=L$, as described by the Hamiltonian
\begin{equation}
\label{q32}
H_0(z;L)=\frac{p^2}{2m}+\U_\mrm{box}(q; L)\,,
\end{equation}
where $U_\mrm{box}(q;L)$ is zero inside the box, and ``infinite'' outside. We further assume that $L=L(t)$ changes with constant rate $u$ for times $t_0\leq t\leq t_1$ and is constant otherwise with $L(t\leq t_0)=L_0$ and $L(t\geq t_1)=L_1$, cf. Fig.~\ref{fig_box}.
Now imagine the aforementioned adiabatic energy shell as a closed loop that is deformed as $L(t)$ is varied with time. Then $H_0(z;L)$ generates motion around this loop, and the auxiliary CD term $H_1(t)=m\dot{L} \cdot \xi=mu\, \xi$ adjusts each trajectory $z$ so that it remains on-shell \cite{jarzynski_2013}, see Fig.~\ref{fig_box}. The dashed lines represent the adiabatic energy shells corresponding to $H_0$ for a particular energy $E$. Notice that particles at $q=L$ hit the hard wall, and are ``boosted'' from one branch to the other. In other words, particles hitting the hard wall with momentum $p$ are reflected at $q=L$ and travel back with $-p$, and so close the loop. The solid lines, $p+mu\,q/L$ and $-p+mu\,q/L$ represent a level surface of the adiabatic invariant $I$ corresponding to the full counterdiabatic Hamiltonian
\begin{equation}
\label{q33}
H(z;L)=\frac{p^2}{2m}+\U_\mrm{box}(q;L)+\frac{u}{L}\,qp\,.
\end{equation}
In the previous discussion we were asking for a set of coordinates $(\bar{q},\bar{p})$ and the corresponding Hamiltonian $\bar{H}(\bar{q},\bar{p},t)$, for which here the solid lines represent the exact solution.
For times $t< t_0$ and $t>t_1$ the energy shells for old and new coordinates are identical. This means that at $t=t_0$ the trajectories $z$ have to ``jump'' from $p=\pm\sqrt{2m E}$ to $\bar{p}=\pm p+mu\, q/L$, where $\bar{q}\equiv q$, and at $t=t_1$ back to the \textit{unperturbed} shell. These jumps are induced by a force
\begin{equation}
\label{q34}
\mathfrak{f}(\bar{q},t)=m\frac{\bar{q} u}{L_0}\,\de{t-t_0}-m\frac{\bar{q} u}{L_1}\,\de{t-t_1}\,,
\end{equation}
which applies ``impulses'' at $t=t_0$ and $t=t_1$. The latter force is the derivative of an auxiliary potential, $\mathfrak{f}(\bar{q},t)=-\pd_{\bar{q}} \U_1(\bar{q},t)$, 
\begin{equation}
\label{q35}
\begin{split}
\U_1(\bar{q},t)&=-\frac{m}{2}\frac{\bar{q}^2 u}{L_0}\,\de{t-t_0}+\frac{m}{2}\frac{\bar{q}^2 u}{L_1}\,\de{t-t_1}\,,\\
&=-\frac{m}{2}\frac{\ddot{L}(t)}{L(t)}\,\bar{q}^2\,,
\end{split}
\end{equation}
which we recognize as the additional potential term in the transformed Hamiltonian $\bar{H}(\bar{q},\bar{p},t)$ \eqref{q18}, with $\gamma(t)=L(t)/L(0)$ and $f(t)=0$.

Therefore, we conclude that the additional harmonic term in the Hamiltonian \eqref{q18} with possible negative spring constant $-m\ddot{\gamma}/\gamma$ is nothing else but the term necessary to facilitate the transfer of the classical trajectories from the energy shells of $H_0$ to those invariant under $H$ \eqref{q17} and eventually $\bar{H}$ \eqref{q18}. 
Interestingly enough, this result agrees with the CD derived in the quantum case for a time-dependent box-like confinement using Lewis-Riesenfeld invariants and reverse engineering of scaling laws \cite{campo_boshier_2012}.

\section{Multi-particle quantum systems\label{sec:multi}}

In the previous section we showed how the auxiliary, classical  term in the counterdiabatic Hamiltonian can be brought into a local form. 
We will next apply this finding to general multi-particle quantum systems. Let us consider the  broad family of many-body systems described by the Hamiltonian
\begin{equation}
\label{q36}
\hat{\mathcal{H}}_0=
\sum_{i=1}^N\left[-\frac{\hbar^2}{2m}\Delta_{\q_i}\!
+\U(\q_i,\mb{\lambda}(t))\right]+\epsilon(t)\sum_{i<j}V(\q_i-\q_j)\,,
\end{equation}
with ${\q_i}\in\mathbb{R}^D$ unless stated otherwise ($D$ denoting the effective dimension of the system), and where $\Delta_{\q_i}$ is the Laplace operator, and $\U(\q,t)$ represents an external trap  whose time-dependence is of the form \eqref{q02}, $\U(\q,\mb{\lambda}(t))=\U_0\left[\left(\q-\f(t)\right)/\gamma(t)\right]/\gamma(t)^2$. As before \eqref{q02}, the trap can be shifted by the time-dependent displacement $\f=\f(t)$ and simultaneously modulated by the scaling factor $\gamma=\gamma(t)$. We further assume that the two-body interaction potential obeys 
\begin{equation}
\label{q37}
{\rm V}(\kappa \q)=\kappa^{-\alpha}\,{\rm V}(\q),
\end{equation}
which includes relevant examples in ultracold gases such as  the  pseudo-potential for contact interactions \cite{yurovsky_2008}, e.g., the Fermi-Huang potential for s-wave scattering for which $\alpha=D$ \cite{huang_1996}.

We define the dimensionless coupling constant $\epsilon(0)=1$ at $t=0$ and consider   
 a stationary state $\Psi(t=0)=\Psi(\q_1,\dots,\q_N;t=0)$, with chemical potential $\mu$, i.e.,  $\hat{\mathcal{H}}_0\Psi=\mu\Psi$. The scale-invariant solution for this multi-particle quantum system, that generalizes the wavefunction for a single degree of freedom discussed earlier, reads
\begin{equation}
\label{q38}
\Psi(t)=\gamma^{-ND/2}
e^{-i\mu \tau/\hbar}
\Psi\left[\frac{\q_1-\f(t)}{\gamma(t)},\dots,\frac{\q_N-\f(t)}{\gamma(t)};0\right],
\end{equation}
where $\tau$ is the time-like variable introduced above \eqref{q28}. 
By substituting the latter ansatz into the many-body Schr\"odinger equation  we find that $\Psi(t)$ is actually the exact time-dependent solution  for the dynamics generated by CD Hamiltonian
\begin{equation}
\begin{split}
\label{q39}
&\gamma^2\hat{\mathcal{H}}=
\sum_{i=1}^N\left[-\frac{\hbar^2}{2m}\Delta_{\tilde{\q}_i}+\U_0(\tilde{\q}_i)\right]\\
&\quad+\epsilon\gamma^{2-\alpha}\sum_{i<j}V(\tilde{\q}_i-\tilde{\q}_j)\\
 &\quad +\sum_{i=1}^N\left[-i\frac{\hbar \partial_{\tau}\f}{\gamma}\cdot\partial_{\tilde{\q}_i}-i\frac{\hbar \partial_{\tau}\gamma}{2\gamma}(\tilde{\q}_i\cdot \partial_{\tilde{\q}_i}+\partial_{\tilde{\q}_i}\cdot\tilde{\q}_i)\right],
\end{split}
\end{equation}
where the scaled spatial coordinate  reads $\tilde{\q}_i=\left(\q_i-\f(t)\right)/\gamma$, as before. 
The  scale-invariant solution to a related classical and restricted problem with $\f(t)=0$ was derived by Perelomov \cite{perelomov_1978}. 
We observe that in an interacting system (with $V\neq 0$), there is an additional consistency condition for the dynamics to be scale-invariant 
\begin{equation}
\label{q42}
\epsilon(t)=\gamma(t)^{\alpha-2}
\end{equation}
given the definition $\epsilon(0)=1$. Generally, inducing a scale-invaraint dynamics in an interacting system requires to tune interaction along the process. 
In  ultracold atom experiments, this is a routine task in the laboratory assisted by means of a Feshbach resonance \cite{timmermans_1999} or a modulation of the transverse confinement in low-dimensional systems \cite{staliunas_2004,campo_epl_2011}.   No interaction tuning is required in processes involving transport exclusively, this is, in protocols for which $\f=\f(t)$ and $\gamma(t)=1$. 
For processes with $\gamma(t)\neq1$, there are relevant scenarios for which $\alpha=2$ and no interaction tuning is required \cite{campo_2011,campo_boshier_2012}.
In addition, for processes with $\gamma(t)\neq1$ and $\alpha\neq2$, whenever the scaling factor remains of order unity along the process, $\gamma(t)\sim\mathcal{O}(1)$,
a high-fidelity quantum driving is achieved even in the absence of interaction tuning, i.e. while keeping $\epsilon(t)=1$ \cite{campo_boshier_2012}.

Provided that the consistency equation Eq. \eqref{q42} is fulfilled (or approximately satisfied) so that $\epsilon\gamma^{2-\alpha}=1$,  
$\hat{\mathcal{I}}=\gamma^2\hat{\mathcal{H}}_0$ becomes a first integrand or constant of motion, and can be identified as an invariant operator $\hat{\mathcal{I}}$ satisfying
\begin{equation}
\label{q41}
\frac{\td\hat{\mathcal{I}}}{\td t}=\frac{\pd\hat{\mathcal{I}}}{\pd t}+\frac{1}{i\hbar}\com{\hat{\mathcal{I}}}{\hat{\mc{H}}}\,,
\end{equation}
and that is the quantum equivalent of the classical, energy-like dynamical invariant \eqref{q27}. 

The third line in Eq.~\eqref{q39} corresponds to the auxiliary CD Hamiltonian  which in the original variables, $(\q_i,\p_i)$, reads
\begin{equation}
\label{q43}
\hat{\mathcal{H}}_1=\sum_{i=1}^N\bigg[\dot{\f}\cdot\p_i+\frac{ \dot{\gamma}}{2\gamma}\{\q_i-\f(t),\p_i\}\bigg].
\end{equation}
Here, the curly brackets denote the anticommutator of two operators $A$ and $B$, $\{A,B\}=A\cdot B+B\cdot A$. In complete analogy to the classical case, the first term is the auxiliary CD term associated with transport along the trajectory $\q=\f(t)$, while the second-one is associated with the expansion. Equation~\eqref{q43} agrees with the single particle expression in Eq.~\eqref{q10} and previous results derived for power-law traps \cite{chen_2011,jarzynski_2013,campo_2013}.

In the previous section we found coordinate transformations, that allowed us to write the CD Hamiltonian for a system with one degree of freedom  in local form. The crucial steps involved  finding a generating function for the coordinate transformation, and a corresponding dynamical invariant. In the following, we will apply the same ideas to the multi-particle Hamiltonian $\hat{\mathcal{H}}$ \eqref{q39}.
The representation in quantum mechanics of the group of linear canonical transformations has been discussed at length by Moshinsky, see e.g. \cite{moshinsky_96}.
We  denote the quantum, multi-particle unitary transformation that plays the role of the classical generating function $F(q,\bar{p},t)$ \eqref{q20} by $\mathcal{U}$. It reads 
\begin{equation}
\label{q44}
\mathcal{U}
=\prod_{i=1}^N\e{\frac{im }{\hbar}\dot{\f}\cdot\q_i+\frac{im \dot{\gamma}}{2\hbar \gamma}\left(\q_i-\f\right)^2-i\frac{m}{2}\int^t_0\td s\,\dot{\f}^2}\,.
\end{equation}
The latter functions transform the ``old'' set of coordinates $(\q_i,\p_i)$ to a new set $(\bar{\q}_i,\bar{\p}_i) $ according to
\begin{subequations}
\label{q45}
\begin{eqnarray}
\q_i&\rightarrow& \bar{\q}_i=\mathcal{U}\q_i\mathcal{U}^{\dag}=\q_i,\\
\p_i&\rightarrow& \bar{\p}_i=\mathcal{U} \p_i\mathcal{U}^{\dag}=\p_i -\frac{m \dot{\gamma}}{\gamma}\left(\q_i-\f\right)-m\, \dot{\f},\\
\hat{\mathcal{H}} &\rightarrow& \hat{\bar{\mathcal{H}}}(t)=\mathcal{U}\hat{\mathcal{H}}(t)\mathcal{U}^{\dag}-i\hbar\, \mathcal{U}\,\partial_t\mathcal{U}^{\dag}.
\end{eqnarray}
\end{subequations}
Here, the new representation of the CD Hamiltonian  becomes
\begin{equation}
\begin{split}
\label{q46}
\hat{\bar{\mathcal{H}}}(t)=&\sum_{i=1}^N\!\left[-\frac{\hbar^2}{2m}\Delta_{\bar{\q}_i} \!+\!\U(\bar{\q}_i,\mb{\lambda}(t))\right]\!+\!\epsilon(t)\sum_{i<j}V(\bar{\q}_i-\bar{\q}_j)\\
&+\sum_{i=1}^N\left[-\frac{m}{2}\frac{\ddot{\gamma}}{\gamma}(\bar{\q}_i-\f)^2 -m\,\ddot{\f}\cdot\bar{\q}_i\right]\,,
 \end{split}
\end{equation}
which is the multi-particle quantum equivalent of the classical Hamiltonian $\bar{H}(\bar{q},\bar{p},t)$ \eqref{q18}. Under this canonical transformation the time-evolution of the initial state is mapped to $\Psi(t) \rightarrow \Phi(t)=\mathcal{U}\Psi(t)$. 
Finally, it follows that the dynamical invariant $\mc{I}$ can be written in new coordinates $(\bar{\q}_i,\bar{\p}_i)$,
\begin{equation}
\label{q47}
\begin{split}
\hat{\mathcal{I}}=&\sum_{i=1}^N
\frac{1}{2m}\left[\gamma(\bar{\p}_i-m\dot{f})-m\dot{\gamma}(\bar{\q}_i-\f)\right]^2\\
&+\sum_{i=1}^N \U_0\left(\frac{\bar{\q}_i-\f}{\gamma}\right)+\sum_{i<j}V\left(\frac{\bar{\q}_i}{\gamma}-\frac{\bar{\q}_j}{\gamma}\right)\,,
\end{split}
\end{equation}
which is equivalent to the second line of Eq.~\eqref{q27}. 

We note that the dynamics governed by $\hat{\bar{\mathcal{H}}}(t)$ induces a phase modulation associated with $\mathcal{U}$ that generally leads to excitations away from the adiabatic trajectory $\Psi(t)$. The nonadiabatic nature of the resulting shortcuts to adiabaticity is exclusively captured by $\mathcal{U}$ (phase modulations), while local correlations function are identical at all times with the adiabatic ones, given that $|\Psi(t)|^2=|\Phi(t)|^2$.
Nonetheless,  $\mathcal{U}$ reduces to the identity and $\Phi(t)=\Psi(t)$ at the beginning and end of the process (e.g. at time $t=\{0,\tau_{\rm F}\}$). Further, it is straightforward to design protocols involving only smooth modulations of the auxiliary counterdiabatic field (requiring no  `impulses') of relevance to experimental realizations, as we shall discuss in section \ref{sec:eng}.
In the following sections the formalism just described will prove useful to engineer shortcuts to adiabaticity for several non-trivial systems.

\section{Local counterdiabatic driving for an arbitrary trapping potential\label{sec:trap}}

The only condition we have imposed in the preceding sections on the time-dependence of the external potential is its scale-invariant form \eqref{q02}. 
To illustrate the generality of our  approach, let $\U(\q,t)$  have  a power series expansion
\begin{equation}
\label{q48}
\U(\q,t)=\sum_{p=0}^{\infty}\alpha_p(t)(\q-\f)^{p},
\end{equation}
where $\alpha_p(t)=\U_0^{(p)}({\bf 0})/p!$ with $\U_0^{(p)}$ denoting the $p$th derivative, and  $\U_0^{(0)}({\bf 0},t)=\U_0({\bf 0},t)$. 
Further, assume the potential to be isotropic, the extension to anisotropic potentials being straightforward.
Given a process governed by the  time-dependent potential \eqref{q48},  we aim at finding a local CD  protocol.

First, let us  impose the form required for scaling laws $\U(\q,t)=\U_0[(\q-\f)/\gamma]/\gamma^2$ \eqref{q02}, which implies the following relationship among the coefficients in the series expansion \eqref{q48}
\begin{equation}
\label{q50}
\alpha_0(t)=\frac{\alpha_0(0)}{\gamma^2} ,\,
\alpha_1(t)=\frac{\alpha_1(0)}{\gamma^3},\dots,\,
\alpha_p(t)=\frac{\alpha_p(0)}{\gamma^{p+2}}.
\end{equation}
Hence, for an arbitrary potential,  the time-modulation in \eqref{q48} can be implemented provided 
that each coefficient $\alpha_p(t)$ can be tuned independently. 
The latter condition leads to the recurrence relation
\begin{equation}
\label{q51}
\frac{\alpha_{p}(t)}{ \alpha_{p}(0)}  =  \left[\frac{\alpha_{p-1}(t)}{\alpha_{p-1}(0)}\right]^{\frac{p+2}{p+1}} =\left[\frac{\alpha_{p-m}(t)}{\alpha_{p-m}(0)}\right]^{\frac{p+2}{p-m+2}},
\end{equation}
where the last exponent results from $\prod_{k=1}^{m}\frac{p-k+3}{p-k+2}=\frac{p+2}{p-m+2}$. The auxiliary potential terms in  \eqref{q46} can be absorbed in the definition 
of the expansion coefficients   
\begin{equation}
\label{q52}
\tilde{\alpha}_p(t)=\alpha_p(t)-\frac{m\ddot{\gamma}}{2\gamma}\delta_{p,2}-m\ddot{f}(\delta_{p,1}+\delta_{p,0}),
\end{equation}
so that the local CD potential is given by $\bar{\U}(\bar{\q},t)=\sum_{p=0}^{\infty}\tilde{\alpha}_p(t)(\bar{\q}-\f)^{p}$, which is the sum of the one-body trapping potential $\U(\q,t)$  and the auxiliary terms in $\hat{\bar{\mathcal{H}}}(t)$. 
The required modulation of the $\tilde{\alpha}_p(t)$ coefficients makes the implementation of CD protocols with non-harmonic traps  particularly amenable to the painting potential technique \cite{henderson_2009}.

\paragraph*{Example -- the quartic trap potential}
As an illustrative example, we consider the quartic potential
\begin{equation}
\label{q53}
\U(\q,t)=\alpha_2(t)(\q-\f)^2+\alpha_4(t)(\q-\f)^4,
\end{equation}
where the time-modulation
\begin{equation}
\label{q54}
\alpha_2(t)=\frac{\alpha_2(0)}{\gamma^4}\quad\mathrm{and}\quad \alpha_4(t)=\frac{\alpha_4(0)}{\gamma^6}
\end{equation}
leads to a scaling of the form \eqref{q48}, associated with a scale-invariant dynamics.
Provided that \eqref{q42} is satisfied, the CD potential, for which $\Phi(t)=\mathcal{U}\Psi(t)$ is the exact solution to the many-body Schr\"odinger equation, is simply given by
\begin{equation}
\label{q55}
\bar{\U}(\bar{\q},t)=-m\ddot{\f}\cdot\bar{\q}+\left(\frac{\alpha_2(0)}{\gamma^4}-\frac{m\ddot{\gamma}}{2\gamma}\right)(\bar{\q}-\f)^2+\frac{\alpha_4(0)}{\gamma^6}(\bar{\q}-\f)^4.
\end{equation}

\begin{table*}[t]
\centering
\footnotesize
\begin{tabular}{llll}
\hline \\[-1ex]
 Name  &  $\U(q,t)$  & Time-dependence & counterdiabatic  modulation $-\ddot{\gamma}/\gamma$\\ [2.pt]
\hline \\[-1ex]

Arbitrary potential  &\hspace{1mm}  $\frac{1}{\gamma^2}\U_0(\frac{q}{\gamma})$             &arbitrary $\gamma(t)$ & \hspace{1mm} $-\frac{\ddot{\gamma}(t)}{\gamma(t)}$\\

Power-law  trap &\hspace{1mm}  $A|q|^b$             &\hspace{1mm} $A(t)=\frac{A(0)}{\gamma^2+b}$ & $\frac{\dot{A}(t)}{(2+b)A(t)}-\frac{3+b}{(2+b)^2}\big[\frac{\dot{A}(t)}{A(t)}\big]^2$\\

Modified P\"oschl-Teller &\hspace{1mm}  $-\frac{\hbar^2}{2m}\alpha^2\frac{\lambda(\lambda-1)}{\cosh^2\alpha q} $             & $\alpha(t)=\frac{\alpha(0)}{\gamma}$& \hspace{1mm}
$\frac{\ddot{\alpha}(t)}{\alpha(t)}-2\big[\frac{\dot{\alpha}(t)}{\alpha(t)}\big]^2$\\
well \cite{flugge_1971}  ($\lambda>1$)        &\hspace{1mm}   &     & \\

P\"oschl-Teller  well \cite{flugge_1971} &\hspace{1mm}  $\frac{\hbar^2}{2m}\alpha^2\left(\frac{\lambda(\lambda-1)}{\cos^2\alpha q} +\frac{\kappa(\kappa-1)}{\sin^2\alpha q}\right)$             &$\alpha(t)=\frac{\alpha(0)}{\gamma}$& \hspace{1mm}
$\frac{\ddot{\alpha}(t)}{\alpha(t)}-2\big[\frac{\dot{\alpha}(t)}{\alpha(t)}\big]^2$\\
 ($\lambda,\kappa>1$)        &\hspace{1mm}   &     & \\

Optical Lattice  &\hspace{1mm}  $A\sin^2(\alpha q)$             & $A(t)=\frac{A(0)}{\gamma^2}, \alpha(t)=\frac{\alpha(0)}{\gamma}$ & \hspace{1mm} $\frac{\dot{A}(t)}{2A(t)}-\frac{3}{4}\big[\frac{\dot{A}(t)}{A(t)}\big]^2=\frac{\ddot{\alpha}(t)}{\alpha(t)}-2\big[\frac{\dot{\alpha}(t)}{\alpha(t)}\big]^2$\\

Gaussian well   &\hspace{1mm}  $-A\exp(-\alpha^2q^2)$             & $A(t)=\frac{A(0)}{\gamma^2}, \alpha(t)=\frac{\alpha(0)}{\gamma}$ & \hspace{1mm} $\frac{\dot{A}(t)}{2A(t)}-\frac{3}{4}\big[\frac{\dot{A}(t)}{A(t)}\big]^2=\frac{\ddot{\alpha}(t)}{\alpha(t)}-2\big[\frac{\dot{\alpha}(t)}{\alpha(t)}\big]^2$\\

Finite square well   &\hspace{1mm}  $-A\Theta(\alpha-q)$             & $A(t)=\frac{A(0)}{\gamma^2}, \alpha(t)=\alpha(0)\gamma$ & \hspace{1mm} $\frac{\dot{A}(t)}{2A(t)}-\frac{3}{4}\big[\frac{\dot{A}(t)}{A(t)}\big]^2=-\frac{\ddot{\alpha}(t)}{\alpha(t)}$\\

Exponential &\hspace{1mm}  $-A\exp(-r/\alpha)$             & $A(t)=\frac{A(0)}{\gamma^2}, \alpha(t)=\alpha(0)\gamma$ & \hspace{1mm}$\frac{\dot{A}(t)}{2A(t)}-\frac{3}{4}\big[\frac{\dot{A}(t)}{A(t)}\big]^2=-\frac{\ddot{\alpha}(t)}{\alpha(t)}$\\

Yukawa &\hspace{1mm}  $-A\frac{\exp(-r/\alpha)}{r/\alpha}$             & $A(t)=\frac{A(0)}{\gamma^2}, \alpha(t)=\alpha(0)\gamma$ & \hspace{1mm}$\frac{\dot{A}(t)}{2A(t)}-\frac{3}{4}\big[\frac{\dot{A}(t)}{A(t)}\big]^2=-\frac{\ddot{\alpha}(t)}{\alpha(t)}$\\

Wood-Saxon  &\hspace{1mm}  $-A\frac{\exp(-r/\alpha)}{1-\exp(r/\alpha)}$             &$A(t)=\frac{A(0)}{\gamma^2}$, $\alpha(t)=\alpha(0)\gamma$ & \hspace{1mm}$\frac{\dot{A}(t)}{2A(t)}-\frac{3}{4}\big[\frac{\dot{A}(t)}{A(t)}\big]^2=-\frac{\ddot{\alpha}(t)}{\alpha(t)}$\\
 ($\alpha\ll f$)        &\hspace{1mm}   &    &  \\

Hulth\'en  &\hspace{1mm}  $-\frac{A}{1+\exp[(r-f)/\alpha]}$             & $A(t)=\frac{A(0)}{\gamma^2}$, $\alpha(t)=\alpha(0)\gamma$ & \hspace{1mm}$\frac{\dot{A}(t)}{2A(t)}-\frac{3}{4}\big[\frac{\dot{A}(t)}{A(t)}\big]^2=-\frac{\ddot{\alpha}(t)}{\alpha(t)}$\\
 ($\alpha\ll f$)        &\hspace{1mm}   &     & \\

Kratzer  &\hspace{1mm}  $-2A\left(\frac{\alpha}{r}-\frac{1}{2}\frac{\alpha^2}{r^2}\right)$             & $A(t)=\frac{A(0)}{\gamma^2}$, $\alpha(t)=\alpha(0)\gamma$& $\frac{\dot{A}(t)}{2A(t)}-\frac{3}{4}\big[\frac{\dot{A}(t)}{A(t)}\big]^2=-\frac{\ddot{\alpha}(t)}{\alpha(t)}$\\
\vspace*{1mm}

Morse &\hspace{1mm}  $A^2+B^2 \exp(-2\alpha q) $             & $A(t)=\frac{A(0)}{\gamma}, B(t)=\frac{B(0)}{\gamma},  \alpha(t)=\frac{\alpha(0)}{\gamma}$ & \hspace{1mm}$\frac{\ddot{X}(t)}{X(t)}-2\big[\frac{\dot{X}(t)}{X(t)}\big]^2$ \\
          &\hspace{1mm}  $- 2B(A+ \alpha/2) \exp(-\alpha q) $ &     &\quad $(X=A,B,\alpha)$ \\


Eckart &\hspace{1mm} $A^2+B^2/A^2 -2B$ coth $\alpha q$ &   $A(t)=\frac{A(0)}{\gamma}, B(t)=\frac{B(0)}{\gamma^2},  \alpha(t)=\frac{\alpha(0)}{\gamma}$ & \hspace{1mm}
$\frac{\ddot{X}(t)}{X(t)}-2\big[\frac{\dot{X}(t)}{X(t)}\big]^2=\frac{\dot{B}(t)}{2B(t)}-\frac{3}{4}\big[\frac{\dot{B}(t)}{B(t)}\big]^2$ \\
$(B > A^2)$ &\hspace{1mm} $+ A(A - \alpha) {\rm cosech}^2 \alpha q$  &  &\quad$(X=A,\alpha)$\\

Scarf I &\hspace{1mm}   $-A^2+(A^2+B^2-A\alpha) {\rm sec}^2 \alpha q$ & $A(t)=\frac{A(0)}{\gamma}, B(t)=\frac{B(0)}{\gamma},  \alpha(t)=\frac{\alpha(0)}{\gamma}$ & \hspace{1mm}  $\frac{\ddot{X}(t)}{X(t)}-2\big[\frac{\dot{X}(t)}{X(t)}\big]^2$\\ 
(trigonometric) & \hspace{1mm} $-B(2A - \alpha) \tan \alpha q\  {\rm sec} \alpha q$ & &\quad  $(X=A,B,\alpha)$\\

Scarf II & \hspace{1mm}  $A^2+(B^2-A^2-A\alpha) {\rm sech}^2\alpha q$  & $A(t)=\frac{A(0)}{\gamma}, B(t)=\frac{B(0)}{\gamma},  \alpha(t)=\frac{\alpha(0)}{\gamma}$ &\hspace{1mm}  $\frac{\ddot{X}(t)}{X(t)}-2\big[\frac{\dot{X}(t)}{X(t)}\big]^2$ \\
(hyperbolic) &\hspace{1mm}  $+B(2A+\alpha){\rm sech}~ \alpha q\tanh \alpha q$  & & \quad $(X=A,B,\alpha)$\\

Generalized P\"{o}schl-Teller &\hspace{1mm} $A^2+(B^2+A^2+A\alpha) {\rm cosech}^2\alpha r$ & $A(t)=\frac{A(0)}{\gamma}, B(t)=\frac{B(0)}{\gamma},  \alpha(t)=\frac{\alpha(0)}{\gamma}$ & \hspace{1mm} $\frac{\ddot{X}(t)}{X(t)}-2\big[\frac{\dot{X}(t)}{X(t)}\big]^2$ \\
$( A < B)$ &\hspace{1mm} $- B(2A+\alpha)$ coth $\alpha r$ cosech $\alpha r$ &  &\quad $(X=A,B,\alpha)$\\

 P\"{o}schl-Teller II &\hspace{1mm} $(A-B)^2-A(A+\alpha){\rm sech}^2\alpha r$ & $A(t)=\frac{A(0)}{\gamma}, B(t)=\frac{B(0)}{\gamma},  \alpha(t)=\frac{\alpha(0)}{\gamma}$ & \hspace{1mm} $\frac{\ddot{X}(t)}{X(t)}-2\big[\frac{\dot{X}(t)}{X(t)}\big]^2$ \\
$( B < A)$ &\hspace{1mm} $+ B(B-\alpha)$ cosech $\alpha r$ & & \quad $(X=A,B,\alpha)$\\

Rosen-Morse I &\hspace{1mm}   $A(A - \alpha)$cosec$^2 \alpha q + 2B$ cot $ \alpha q$ &  $A(t)=\frac{A(0)}{\gamma}, B(t)=\frac{B(0)}{\gamma^2},  \alpha(t)=\frac{\alpha(0)}{\gamma}$ & \hspace{1mm} $\frac{\ddot{X}(t)}{X(t)}-2\big[\frac{\dot{X}(t)}{X(t)}\big]^2=\frac{\dot{B}(t)}{2B(t)}-\frac{3}{4}\big[\frac{\dot{B}(t)}{B(t)}\big]^2$ \\ 
(trigonometric) & \hspace{1mm} $-A^2 + B^2/A^2$ \vspace*{1mm} &  & \quad $(X=A,\alpha)$\\
$(0 \leq \alpha q \leq \pi)$ & \vspace*{1mm} &  &\\


Rosen-Morse II &\hspace{1mm} $A^2+B^2/A^2 - A(A +\alpha) {\rm sech}^2 \alpha q$ &
$A(t)=\frac{A(0)}{\gamma}, B(t)=\frac{B(0)}{\gamma^2},  \alpha(t)=\frac{\alpha(0)}{\gamma}$ &  \hspace{1mm}$\frac{\ddot{X}(t)}{X(t)}-2\big[\frac{\dot{X}(t)}{X(t)}\big]^2=\frac{\dot{B}(t)}{2B(t)}-\frac{3}{4}\big[\frac{\dot{B}(t)}{B(t)}\big]^2$ \\  
(hyperbolic) $(B < A^2)$ & \hspace{1mm}+ 2$B$ tanh $\alpha q$ &  &\quad $(X=A,\alpha)$\\

\hline
\end{tabular}
\caption{List of counterdiabatic driving  schemes for potentials which acquire the form of $\U(q,t)$ in Eq.~\eqref{q02} under the indicated time-dependence of the parameters $A$, $B$, and $\alpha$, including well-known shape-invariant potentials in supersymmetric quantum mechanics. 
A local CD protocol for expansions and compressions ($\f=0$) is induced by the potential $\bar{\U}(\bar{q},t)=\U(\bar{q},t)-m\ddot{\gamma}\bar{q}^2/(2\gamma)$.
 When meaningful, the same potentials can be used as a two-body potential ${\rm V}$ with $\alpha=2$. A single degree of freedom is considered for clarity; the extension to higher dimensions under cylindrical or spherical symmetry is straightforward following \cite{campo_boshier_2012}.
The range of potentials is $- \infty \leq q \leq \infty, 0 \leq r \leq \infty$, unless stated otherwise. The family of power-law potentials \cite{jarzynski_2013,campo_2013} includes the harmonic case ($b=2$) \cite{campo_2011} and the infinite square well ($b=\infty$) \cite{campo_boshier_2012}.}
\label{SIP}
\end{table*}

In many instances, the  coefficients in the set $\{\alpha_p(0)\}_p$ associated with the power-series \eqref{q48} are inter-related, and the required time-dependence of the potential can be brought into the form of Eq.~\eqref{q27}  by direct inspection or a scaling analysis. A list of examples and the associated CD protocols  is provided in Table~\ref{SIP}, which includes among others the family of shape-invariant potentials  in supersymmetric quantum mechanics \cite{cooper_1995}, as well as common potentials in atom optics, such as several types of wells and optical lattices.

\section{Counderdiabatic driving of nonlinear systems\label{sec:nonlin}}

The original formulation of CD  is restricted to linear systems. However, it was recently shown that it can be  generalized to non-linear systems undergoing scale-invariant expansions and compressions \cite{campo_2013}. The approach developed in the previous sections allows us  also to treat nonlinear systems. 

Typically, nonlinear (``quantum'') systems are described by effective evolution equations derived within a mean-field approach. 
A prominent example is the  description of Bose-Einstein condensates, where scale-invariant dynamics is of great relevance to  time-of-flight measurements \cite{castin_1996,kagan_1996,egusquiza_2011}. The dynamics is described by the time-dependent Gross-Pitaevskii equation (TDGPE) governing the (normalized) wavefunction $\Psi(\q,t)$ of a Bose-Einstein condensate 
\begin{equation}
\label{q56}
i\hbar\,\partial_t\Psi(\q,t)=\left[-\frac{\hbar^2}{2m}\Delta_{\q}+\U(\q,t)+g_{D}|\Psi(\q,t)|^2\right]\Psi(\q,t)\,,
\end{equation}
where $\U(\q,t)$ shall again be of the scale-invariant form \eqref{q02}. 
This nonlinear Schr\"odinger equation can be obtained from the model describing a D-dimensional many-body Bose gas with regularized contact interactions (the Fermi-Huang pseudopotential) using a mean-field approximation (e.g.  assuming that the many-body wavefunction takes the form of a Hartree product). 
We can then expect that a protocol derived in the previous section, exact for the many-body description, should carry over the TDGPE.
Indeed, a stationary state $\Psi$ at $t=0$ with chemical potential $\mu$ can be forced to obey a scale-invariant  ansatz of the form  $\Psi(\q,t)=\e{-i\mu \tau(t)/\hbar} \gamma^{-D/2}\Psi[(\q-\f)/\gamma,t=0]$. This  ansatz is the exact solution to a counterdiabatic nonlinear evolution equation,  the modified non-local TDGPE
\begin{equation}
\label{q57}
\begin{split}
i\hbar\,\partial_t\Psi(\q,t)=&\bigg[-\frac{\hbar^2}{2m}\Delta_{\q}+\U(\q,t)+\dot{\f}\cdot \p\\
& +\frac{\dot{\gamma}}{2\gamma}\{\q-\f,\p\}+g_{D}|\Psi(\q,t)|^2\bigg]\Psi(\q,t),
\end{split}
\end{equation}
provided that the interaction coupling strength is tuned according to $g_D=g_D(t=0)\gamma^{D-2}$. 
We observe that this time-dependence of the nonlinear interactions agrees with that required of the D-dimensional Bose gas and can be achieved using, for instance, by tuning a magnetic field through a Feshbach resonance \cite{timmermans_1999}. Alternatively, in low-dimensional quantum gases ($D=1,2$) it can be implemented by modulating the transverse confinement \cite{staliunas_2004,campo_epl_2011}. 

As before, the modified TDGPE can be brought into local form by applying the canonical time-dependent transformation $\mc{U}$ \eqref{q44} ($N=1$), which leads to
\begin{equation}
\label{q58}
\begin{split}
i\hbar\,\partial_t\Phi(\bar{\q},t)&=\bigg[-\frac{\hbar^2}{2m}\Delta_{\bar{\q}}+\U(\bar{\q},t)-\frac{m}{2}\frac{\ddot{\gamma}}{\gamma}(\bar{\q}-\f)^2\\
&  -m\ddot{\f}\cdot\bar{\q}+g_{D}|\Phi(\bar{\q},t)|^2\bigg]\Phi(\bar{\q},t),
\end{split}
\end{equation}
with exclusively local potential terms, and $\Phi=\mathcal{U}\Psi$.

It is worth mentioning that Kundu \cite{kundu_2009} has shown that  inhomogeneous nonlinear Schr\"odinger equations of this class are equivalent to the standard homogeneous nonlinear Schr\"odinger equation, and hence admit a zero-curvature representation, explaining their integrability \cite{fadeev_2007}.

Furthermore, CD can be applied as well to other examples of nonlinear evolution, for which a non zero-curvature representation has not been found to date. A relevant instance is the mean-field theory developed  by Kolomeisky {\it et al.} \cite{kolomeisky_2000,kolomeisky_1992} which accurately describes the ground state density profile of one-dimensional bosons with hard-core contact interactions, i.e.,  a Tonks-Girardeau gas \cite{girardeau_1960} (and its dual system under Bose-Fermi duality, a one-dimensional spin-polarized Fermi gas \cite{girardeau_2000}), up to spatial anti-bunching \footnote{The exact many-body quantum system describing a Tonks-Girardeau gas ($N$ one-dimensional bosons with contact interactions of infinite amplitude) is exactly solvable by means of the Bose-Fermi duality as discussed by Girardeau \cite{girardeau_1960}. It is remarkable that the classical integrability condition in terms of the zero-curvature representation for the mean-field Kolomeisky equation describing this system has not yet been found.}. 
The time-dependent version of the Kolomeisky equation reads,
\begin{equation}
\label{q59}
\begin{split}
i\hbar\,\pd_t \Psi(q,t)&=\bigg[-\frac{\hbar^2}{2m}\pd_q^2+\U(q,t)\\
&\quad\quad+\frac{\pi^2\hbar^2}{2m}|\Psi(q,t)|^4\bigg]\Psi(q,t),
\end{split}
\end{equation}
where $\Psi(q,0)=\sqrt{n(q,0)}$ is the square root of the initial density profile. 
CD protocols can be directly obtained from the one-dimensional version of Eqs.~(\ref{q57}) and (\ref{q58}), respectively,  by replacing $g_{1}|\Psi(\q,t)|^2$ by $\pi^2\hbar^2/2m\,|\Psi(q,t)|^4$ (recall that $|\Phi(q,t)|=|\Psi(q,t)|$). Note that the quintic non-linearity  arises from the repulsive contact interactions and can be considered as a potential term with $\alpha=2$. As a result no interaction tuning is required, provided that the gas remains in the Tonks-Girardeau regime. 
It is worth emphasizing that, as a mean-field theory, the model by Kolomeisky \etal \eqref{q59}  overestimates phase coherence. Therefore, while it properly describes the scale-invariant  dynamics in a time-dependent harmonic trap \cite{kim_2003}, it fails to accurately account for processes involving interference such as splitting and recombination \cite{girardeau_2000}. 
Nonetheless, the Kolomeisky equation has been successfully applied to describe the formation of shock waves and it is accurate as long as changes in the density occur on a length scale larger than the ``Fermi'' length \cite{damski_2004}. As a result, under scale-invariant driving,  protocols derived from \eqref{q59} agree with those designed using an exact many-body treatment  \cite{campo_2011,campo_boshier_2012}. 

We close this section by mentioning that other non-linear processes that can be assisted by CD include the (mean-field) growth dynamics of a Bose-Einstein condensate \cite{ozcakmakli_2012}.
Nonetheless, phase fluctuations in the newborn condensate are expected to result in the formation of solitons \cite{lamporesi_2013} or  vortices \cite{weiler_2008}, depending in the dimensionality, as dictated by the Kibble-Zurek mechanism \cite{delcampo_2013,delcampo_zurek_2013}.

\section{Counterdiabatic driving and reverse engineering: scaling laws\label{sec:scale}}

In the previous discussion, we showed that CD can be used to enforce scale-invariant dynamics in which the scaling factor follows the adiabatic trajectory.
However, there are more general scaling laws, which are associated with an invariant of motion, provided that a set of consistency equations is satisfied \cite{gritsev_2010}.
Knowledge of these paves the way to engineering shortcuts to adiabatic scale-invariant processes.
For a large family of many-particle systems in a time-dependent harmonic trap such design was reported in  \cite{campo_2011}, 
extending previous results for the single-particle Schr\"odinger equation \cite{chen_2010} as well as the Gross-Pitaevskii equation describing Bose-Einstein condensates in the mean-field \cite{muga_2009}.
These results have been further extended to time-dependent box-like  confinements \cite{campo_boshier_2012} and  arbitrary power-law potentials  \cite{jarzynski_2013,campo_2013}.
We next present the general scaling laws associated with a family of Hamiltonians which include all the aforementioned results. Having done so, we shall 
establish their explicit relation with CD. 
Let us consider the Hamiltonian \eqref{q36} in which harmonic and linear terms in $\q_i$ are explicitly written
\begin{equation}
\label{q60}
\begin{split}
&\hat{\mathcal{H}}_0= 
\sum_{i=1}^N\left[-\frac{\hbar^2}{2m}\Delta_{\q_i}+\U(\q_i,t)\right]+\epsilon(t)\sum_{i<j}V(\q_i-\q_j)\\
&+\sum_{i=1}^N\left[\frac{m}{2}\om^2(t)[\q_i-\f(t)]^2+m\,\F(t)\cdot\q_i
\right]\,,
\end{split}
\end{equation}
with an arbitrary modulation of the coefficients $\om(t)$ (harmonic trap frequency) and  $\F(t)$.
A stationary state $\Psi$ of the system \eqref{q60} at $t=0$, follows a scale-invariant evolution 
\begin{equation}
\label{q61}
\Phi\left(\{\q_i\},t\right)=e^{-\frac{i}{\hbar}\int^t_0\frac{m}{2}\dot{\f}^2 dt'}e^{i\sum_{i=1}^N\!\!\frac{ m\dot{\gamma}}{2\gamma\hbar}(\q_i-\f)^2+\frac{im}{\hbar}\dot{\f}\cdot\q_i}
\Psi(t)\,,
\end{equation}
with $\Psi(t)$ given by Eq.~\eqref{q38}, whenever the following consistency conditions are satisfied 
\begin{equation}
\label{q62}
\om^2(t)=\frac{\om_0^2}{\gamma^4}- \frac{\ddot{\gamma}}{\gamma},\quad \F(t) = -\ddot{\f},\quad \epsilon(t)  =  \gamma^{\alpha -2},
\end{equation}
 with $\om_0=\om(0)$, and satisfying $\gamma=\gamma(t)$ the boundary conditions $\gamma(0)=1$ and $\dot{\gamma}(0)=0$. 

Generally, the resulting dynamics are not adiabatic. Only in the adiabatic limit, where $\ddot{\gamma}\rightarrow 0$ and $\ddot{\f}\rightarrow 0$ in Eq.~\eqref{q62}, we find that the solution for the scaling factor takes the form $ \om^2(t)=\om_0^2/\gamma^4$ and $\F(t)=0$. Nonetheless, counterdiabatic driving provides means to induce finite-time evolution \eqref{q61} that effectively follows the adiabatic trajectory of the scaling factor. The frequency of the trap is to be replaced by \cite{campo_2013,campo_2011}
\begin{equation}
\label{q64}
\om^2(t)\rightarrow\om^2(t)- \frac{\ddot{\gamma}}{\gamma}=\om^2(t)-\frac{3}{4}\frac{\dot{\om}^2}{\om^2}+\frac{1}{2}\frac{\ddot{\om}}{\om},
\end{equation}
while  the modulation of the linear term remains $\F(t)\rightarrow -\ddot{\f}$.

\section{Engineering shortcuts to adiabaticity assisted by smooth counterdiabatic fields\label{sec:eng}}

In the last part of the present analysis we shall illustrate how the time dependence of the control parameters is to be designed to engineer shortcuts to adiabaticity based on CD, without the requirement of impulse auxiliary fields. Assume we wish to find a shortcut to an adiabatic expansion or compression by changing the scaling factor $\gamma$ from an initial value $\gamma(t=0)=1$ to a final value $\gamma_{\rm F}$ at $t=\tau_{\rm F}$, while at the same time transporting the system by shifting the external trapping potential from  $\f(t=0)=0$ to $\f(\tau_\mrm{F})=\f_{\rm F}$. Further, let us  impose that  the auxiliary Hamiltonian is switched on at $t=0$ and switched off at $t=\tau_\mrm{F}$, cf. the example in Sec.~\ref{sec:gen}, i.e. $\hat{\mathcal{H}}=\hat{\mathcal{H}}_0$ at $t=\{0,\tau_{\rm F}\}$.
We have seen earlier that the  auxiliary nonlocal term $\hat{\mathcal{H}}_1$ \eqref{q43} induces the adiabatic dynamics along the instantaneous eigenstates of the system Hamiltonian $\hat{\mathcal{H}}_0$, \eqref{q36}.  The time-dependent coefficient of \eqref{q43} is governed by the rate of change of the scaling factor $\gamma$ and the shift function $\f$. Therefore, for it to vanish at $t=\{0,\tau_{\rm F}\}$ the following boundary conditions are required:
\begin{equation}
\label{q65}
\begin{split}
\gamma(0)=1, &\quad \dot{\gamma}(0)=0,\\
\gamma(\tau_{\rm F})=\gamma_{\rm F},&\quad \dot{\gamma}(\tau_{\rm F})=0,
\end{split}
\end{equation}
and
\begin{equation}
\label{q66}
\begin{split}
\f(0)= {\bf 0}, &\quad \dot{\f}(0)={\bf 0},\\
\f(\tau_{\rm F})=\f_F, &\quad \dot{\f}(\tau_{\rm F})={\bf 0},
\end{split} 
\end{equation}
which can be used to determine an interpolating ansatz, e.g.,
\begin{equation}
\label{q67}
\begin{split}
\gamma(t)&=1+3(\gamma_{\rm F}-1)\frac{t^2}{\tau_{\rm F}^2}+2(\gamma_{\rm F}-1)\frac{t^3}{\tau_{\rm F}^3},\\
\f(t)&=\left(3\frac{t^2}{\tau_{\rm F}^2}-2\frac{t^3}{\tau_{\rm F}^3}\right)\,\f_{\rm F}.
\end{split}
\end{equation}
Alternatively, we have seen that the local counterdiabatic driving protocol \eqref{q46} leads to the time-evolution $\Phi(t)=\mathcal{U}\Psi(t) $. Imposing  
$\hat{\bar{\mathcal{H}}}=\hat{\mathcal{H}}_0$ at $t=\{0,\tau_{\rm F}\}$ and demanding the initial and final state to be stationary, so that $\Phi=\Psi$, leads to \eqref{q65} and \eqref{q66} supplemented by 
\begin{equation}
\label{q68}
\begin{split}
\ddot{\gamma}(0)=0&\quad \ddot{\gamma}(\tau_{\rm F})=0,\\
 \ddot{\f}(0)={\bf 0}&\quad \ddot{\f}(\tau_{\rm F})={\bf 0},
\end{split}
\end{equation}
which are satisfied by an interpolating ansatz such as
\begin{equation}
\label{q69}
\begin{split}
\gamma(t)&=1+10(\gamma_{\rm F}-1)\frac{t^3}{\tau_{\rm F}^3}-15(\gamma_{\rm F}-1)\frac{t^4}{\tau_{\rm F}^4}+6(\gamma_{\rm F}-1)\frac{t^5}{\tau_{\rm F}^5},\\
\f(t)&=\left(10\frac{t^3}{\tau_{\rm F}^3}-15\frac{t^4}{\tau_{\rm F}^4}+6\frac{t^5}{\tau_{\rm F}^5}\right)\,\f_{\rm F}.
\end{split}
\end{equation}
In effect, Eq.~\eqref{q69} provides a ``recipe'' of how to engineer shortcuts to adiabaticity in expansions and transport processes. 

\begin{figure}
\includegraphics[width=0.9\linewidth]{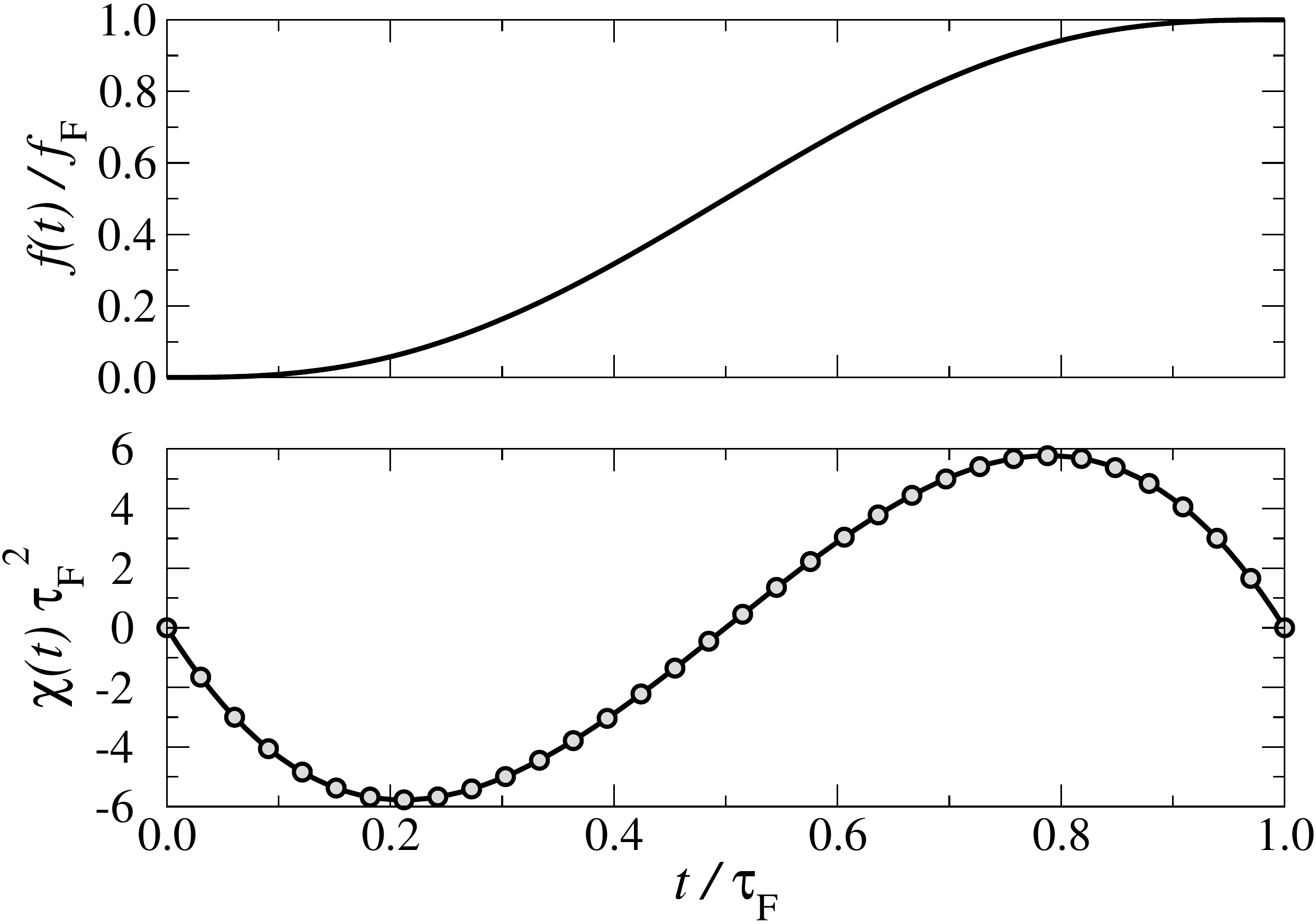}
\caption{\label{transport} {\bf Shortcut to adiabatic transport by local counterdiabatic driving.} 
Whenever the trapping potential $\U(\q,t)=\U_0(\q-f(t)\hat{{\bf n}})$ (with $\gamma(t)=1$), the auxiliary counterdiabatic term  with ${\bf F}=-\ddot{\f}_{\rm F}=:\f_{\rm F}\chi(t)$ 
induces the self-similar evolution $\Phi(t)=\mathcal{U}\Psi(t)$, which at $t=\{0,\tau_{\rm F}\}$ reduces to $\Psi(t)=\Phi(t)$ and the auxiliary term vanishes. 
CD guarantees that the density profile of the system is centered at all times at $\q=f(t)\hat{{\bf n}}$. This protocol holds exactly for an arbitrary trapping potential $\U_0(\q)$ and is valid for an arbitrary single-particle, non-linear or many-body system, requiring no modulation of the coupling constant $\epsilon(t)$ in the case of interacting systems, i.e., $\epsilon(t)=1$. 
}
\end{figure}
It is instructive to consider the amplitude $\chi(t)$ of the counterdiabatic term ${\bf F}=-\ddot{\f}_{\rm F}\equiv\f_{\rm F}\chi(t)$ to assist transport in terms of the dimensionless  time $s\equiv t/\tau_{\rm F}$,
\beqa
\chi(t)=- 60s\,(2s^2-3s+1)\,\tau_{\rm F}^2.
\eeqa
We note that the function $\chi(t)<0$ in the interval $(0,\tau_{\rm F}/2)$, so that the CD term speeds up the wavepacket towards the target position $\f_{\rm F}$. In the subsequent stage $(\tau_{\rm F}/2,\tau_{\rm F})$, $\chi(t)>0$ so that the CD term decelerates the translation until the evolving state is centered at $\q=\f_{\rm F}$, and the CD term vanishes.
The extrema ${\bf F}_{\pm}=\pm\f_{\rm F}\,45/8\tau_{\rm F}^2$ are reached at $t=\tau_{\rm F}/6\,(3\pm\sqrt{3})$.
For a transport function $\f(t)=f(t)\hat{{\bf n}}$ in the direction of an arbitrary unit vector $\hat{{\bf n}}$. Figure~\ref{transport} shows the time-dependence of both $f(t)$ and $\chi(t)$.
CD provides an exact STA to transport process, directly applicable to many-body systems such as trapped ion chains, without the restrictions to pertubative treatments in previous proposals \cite{torrontegui_2011,palmero_2013}. Alternative transport functions can be designed using optimal control theory \cite{Chen_2011x}.
\begin{figure}
\includegraphics[width=0.9\linewidth]{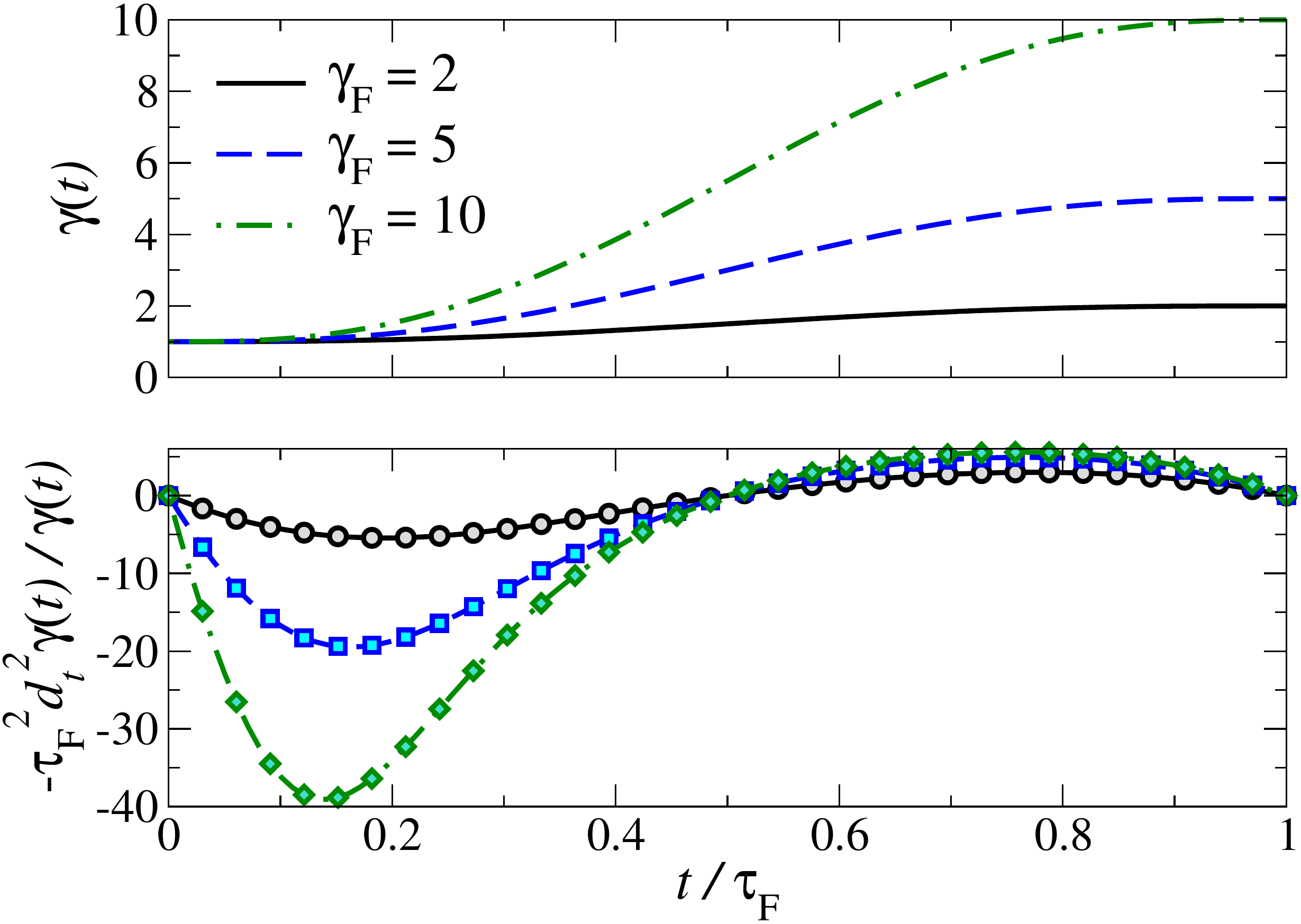}
\caption{\label{expansion} {\bf Shortcut to an adiabatic expansion by local counterdiabatic driving.}
Whenever the trapping potential $\U(\q,t)=\U_0(\q/\gamma(t))/\gamma(t)^2$ (with $\f(t)=0$), the auxiliary counterdiabatic term  modulated by $-m\ddot{\gamma}/\gamma(t)$ 
induces a non-adiabatic  self-similar  dynamics which reduces to the target state $\Psi(t)$ at $t=\{0,\tau_{\rm F}\}$,  when the auxiliary term vanishes. 
This protocol holds exactly for an arbitrary trapping potential $\U_0(\q)$ and can be applied to an arbitrary single-particle, non-linear or many-body systems.  
In the case of interacting systems, it requires a modulation of the coupling constant $\epsilon(t)$ as in Eq. \eqref{q42}. 
}
\end{figure}

The time dependence of the CD term to assist  expansions  scales in a similar way with the quench rate, this is, $-m\ddot{\gamma}\bar{\q}_i^2/\gamma\propto\tau^2$. 
In an early stage $(0,\tau_{\rm F}/2)$, the gain in the speedup of the nonadiabatic expansion is provided  by the auxiliary inverted harmonic potential, when $\ddot{\gamma}>0$, see Fig.~\ref{expansion}. The quickly expanding cloud is then slowed down ($\ddot{\gamma}<0$) by the CD term during the time interval $(\tau_{\rm F}/2,\tau_{\rm F})$, and it comes to rest as the CD term  vanishes  ($\ddot{\gamma}\rightarrow 0^-$) and the completion of the process is approached as $t\rightarrow\tau_{\rm F}$.

We point out that alternative  trajectories can be designed which are optimal according to a variety of criteria, such as minimum mean energy or operating time with fixed resources, using optimal control theory.
Stefanatos \cite{stefanatos_2013} has presented  time-optimal  protocols for the scale-invariant expansion dynamics in a time-dependent box. As we have seen,  the auxiliary driving potential in a shortcut to an adiabatic expansion for this particular example \cite{campo_boshier_2012} takes the general form associated with local CD protocols \cite{campo_2013}. Hence, the results in  \cite{stefanatos_2013} apply generally to CD scale-invariant dynamics. Similarly,  optimal trajectories for efficient transport \cite{Chen_2011x} can be adopted for scale-invariant driving with $\f(t)\neq 0$. Furthermore, an analysis of the Ehrenfest dynamics \cite{choi_2013} in CD protocols can be used as a guideline to engineer modulations of both $\gamma(t)$ and $\f(t)$.

\section{Concluding remarks\label{sec:con}}

A unifying framework has been introduced to design shortcuts to adiabaticity in both classical and quantum systems for scale-invariant processes, such as expansions, compressions, and transport. The dynamical symmetry in these processes provides the leverage with respect to the original approach to counterdiabatic driving, which demands knowledge of the spectral properties of the instantaneous Hamiltonian of the system \cite{demirplak_rice_2003,demirplak_rice_2005,berry_2009}.

In particular, we  found a closed form expression for the auxiliary term in the counterdiabatic Hamiltonian, and proposed a framework that allows to rewrite this expression in local form, which is of relevance to experimental realizations. The formalism of generating functions provides a simple, and intuitive way to find the canonical transformations that achieve this goal.

These findings were used to construct driving protocols mimicking adiabatic dynamics  for multi-particle quantum systems with arbitrary trapping potentials, as illustrated in non-harmonic examples. As an upshot, the requirements to speed up finite-time thermodynamic processes are greatly loosened. We envision applications of these ideas in the development of optimal  cooling schemes \cite{salamon_2008,salamon_2009,hoffmann_2011,choi_2011,salamon_2012,yuce_2012}, and their experimental implementation \cite{schaff_2010}. Our results also facilitate the realization of friction-free quantum pistons \cite{campo_boshier_2012,stefanatos_2013,dimartino_2013} and superadiabatic engines \cite{campo_2013b,deng_2013} by relaxing the restrictions to the shape of the confining potential and the nature of the working medium.

Remarkably, these protocols are not restricted  to non-interacting systems. In this context, we have illustrated the realization of shortcuts to adiabaticity in  systems described by non-linear equations of motion such as the Gross-Pitaevskii equation and higher-order non-linear Schr\"odinger equations. By doing so, we have shown that it is possible to  perform a fast counterdiabatic decompression (compression) of an interacting Bose-Einstein condensate in which the final state is free from excitations, providing a new route to the previous use of shortcuts to adiabaticity in the laboratory \cite{schaff_2011}.  
Further, we have  shown that experimentally-realizable countertiabatic driving schemes can be  applied to a wide variety of strongly-correlated many-body systems, broadening the applicability of  fast transport single-particle protocols \cite{torrontegui_2011,negretti_2013} explored in trapped-ion experiments \cite{bowler_2012,walther_2012}, without resorting on the validity of perturbative methods. The applicability of shortcuts to adiabaticity to many-body quantum fluids paves the way to the realization of a quantum dynamical microscope, in which a quantum fluid cloud can be   scaled up while preserving quantum correlations \cite{campo_2011}. Spatially zooming up the quantum state in these systems constitutes an interesting complementary alternative to other imaging techniques based on long expansion times \cite{lamporesi_2013} and large numerical aperture optics \cite{bakr_2009,endres_2011}.

\acknowledgments{It is a pleasure to thank  Bogdan Damski, I\~nigo L. Egusquiza, Emilie Passemar, Ayoti Patra, and Nikolai Sinitsyn for stimulating discussions and comments on the manuscript. SD and CJ acknowledge support from the National Science Foundation (USA) under grant DMR-1206971. This research is further supported by the U.S Department of Energy through the LANL/LDRD Program and a  LANL J. Robert Oppenheimer fellowship (AdC).
}

\appendix

\section{Scale-invariant eigenfunctions\label{sec:scalef}}

In this appendix we prove the scaling law for wave-functions used in Sec.~\ref{sec:xp}. Let $\psi_n^0(q)$ be an eigenfunction of the Hamiltonian $\hat{H}_0$ with
\begin{equation}
\label{a1}
\begin{split}
\hat{H}_0(1)\,\psi_n^0(q)&=E_n\,\psi_n^0(q)\,,\\
\hat{H}_0(1)&=-\frac{\hbar^2}{2m}\,\pd^2_q+\U_0(q)\,,
\end{split}
\end{equation}
then $\psi_n(q,\gamma)=\alpha(\gamma)\,\psi_n^0\left(q/\gamma\right)$ is an eigenfunction of $\hat{H}_0(\gamma)$ with
\begin{equation}
\label{a2}
\hat{H}_0(\gamma)=-\frac{\hbar^2}{2m}\,\pd^2_q+\frac{1}{\gamma^2}\,\U_0\left(q/\gamma\right)\,,
\end{equation}
where $\alpha(\gamma) $ is a normalization. The latter can be proven by direct evaluation. Consider
\begin{equation}
\label{a3}
\begin{split}
\hat{H}_0(\gamma)\,\psi_n(q,\gamma)&= -\frac{\hbar^2}{2 m}\,\pd_q^2 \alpha(\gamma)\,\psi_n^0\left(q/\gamma\right)\\
&+\frac{1}{\gamma^2}\,\U_0\left(q/\gamma\right)\alpha(\gamma)\,\psi_n^0\left(q/\gamma\right)\,,
\end{split}
\end{equation}
which can be written  as
\begin{equation}
\label{a4}
H_0(\gamma)\psi_n(q,\gamma)=\frac{\alpha(\gamma)}{\gamma^2}\,E_n\, \psi_n^0(\sigma)=\frac{E_n}{\gamma^2}\,\psi_n(q,\gamma)\,,
\end{equation}
in terms of $\sigma=q/\gamma$. It follows that if $E_n$ is an eigenvalue of $H_0$ then $E_n/\gamma^2$ is an eigenvalue of $H_0(\gamma)$ \eqref{a2}. The prefactor $\alpha(\gamma)=\gamma^{-1/2}$ can be determined from the normalization of $\psi_m(q,\gamma)$ 
\begin{equation}
\label{a5}
\begin{split}
1&=\int\td q\,[\psi_n(q,\gamma)]^2\,,\\
&=[\alpha(\gamma)]^2\,\int\td q\,\left[\psi_n\left(q/\gamma\right)\right]^2\,,\\
&=[\alpha(\gamma)]^2\,\gamma\,\int\td\sigma\,[\psi_n(\sigma)]^2=[\alpha(\gamma)]^2\,\gamma\,.
\end{split}
\end{equation}
In the above Sec.~\ref{sec:xp} we use a slightly more general result, where we allowed additionally for transport along $q$. The validity can be easily checked by replacing $q\rightarrow q-f$ everywhere in the latter proof.

\section{Illustrative example -- The Morse oscillator\label{sec:morse}}

This appendix is dedicated to an illustration of the classical results put forward in Ref.~\cite{jarzynski_2013}. Recall the classical counterdiabatic Hamiltonian,
\begin{equation}
\label{b01}
H(z,t)=H_0(z;\mb{\lambda}(t)) + \dot{\mb{\lambda}} \cdot \mb{\xi}(z,\mb{\lambda}(t))\,,
\end{equation}
where $H_1(t)=\dot{\mb{\lambda}} \cdot \mb{\xi}(\mb{\lambda}(t)) $ is the counterdiabatic term.

It was further shown that the generator $\mb{\xi}$ satisfies,
\begin{equation}
\label{b02}
\mc{\xi}(z_b;\mb{\lambda})-\mc{\xi}(z_a;\mb{\lambda})=\int_a^b \td t\,\nabla\tilde{H}_0(z(t);\mb\lambda)\,,
\end{equation}
where $z_a$ and $z_b$ are two points on the same energy shell of $H_0(z,\mb{\lambda})$, and $z(t)$ is a trajectory that evolves under $H_0$ from $z_a$ to $z_b$. For the sake of simplicity of notation, we denote in the following the gradient with respect to the control parameter $\mb{\lambda}$ by simply $\nabla$. In Eq.~\eqref{b02} $\nabla\tilde{H}_0=\nabla H_0-\la \nabla H_0\ra_{H_0,\mb{\lambda}}$, where $\la \dots\ra_{E,\mb{\lambda}}$ is the microcanonical average with
\begin{equation}
\label{b03}
\la\dots\ra_{E,\mb{\lambda}}\equiv \frac{1}{\pd_E\Omega}\,\int\td z\,\de{E-H_0(z,\mb{\lambda})}\,.
\end{equation} 
Finally, $\Omega(E,\mb{\lambda})$ is the phase space volume enclosed by the energy shell $E$,
\begin{equation}
\label{b04}
\Omega(E,\mb{\lambda})\equiv\int\td z\,\Theta(E-H_0(z,\mb{\lambda}))\,,
\end{equation}
and we have
\begin{equation}
\label{b05}
\nabla E(\Omega,\mb{\lambda})=-\frac{\nabla \Omega(E,\mb{\lambda})}{\pd_E\,\Omega(E,\mb{\lambda})}=\la\nabla H_0\ra_{E,\mb{\lambda}}\,.
\end{equation}
The latter formalism can be used to calculate explicit expression for the generator $\mb{\xi}$.

We proceed by illustrating the above findings for a completely analytical solvable system, namely the parametric Morse oscillator. The Morse potential can be written as \cite{flugge_1971},
\begin{equation}
\label{b06}
\U(q)=\U_m\,\left[\e{-2\beta q}-2\e{-\beta q}\right]\,.
\end{equation}
In the remainder of this appendix we compute $\xi$ \eqref{b01} explicitly for three different driving protocols. We start with scale invariant driving, before we vary either only the potential width $\beta$ or the potential depth $\U_m$.

\begin{figure}
\includegraphics[width=0.9\linewidth]{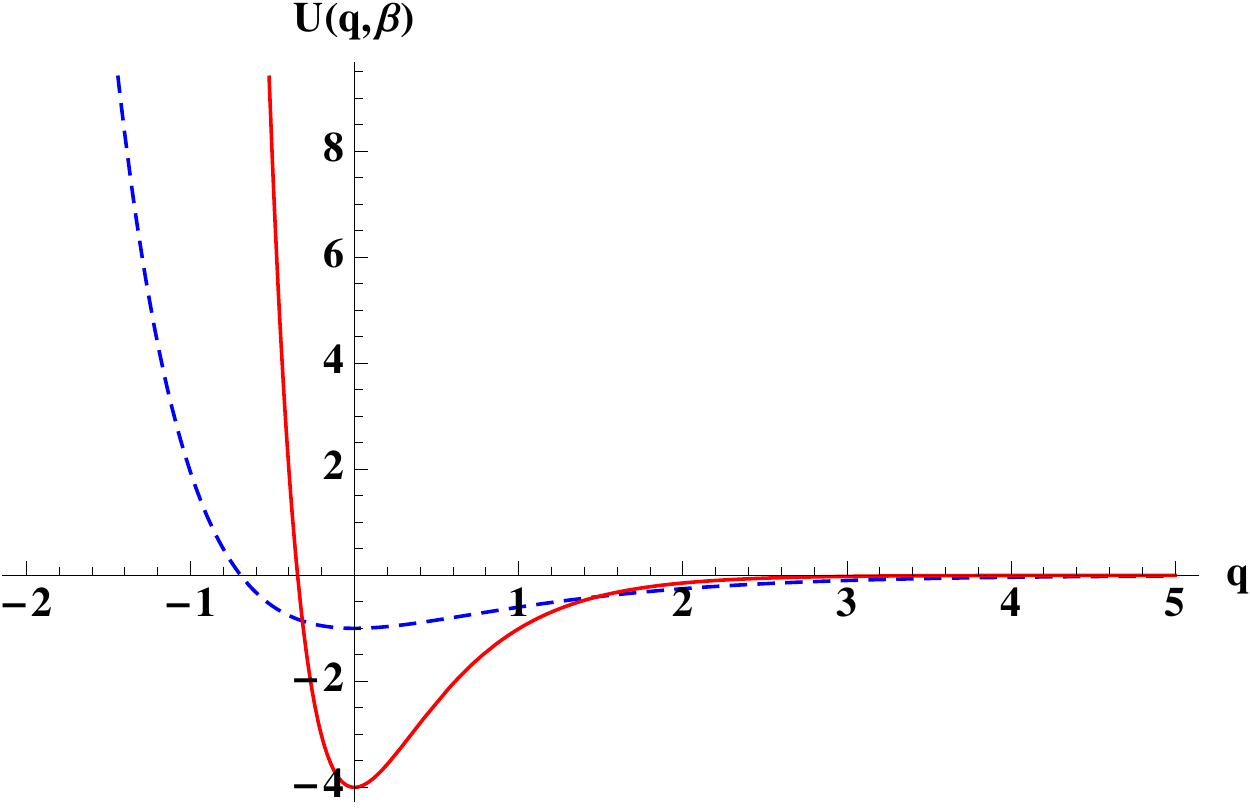}
\caption{\label{fig1} {\bf Scale-invariant counterdiabatic driving of the Morse oscillator} Morse potential $U(q,\beta)$ \eqref{b07} for $\gamma=1$ (blue, dashed line) and $\gamma=1/2$ (red, solid line).}
\end{figure}

\begin{figure}
\includegraphics[width=0.9\linewidth]{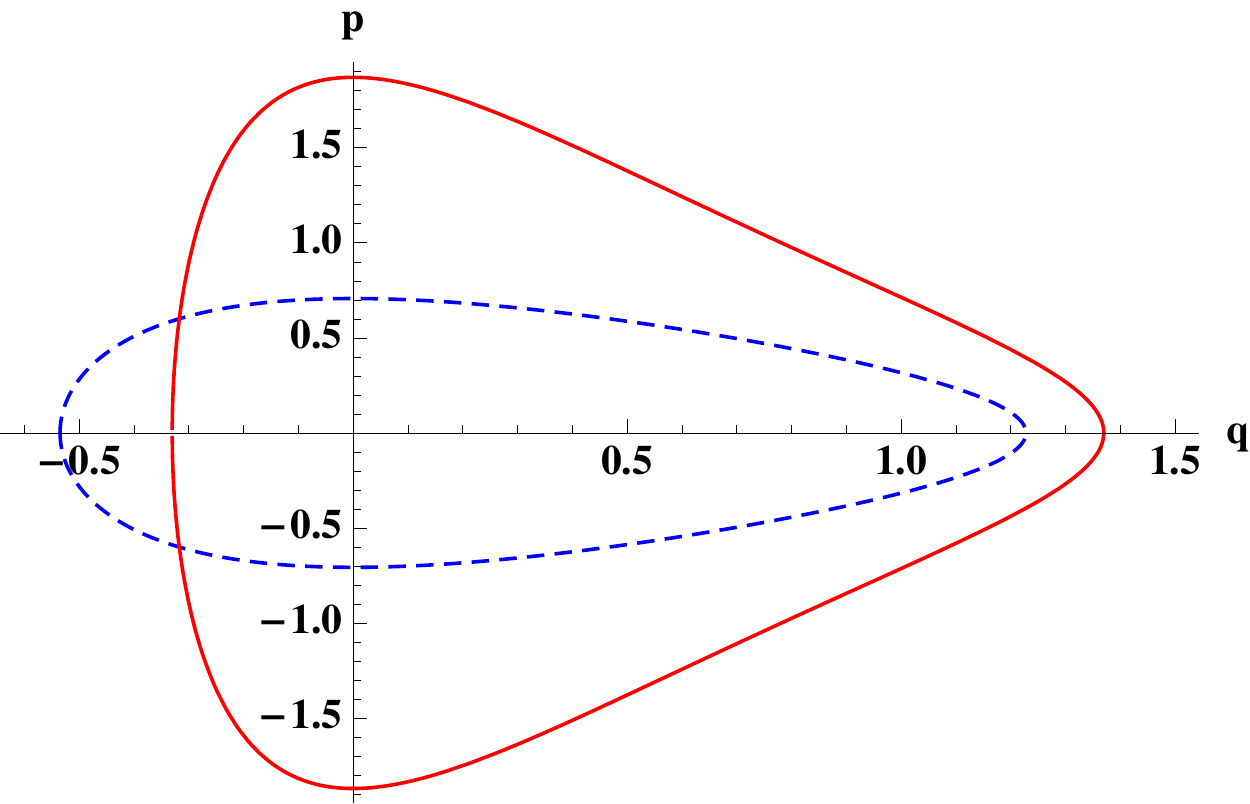}
\caption{\label{fig2} Phase space volume $\Omega(E,\gamma)$ corresponding to $\U(q,\gamma)$ \eqref{b07} for $E=-1/2$ and $\gamma=1$ (blue, dashed line) and $\gamma=1/2$ (red, solid line). }
\end{figure}

\paragraph*{Scale-invariant parameterization}
For scale-invariant driving $\U(q)$ \eqref{b06} takes the form,
\begin{equation}
\label{b07}
\U(q,\gamma(t))=\frac{U_m}{\gamma^2(t)} \left[\e{-2 \frac{\beta\,q}{\gamma(t)}}-2\e{-\frac{\beta\,q}{\gamma(t)}}\right]\,.
\end{equation}
For the sake of simplicity and to avoid clutter, we work in units where $U_m=1$ and $\beta=1$. In Fig.~\ref{fig1}, $\U(q,\gamma)$ is shown for two different values of $\gamma$.
For the present case Eq.~\eqref{b02} simplifies to read
\begin{equation}
\label{b08}
\xi(z_b;\,\gamma)-\xi(z_a;\,\gamma)=\int\limits_a^b\td t\,\left(\pd_\gamma H_0-\la\pd_\gamma H_0\ra \right)\,,
\end{equation}
where $z$ is again the phase space variable. In Fig.~\ref{fig2} we plot two energy shells corresponding to the same energy but different values of $\gamma$, whose volume is the microcanonical partition function $\Omega(E,\gamma) $.

As a first step we have to calculate the phase space volume $\Omega(E,\gamma)$, which is given by
\begin{equation}
\label{b09}
\begin{split}
\Omega(E,\gamma)&=\int\limits_{q_1}^{q_2}\td q\,p(q,E)\,,\\
& =\int\limits_{q_1}^{q_2}\td q\,\sqrt{2\mu \left(E-\U(q,\gamma)\right)}\,,
\end{split}
\end{equation}
where $p(q,E)$ is the momentum, and $p(q_{1,2},E)=0$. Note that $\Omega(E,\gamma)$ is only finite for bound state, that means $0\geq E\geq -\gamma^{-2}$. For the sake of simplicity we continue in units, where the mass $m=1/2$. Then the zeros of the momentum are given by,
\begin{equation}
\label{b10}
\begin{split}
q_1&=-\gamma\, \lo{1+\sqrt{1+E\,\gamma^2}}\,,\\
q_2&=-\gamma\, \lo{1-\sqrt{1+E\,\gamma^2}}\,.
\end{split}
\end{equation}
The phase space volume is obtained by calculating the integral in Eq.~\eqref{b09}. We have,
\begin{equation}
\label{b11}
\Omega(E,\gamma)=2\pi\left(1-\sqrt{-E\,\gamma^2}\right)\,.
\end{equation}
Accordingly, we obtain with Eq.~\eqref{b08},
\begin{equation}
\label{b12}
\dot{\xi}=2 p^2/\gamma +2 q\gamma^{-4}\,\e{-2 q/\lambda}-2 q\gamma^{-4}\,\e{-q/\gamma}\,.
\end{equation}
Finally, $\xi$ is obtained by either integrating Eq.~\eqref{b08} or by solving the partial differential equation
\begin{equation}
\label{b13}
\dot{\xi}=-\pd_q \xi\,\pd_p H_0+\pd_p \xi\,\pd_q H_0=-\{\xi, H_0\}\,,
\end{equation}
where $\{.,.\}$ denotes again the Poisson bracket. The right hand side of Eq.~\eqref{b13} can be explicitly written as,
\begin{equation}
\label{b14}
\begin{split}
-\{\xi, H_0\}&=2 p\,\pd_q\xi +2 \gamma^{-3}\,\e{-2 q/\gamma}\,\pd_p \xi\\
&-2\gamma^{-3}\,\e{- q/\gamma}\,\pd_p\xi\,.
\end{split}
\end{equation}
One easily convinces oneself (almost by inspection) that a solution to Eq.~\eqref{b13} is given by,
\begin{equation}
\label{b15}
\xi=q p/\gamma\,,
\end{equation}
which is a solution to Eq.~\eqref{b15}. The latter result is in perfect agreement with the above analysis that led to Eq.~\eqref{q13}.

\paragraph*{Arbitrary parameterization}
Scale-invariant dynamics have proven to be theoretically useful as they allow the computation of the counterdiabatic term $\xi$ in closed form. From an experimental point of view, however, it might be more relevant to vary the two parameters, the potential depth $\U_m$ and the width $\beta$, independently. Therefore, we continue our discussion with examples where we vary only $\U_m$ or $\beta$, while the other parameter is kept constant. 

\paragraph*{Time-dependent width}

We continue by considering a Morse potential of the form
\begin{equation}
\label{b16}
\U_w(q,\beta(t))=\e{-2\beta(t) q}-2\e{-\beta(t) q}\,,
\end{equation}
where we set again for the sake of simplicity $U_m=1$. This choice of units it not necessary, but convenient as it drastically reduces the clutter in the formulas.
%
As before we need to compute $\Omega(E,\beta)$ first.
%
With the zeros of the momentum $p(q,E)$ being given by
\begin{equation}
\label{b17}
\begin{split}
q_1&=-\frac{1}{\beta}\,\lo{1+\sqrt{1+E}}\,,\\
q_2&=-\frac{1}{\beta}\,\lo{1-\sqrt{1+E}}\,,
\end{split}
\end{equation}
we obtain
\begin{equation}
\label{b18}
\Omega_w(E,\beta)=2\pi\,\left(\frac{1}{\beta^2}-\frac{\sqrt{-E}}{\beta}\right)\,.
\end{equation}
Accordingly we have for $\dot{\xi}$
\begin{widetext}
\begin{equation}
\label{b19}
\dot{\xi}_w=-2 p^2/\beta-2\e{-2\beta q}/\beta+4\e{-\beta q}/\beta -2 q\e{-2\beta q}+2 q \e{-\beta q} + 4\,\sqrt{-p^2-\e{-2\beta q}+2\e{\beta q} }/\beta^2.
\end{equation}
After a few lines a simplified expression for $\xi$ can be found, which reads,
\begin{equation}
\label{b20}
\xi_w=\frac{p}{\beta^2}+\frac{qp}{\beta}-\frac{1}{\beta^2}\,\arctan\left(\frac{1-\e{-\beta q}}{p}\right)-\frac{2}{\beta^3}\,\mrm{arccot}\left(\frac{p\,\e{-\beta q}\,\sqrt{-H_0(q,p)}}{1+\e{\beta q}\,H_0(q,p)}\right)\,.
\end{equation}
We observe that the second term in Eq.~\eqref{b18} is identical to the counterdiabatic term for scale-invariant dynamics. Moreover, $H_0$ is the ``unperturbed'' Hamiltonian as a function of the phase space variables $p$ and $q$. Note also that the latter expression is valid for all bound states, i.e. $E=H_0(q,p)\leq 0$.

\paragraph*{Time-dependent depth}

As a final example let us consider a Morse potential, whose depth $\U_m$ is varied, while its width is kept constant. We have,
\begin{equation}
\label{b21}
\U_d(q,\U_m(t))=\U_m(t)\,\left(\e{-2 q}-2\e{-q}\right)\,,
\end{equation}
and where we this time set $\beta=1$.
As before we need the zeros of the momentum, which read here
\begin{equation}
\label{b22}
q_1=\lo{\frac{-\U_m+\sqrt{\U_m\left(E+\U_m\right)}}{E}}\quad \mathrm{and}\quad q_2=\lo{-\frac{\U_m+\sqrt{\U_m\left(E+\U_m\right)}}{E}}\,,
\end{equation}
with which we obtain,
\begin{equation}
\label{b23}
\Omega_d(E,\U_m)=2\pi \left(\sqrt{\U_m}-\sqrt{-E}\right)\,.
\end{equation}
In complete analogy to the previous two examples we compute $\dot{\xi}$, which reads here,
\begin{equation}
\label{b24}
\dot{\xi}_d=\e{-2 q}-2\e{-q}+\sqrt{-p^2-\U_m\left(\e{-2 q}+2\e{ q}\right)}/\sqrt{\U_m}\,. 
\end{equation}
Although the expression for $\dot{\xi}$ appears to be simpler, the integral, $\xi$, is much more involved. After several lines of simplification we obtain,

\begin{equation}
\label{b25}
\begin{split}
&\xi_d=-\frac{p}{2 \U_m}\\
&-\frac{1}{2\sqrt{\U_m}}\,\arctan{\left(\frac{p\e{q}\left[\e{q} H_0+\U_m+\left(\e{q}-1\right)\sqrt{-\U_m\,H_0}\right]}{\e{2q}\sqrt{-H_0^3}+\e{q}\left(\e{q}-1\right)\sqrt{\U_m}\,H_0+\left(1-2 \e{q}\right)\U_m\,\sqrt{-H_0}+\left(\e{q}-1\right)\sqrt{\U_m^3}}\right)}\,.
\end{split}
\end{equation}
By comparing the closed form expressions for the counterdiabatic term Eqs.~\eqref{b15}, \eqref{b20}, and \eqref{b25} it becomes obvious how much scale-invariant driving simplifies the situation. Whereas for scale-invariant driving $H_1$ can be brought into local form with the help of an appropriate coordinate transformation, this seems hardly feasible in the general case.

\end{widetext}

\bibliography{bib_11}

\begin{thebibliography}{93}%
\makeatletter
\providecommand \@ifxundefined [1]{%
 \@ifx{#1\undefined}
}%
\providecommand \@ifnum [1]{%
 \ifnum #1\expandafter \@firstoftwo
 \else \expandafter \@secondoftwo
 \fi
}%
\providecommand \@ifx [1]{%
 \ifx #1\expandafter \@firstoftwo
 \else \expandafter \@secondoftwo
 \fi
}%
\providecommand \natexlab [1]{#1}%
\providecommand \enquote  [1]{``#1''}%
\providecommand \bibnamefont  [1]{#1}%
\providecommand \bibfnamefont [1]{#1}%
\providecommand \citenamefont [1]{#1}%
\providecommand \href@noop [0]{\@secondoftwo}%
\providecommand \href [0]{\begingroup \@sanitize@url \@href}%
\providecommand \@href[1]{\@@startlink{#1}\@@href}%
\providecommand \@@href[1]{\endgroup#1\@@endlink}%
\providecommand \@sanitize@url [0]{\catcode `\\12\catcode `\$12\catcode
  `\&12\catcode `\#12\catcode `\^12\catcode `\_12\catcode `\%12\relax}%
\providecommand \@@startlink[1]{}%
\providecommand \@@endlink[0]{}%
\providecommand \url  [0]{\begingroup\@sanitize@url \@url }%
\providecommand \@url [1]{\endgroup\@href {#1}{\urlprefix }}%
\providecommand \urlprefix  [0]{URL }%
\providecommand \Eprint [0]{\href }%
\@ifxundefined \urlstyle {%
  \providecommand \doi  [0]{\begingroup \@sanitize@url \@doi}%
  \providecommand \@doi [1]{\endgroup \@@startlink {\doibase
  #1}doi:\discretionary {}{}{}#1\@@endlink }%
}{%
  \providecommand \doi  [0]{doi:\discretionary{}{}{}\begingroup
  \urlstyle{rm}\Url }%
}%
\providecommand \doibase [0]{http://dx.doi.org/}%
\providecommand \Doi [0]{\begingroup \@sanitize@url \@Doi }%
\providecommand \@Doi  [1]{\endgroup\@@startlink{\doibase#1}\@@Doi}%
\providecommand \@@Doi [1]{#1\@@endlink}%
\providecommand \selectlanguage [0]{\@gobble}%
\providecommand \bibinfo  [0]{\@secondoftwo}%
\providecommand \bibfield  [0]{\@secondoftwo}%
\providecommand \translation [1]{[#1]}%
\providecommand \BibitemOpen [0]{}%
\providecommand \bibitemStop [0]{}%
\providecommand \bibitemNoStop [0]{.\EOS\space}%
\providecommand \EOS [0]{\spacefactor3000\relax}%
\providecommand \BibitemShut  [1]{\csname bibitem#1\endcsname}%
\bibitem [{\citenamefont {Collins}\ \emph {et~al.}(1997)\citenamefont
  {Collins}, \citenamefont {Zettl}, \citenamefont {Bando}, \citenamefont
  {Thess},\ and\ \citenamefont {Smalley}}]{collins_1997}%
  \BibitemOpen
  \bibfield  {author} {\bibinfo {author} {\bibfnamefont {Ph.~G.}\ \bibnamefont
  {Collins}}, \bibinfo {author} {\bibfnamefont {A.}~\bibnamefont {Zettl}},
  \bibinfo {author} {\bibfnamefont {H.}~\bibnamefont {Bando}}, \bibinfo
  {author} {\bibfnamefont {A.}~\bibnamefont {Thess}}, \ and\ \bibinfo {author}
  {\bibfnamefont {R.~E.}\ \bibnamefont {Smalley}},\ }\bibfield  {title}
  {\enquote {\bibinfo {title} {Nanotube nanodevice},}\ }\href@noop {}
  {\bibfield  {journal} {\bibinfo  {journal} {Science},\ }\textbf {\bibinfo
  {volume} {278}},\ \bibinfo {pages} {100} (\bibinfo {year}
  {1997})}\BibitemShut {NoStop}%
\bibitem [{\citenamefont {Ming}\ \emph {et~al.}(2003)\citenamefont {Ming},
  \citenamefont {Huang}, \citenamefont {Zorman}, \citenamefont {Mehregany},\
  and\ \citenamefont {Roukes}}]{huang_2003}%
  \BibitemOpen
  \bibfield  {author} {\bibinfo {author} {\bibfnamefont {X.}~\bibnamefont
  {Ming}}, \bibinfo {author} {\bibfnamefont {H.}~\bibnamefont {Huang}},
  \bibinfo {author} {\bibfnamefont {C.~A.}\ \bibnamefont {Zorman}}, \bibinfo
  {author} {\bibfnamefont {M.}~\bibnamefont {Mehregany}}, \ and\ \bibinfo
  {author} {\bibfnamefont {M.~L.}\ \bibnamefont {Roukes}},\ }\bibfield  {title}
  {\enquote {\bibinfo {title} {Nanoelectromechanical systems: {N}anodevice
  motion at microwave frequencies},}\ }\href@noop {} {\bibfield  {journal}
  {\bibinfo  {journal} {Nature},\ }\textbf {\bibinfo {volume} {421}},\ \bibinfo
  {pages} {496} (\bibinfo {year} {2003})}\BibitemShut {NoStop}%
\bibitem [{\citenamefont {Kane}(1998)}]{kane_1998}%
  \BibitemOpen
  \bibfield  {author} {\bibinfo {author} {\bibfnamefont {B.~E.}\ \bibnamefont
  {Kane}},\ }\bibfield  {title} {\enquote {\bibinfo {title} {A silicon-based
  nuclear spin quantum computer},}\ }\href@noop {} {\bibfield  {journal}
  {\bibinfo  {journal} {Nature},\ }\textbf {\bibinfo {volume} {393}},\ \bibinfo
  {pages} {133} (\bibinfo {year} {1998})}\BibitemShut {NoStop}%
\bibitem [{\citenamefont {Kinoshita}\ \emph {et~al.}(2006)\citenamefont
  {Kinoshita}, \citenamefont {Wenger},\ and\ \citenamefont
  {Weiss}}]{kinoshita_2006}%
  \BibitemOpen
  \bibfield  {author} {\bibinfo {author} {\bibfnamefont {T.}~\bibnamefont
  {Kinoshita}}, \bibinfo {author} {\bibfnamefont {T.}~\bibnamefont {Wenger}}, \
  and\ \bibinfo {author} {\bibfnamefont {D.~S.}\ \bibnamefont {Weiss}},\
  }\bibfield  {title} {\enquote {\bibinfo {title} {A quantum {N}ewton's
  cradle},}\ }\href@noop {} {\bibfield  {journal} {\bibinfo  {journal}
  {Nature},\ }\textbf {\bibinfo {volume} {440}},\ \bibinfo {pages} {900}
  (\bibinfo {year} {2006})}\BibitemShut {NoStop}%
\bibitem [{\citenamefont {Avron}\ \emph {et~al.}(1988)\citenamefont {Avron},
  \citenamefont {Raveh},\ and\ \citenamefont {Zur}}]{avron_1988}%
  \BibitemOpen
  \bibfield  {author} {\bibinfo {author} {\bibfnamefont {J.~E.}\ \bibnamefont
  {Avron}}, \bibinfo {author} {\bibfnamefont {A.}~\bibnamefont {Raveh}}, \ and\
  \bibinfo {author} {\bibfnamefont {B.}~\bibnamefont {Zur}},\ }\bibfield
  {title} {\enquote {\bibinfo {title} {Adiabatic quantum transport in multiply
  connected systems},}\ }\href@noop {} {\bibfield  {journal} {\bibinfo
  {journal} {\RMP},\ }\textbf {\bibinfo {volume} {60}},\ \bibinfo {pages} {873}
  (\bibinfo {year} {1988})}\BibitemShut {NoStop}%
\bibitem [{\citenamefont {Kr\'{a}l}\ \emph {et~al.}(2007)\citenamefont
  {Kr\'{a}l}, \citenamefont {Thanopulos},\ and\ \citenamefont
  {Shapiro}}]{kral_2007}%
  \BibitemOpen
  \bibfield  {author} {\bibinfo {author} {\bibfnamefont {P.}~\bibnamefont
  {Kr\'{a}l}}, \bibinfo {author} {\bibfnamefont {I.}~\bibnamefont
  {Thanopulos}}, \ and\ \bibinfo {author} {\bibfnamefont {M.}~\bibnamefont
  {Shapiro}},\ }\bibfield  {title} {\enquote {\bibinfo {title} {Coherently
  controlled adiabatic passage},}\ }\href@noop {} {\bibfield  {journal}
  {\bibinfo  {journal} {\RMP},\ }\textbf {\bibinfo {volume} {79}},\ \bibinfo
  {pages} {53} (\bibinfo {year} {2007})}\BibitemShut {NoStop}%
\bibitem [{\citenamefont {Giovannetti}\ \emph {et~al.}(2006)\citenamefont
  {Giovannetti}, \citenamefont {Lloyd},\ and\ \citenamefont {Maccone}}]{gio06}%
  \BibitemOpen
  \bibfield  {author} {\bibinfo {author} {\bibfnamefont {V.}~\bibnamefont
  {Giovannetti}}, \bibinfo {author} {\bibfnamefont {S.}~\bibnamefont {Lloyd}},
  \ and\ \bibinfo {author} {\bibfnamefont {L.}~\bibnamefont {Maccone}},\
  }\bibfield  {title} {\enquote {\bibinfo {title} {Quantum metrology},}\
  }\href@noop {} {\bibfield  {journal} {\bibinfo  {journal} {\PRL},\ }\textbf
  {\bibinfo {volume} {96}},\ \bibinfo {pages} {010401} (\bibinfo {year}
  {2006})}\BibitemShut {NoStop}%
\bibitem [{\citenamefont {Andresen}\ \emph {et~al.}(1984)\citenamefont
  {Andresen}, \citenamefont {Salamon},\ and\ \citenamefont
  {Berry}}]{andresen_1984}%
  \BibitemOpen
  \bibfield  {author} {\bibinfo {author} {\bibfnamefont {B.}~\bibnamefont
  {Andresen}}, \bibinfo {author} {\bibfnamefont {P.}~\bibnamefont {Salamon}}, \
  and\ \bibinfo {author} {\bibfnamefont {R.~S.}\ \bibnamefont {Berry}},\
  }\bibfield  {title} {\enquote {\bibinfo {title} {Thermodynamics in finite
  time},}\ }\href@noop {} {\bibfield  {journal} {\bibinfo  {journal} {Phys.
  Today},\ }\textbf {\bibinfo {volume} {37}},\ \bibinfo {pages} {62} (\bibinfo
  {year} {1984})}\BibitemShut {NoStop}%
\bibitem [{\citenamefont {Trabesinger}(2012)}]{trabesinger_2012}%
  \BibitemOpen
  \bibfield  {author} {\bibinfo {author} {\bibfnamefont {A.}~\bibnamefont
  {Trabesinger}},\ }\bibfield  {title} {\enquote {\bibinfo {title} {Quantum
  simulation},}\ }\href@noop {} {\bibfield  {journal} {\bibinfo  {journal}
  {Nature Physics},\ }\textbf {\bibinfo {volume} {8}},\ \bibinfo {pages} {263}
  (\bibinfo {year} {2012})}\BibitemShut {NoStop}%
\bibitem [{\citenamefont {Nielsen}\ and\ \citenamefont
  {Chuang}(2000)}]{nielsen_00}%
  \BibitemOpen
  \bibfield  {author} {\bibinfo {author} {\bibfnamefont {M.~A.}\ \bibnamefont
  {Nielsen}}\ and\ \bibinfo {author} {\bibfnamefont {I.~L.}\ \bibnamefont
  {Chuang}},\ }\href@noop {} {\emph {\bibinfo {title} {Quantum Computation and
  Quantum Information}}}\ (\bibinfo  {publisher} {Cambridge University Press},\
  \bibinfo {address} {Cambridge, UK},\ \bibinfo {year} {2000})\BibitemShut
  {NoStop}%
\bibitem [{\citenamefont {Messiah}(1966)}]{messiah_1966}%
  \BibitemOpen
  \bibfield  {author} {\bibinfo {author} {\bibfnamefont {A.}~\bibnamefont
  {Messiah}},\ }\href@noop {} {\emph {\bibinfo {title} {Quantum Mechanics}}}\
  (\bibinfo  {publisher} {John Wiley},\ \bibinfo {address} {New York},\
  \bibinfo {year} {1966})\BibitemShut {NoStop}%
\bibitem [{\citenamefont {Torrontegui}\ \emph {et~al.}(2013)\citenamefont
  {Torrontegui}, \citenamefont {S.~Ib\'a\~nez}, \citenamefont
  {Mart\'inez-Garaot}, \citenamefont {Modugno}, \citenamefont {del Campo},
  \citenamefont {Gu\'ery-Odelin}, \citenamefont {Ruschhaupt}, \citenamefont
  {Chen},\ and\ \citenamefont {Muga}}]{torrontegui_2013}%
  \BibitemOpen
  \bibfield  {author} {\bibinfo {author} {\bibfnamefont {E.}~\bibnamefont
  {Torrontegui}}, \bibinfo {author} {\bibfnamefont {S.}~\bibnamefont
  {S.~Ib\'a\~nez}}, \bibinfo {author} {\bibfnamefont {S.}~\bibnamefont
  {Mart\'inez-Garaot}}, \bibinfo {author} {\bibfnamefont {M.}~\bibnamefont
  {Modugno}}, \bibinfo {author} {\bibfnamefont {A.}~\bibnamefont {del Campo}},
  \bibinfo {author} {\bibfnamefont {D.}~\bibnamefont {Gu\'ery-Odelin}},
  \bibinfo {author} {\bibfnamefont {A.}~\bibnamefont {Ruschhaupt}}, \bibinfo
  {author} {\bibfnamefont {Xi}~\bibnamefont {Chen}}, \ and\ \bibinfo {author}
  {\bibfnamefont {J.~G.}\ \bibnamefont {Muga}},\ }\bibfield  {title} {\enquote
  {\bibinfo {title} {Chapter 2 -- shortcuts to adiabaticity},}\ }\href@noop {}
  {\bibfield  {journal} {\bibinfo  {journal} {Adv. At. Mol. Opt. Phys.},\
  }\textbf {\bibinfo {volume} {62}},\ \bibinfo {pages} {117} (\bibinfo {year}
  {2013})}\BibitemShut {NoStop}%
\bibitem [{\citenamefont {Chen}\ \emph {et~al.}(2010)\citenamefont {Chen},
  \citenamefont {Ruschhaupt}, \citenamefont {Schmidt}, \citenamefont {del
  Campo}, \citenamefont {Gu\'ery-Odelin},\ and\ \citenamefont
  {Muga}}]{chen_2010}%
  \BibitemOpen
  \bibfield  {author} {\bibinfo {author} {\bibfnamefont {X.}~\bibnamefont
  {Chen}}, \bibinfo {author} {\bibfnamefont {A.}~\bibnamefont {Ruschhaupt}},
  \bibinfo {author} {\bibfnamefont {S.}~\bibnamefont {Schmidt}}, \bibinfo
  {author} {\bibfnamefont {A.}~\bibnamefont {del Campo}}, \bibinfo {author}
  {\bibfnamefont {D.}~\bibnamefont {Gu\'ery-Odelin}}, \ and\ \bibinfo {author}
  {\bibfnamefont {J.~G.}\ \bibnamefont {Muga}},\ }\bibfield  {title} {\enquote
  {\bibinfo {title} {Fast optimal frictionless atom cooling in harmonic traps:
  Shortcut to adiabaticity},}\ }\href@noop {} {\bibfield  {journal} {\bibinfo
  {journal} {\PRL},\ }\textbf {\bibinfo {volume} {104}},\ \bibinfo {pages}
  {063002} (\bibinfo {year} {2010})}\BibitemShut {NoStop}%
\bibitem [{\citenamefont {Chen}\ \emph
  {et~al.}(2011){\natexlab{a}}\citenamefont {Chen}, \citenamefont
  {Torrontegui},\ and\ \citenamefont {Muga}}]{chen_2011}%
  \BibitemOpen
  \bibfield  {author} {\bibinfo {author} {\bibfnamefont {X.}~\bibnamefont
  {Chen}}, \bibinfo {author} {\bibfnamefont {E.}~\bibnamefont {Torrontegui}}, \
  and\ \bibinfo {author} {\bibfnamefont {J.~G.}\ \bibnamefont {Muga}},\
  }\bibfield  {title} {\enquote {\bibinfo {title} {{Lewis-Riesenfeld}
  invariants and transitionless quantum driving},}\ }\href@noop {} {\bibfield
  {journal} {\bibinfo  {journal} {\PRA},\ }\textbf {\bibinfo {volume} {83}},\
  \bibinfo {pages} {062116} (\bibinfo {year} {2011}{\natexlab{a}})}\BibitemShut
  {NoStop}%
\bibitem [{\citenamefont {del Campo}(2011){\natexlab{a}}}]{campo_2011}%
  \BibitemOpen
  \bibfield  {author} {\bibinfo {author} {\bibfnamefont {A.}~\bibnamefont {del
  Campo}},\ }\bibfield  {title} {\enquote {\bibinfo {title} {Frictionless
  quantum quenches in ultracold gases: {A} quantum-dynamical microscope},}\
  }\href@noop {} {\bibfield  {journal} {\bibinfo  {journal} {\PRA},\ }\textbf
  {\bibinfo {volume} {84}},\ \bibinfo {pages} {031606(R)} (\bibinfo {year}
  {2011}{\natexlab{a}})}\BibitemShut {NoStop}%
\bibitem [{\citenamefont {del Campo}\ and\ \citenamefont
  {Boshier}(2012)}]{campo_boshier_2012}%
  \BibitemOpen
  \bibfield  {author} {\bibinfo {author} {\bibfnamefont {A.}~\bibnamefont {del
  Campo}}\ and\ \bibinfo {author} {\bibfnamefont {M.~G.}\ \bibnamefont
  {Boshier}},\ }\bibfield  {title} {\enquote {\bibinfo {title} {Shortcuts to
  adiabaticity in a time-dependent box},}\ }\href@noop {} {\bibfield  {journal}
  {\bibinfo  {journal} {Sci. Rep.},\ }\textbf {\bibinfo {volume} {2}},\
  \bibinfo {pages} {648} (\bibinfo {year} {2012})}\BibitemShut {NoStop}%
\bibitem [{\citenamefont {Masuda}\ and\ \citenamefont
  {Nakamura}(2010)}]{masuda_2010}%
  \BibitemOpen
  \bibfield  {author} {\bibinfo {author} {\bibfnamefont {S.}~\bibnamefont
  {Masuda}}\ and\ \bibinfo {author} {\bibfnamefont {K.}~\bibnamefont
  {Nakamura}},\ }\bibfield  {title} {\enquote {\bibinfo {title} {Fast-forward
  of adiabatic dynamics in quantum mechanics},}\ }\href@noop {} {\bibfield
  {journal} {\bibinfo  {journal} {Proc. R. Soc. A},\ }\textbf {\bibinfo
  {volume} {466}},\ \bibinfo {pages} {1135} (\bibinfo {year}
  {2010})}\BibitemShut {NoStop}%
\bibitem [{\citenamefont {Masuda}\ and\ \citenamefont
  {Nakamura}(2011)}]{masuda_2011}%
  \BibitemOpen
  \bibfield  {author} {\bibinfo {author} {\bibfnamefont {S.}~\bibnamefont
  {Masuda}}\ and\ \bibinfo {author} {\bibfnamefont {K.}~\bibnamefont
  {Nakamura}},\ }\bibfield  {title} {\enquote {\bibinfo {title} {Acceleration
  of adiabatic quantum dynamics in electromagnetic fields},}\ }\href@noop {}
  {\bibfield  {journal} {\bibinfo  {journal} {Phys. Rev. A},\ }\textbf
  {\bibinfo {volume} {84}},\ \bibinfo {pages} {043434} (\bibinfo {year}
  {2011})}\BibitemShut {NoStop}%
\bibitem [{\citenamefont {Torrontegui}\ \emph {et~al.}(2012)\citenamefont
  {Torrontegui}, \citenamefont {Mart{\'i}nez-Garaot}, \citenamefont
  {Ruschhaupt},\ and\ \citenamefont {Muga}}]{torrontegui_2012}%
  \BibitemOpen
  \bibfield  {author} {\bibinfo {author} {\bibfnamefont {E.}~\bibnamefont
  {Torrontegui}}, \bibinfo {author} {\bibfnamefont {S.}~\bibnamefont
  {Mart{\'i}nez-Garaot}}, \bibinfo {author} {\bibfnamefont {A.}~\bibnamefont
  {Ruschhaupt}}, \ and\ \bibinfo {author} {\bibfnamefont {J.~G.}\ \bibnamefont
  {Muga}},\ }\bibfield  {title} {\enquote {\bibinfo {title} {Shortcuts to
  adiabaticity: Fast-forward approach},}\ }\href@noop {} {\bibfield  {journal}
  {\bibinfo  {journal} {\PRA},\ }\textbf {\bibinfo {volume} {86}},\ \bibinfo
  {pages} {013601} (\bibinfo {year} {2012})}\BibitemShut {NoStop}%
\bibitem [{\citenamefont {Demirplak}\ and\ \citenamefont
  {Rice}(2003)}]{demirplak_rice_2003}%
  \BibitemOpen
  \bibfield  {author} {\bibinfo {author} {\bibfnamefont {M.}~\bibnamefont
  {Demirplak}}\ and\ \bibinfo {author} {\bibfnamefont {S.~A.}\ \bibnamefont
  {Rice}},\ }\bibfield  {title} {\enquote {\bibinfo {title} {Adiabatic
  population transfer with control fields},}\ }\href@noop {} {\bibfield
  {journal} {\bibinfo  {journal} {J. Chem. Phys. A},\ }\textbf {\bibinfo
  {volume} {107}},\ \bibinfo {pages} {9937} (\bibinfo {year}
  {2003})}\BibitemShut {NoStop}%
\bibitem [{\citenamefont {Demirplak}\ and\ \citenamefont
  {Rice}(2005)}]{demirplak_rice_2005}%
  \BibitemOpen
  \bibfield  {author} {\bibinfo {author} {\bibfnamefont {M.}~\bibnamefont
  {Demirplak}}\ and\ \bibinfo {author} {\bibfnamefont {S.~A.}\ \bibnamefont
  {Rice}},\ }\bibfield  {title} {\enquote {\bibinfo {title} {Assisted adiabatic
  passage revisited},}\ }\href@noop {} {\bibfield  {journal} {\bibinfo
  {journal} {J. Chem. Phys. B},\ }\textbf {\bibinfo {volume} {109}},\ \bibinfo
  {pages} {6838} (\bibinfo {year} {2005})}\BibitemShut {NoStop}%
\bibitem [{\citenamefont {Berry}(2009)}]{berry_2009}%
  \BibitemOpen
  \bibfield  {author} {\bibinfo {author} {\bibfnamefont {M.}~\bibnamefont
  {Berry}},\ }\bibfield  {title} {\enquote {\bibinfo {title} {Transitionless
  quantum driving},}\ }\href@noop {} {\bibfield  {journal} {\bibinfo  {journal}
  {J. Phys. A: Math. Theor.},\ }\textbf {\bibinfo {volume} {42}},\ \bibinfo
  {pages} {365303} (\bibinfo {year} {2009})}\BibitemShut {NoStop}%
\bibitem [{\citenamefont {Berry}(1984)}]{berry_1984}%
  \BibitemOpen
  \bibfield  {author} {\bibinfo {author} {\bibfnamefont {M.~V.}\ \bibnamefont
  {Berry}},\ }\bibfield  {title} {\enquote {\bibinfo {title} {Quantal phase
  factors accompanying adiabatic changes},}\ }\href@noop {} {\bibfield
  {journal} {\bibinfo  {journal} {Proc. R. Soc. A},\ }\textbf {\bibinfo
  {volume} {392}},\ \bibinfo {pages} {45} (\bibinfo {year} {1984})}\BibitemShut
  {NoStop}%
\bibitem [{\citenamefont {Demirplak}\ and\ \citenamefont
  {Rice}(2008)}]{demirplak_rice_2008}%
  \BibitemOpen
  \bibfield  {author} {\bibinfo {author} {\bibfnamefont {M.}~\bibnamefont
  {Demirplak}}\ and\ \bibinfo {author} {\bibfnamefont {S.~A.}\ \bibnamefont
  {Rice}},\ }\bibfield  {title} {\enquote {\bibinfo {title} {On the
  consistency, extremal, and global properties of counterdiabatic fields},}\
  }\href@noop {} {\bibfield  {journal} {\bibinfo  {journal} {J. Chem. Phys.},\
  }\textbf {\bibinfo {volume} {129}},\ \bibinfo {pages} {154111} (\bibinfo
  {year} {2008})}\BibitemShut {NoStop}%
\bibitem [{\citenamefont {del Campo}\ \emph {et~al.}(2012)\citenamefont {del
  Campo}, \citenamefont {Rams},\ and\ \citenamefont
  {Zurek}}]{campo_rams_zurek_2012}%
  \BibitemOpen
  \bibfield  {author} {\bibinfo {author} {\bibfnamefont {A.}~\bibnamefont {del
  Campo}}, \bibinfo {author} {\bibfnamefont {M.~M.}\ \bibnamefont {Rams}}, \
  and\ \bibinfo {author} {\bibfnamefont {W.~H.}\ \bibnamefont {Zurek}},\
  }\bibfield  {title} {\enquote {\bibinfo {title} {Assisted finite-rate
  adiabatic passage across a quantum critical point: {E}xact solution for the
  quantum {I}sing model},}\ }\href@noop {} {\bibfield  {journal} {\bibinfo
  {journal} {\PRL},\ }\textbf {\bibinfo {volume} {109}},\ \bibinfo {pages}
  {115703} (\bibinfo {year} {2012})}\BibitemShut {NoStop}%
\bibitem [{\citenamefont {Bason}\ \emph {et~al.}(2011)\citenamefont {Bason},
  \citenamefont {Viteau}, \citenamefont {Malossi}, \citenamefont {Huillery},
  \citenamefont {Arimondo}, \citenamefont {Ciampini}, \citenamefont {Fazio},
  \citenamefont {Giovannetti}, \citenamefont {Mannella},\ and\ \citenamefont
  {Morsch}}]{bason_2011}%
  \BibitemOpen
  \bibfield  {author} {\bibinfo {author} {\bibfnamefont {M.~G.}\ \bibnamefont
  {Bason}}, \bibinfo {author} {\bibfnamefont {M.}~\bibnamefont {Viteau}},
  \bibinfo {author} {\bibfnamefont {N.}~\bibnamefont {Malossi}}, \bibinfo
  {author} {\bibfnamefont {P.}~\bibnamefont {Huillery}}, \bibinfo {author}
  {\bibfnamefont {E.}~\bibnamefont {Arimondo}}, \bibinfo {author}
  {\bibfnamefont {D.}~\bibnamefont {Ciampini}}, \bibinfo {author}
  {\bibfnamefont {R.}~\bibnamefont {Fazio}}, \bibinfo {author} {\bibfnamefont
  {V.}~\bibnamefont {Giovannetti}}, \bibinfo {author} {\bibfnamefont
  {R.}~\bibnamefont {Mannella}}, \ and\ \bibinfo {author} {\bibfnamefont
  {O.}~\bibnamefont {Morsch}},\ }\bibfield  {title} {\enquote {\bibinfo {title}
  {High-fidelity quantum driving},}\ }\href@noop {} {\bibfield  {journal}
  {\bibinfo  {journal} {Nature Physics},\ }\textbf {\bibinfo {volume} {8}},\
  \bibinfo {pages} {147} (\bibinfo {year} {2011})}\BibitemShut {NoStop}%
\bibitem [{\citenamefont {Zhang}\ \emph {et~al.}(2013)\citenamefont {Zhang},
  \citenamefont {Hyun~Shim}, \citenamefont {Niemeyer}, \citenamefont
  {Taniguchi}, \citenamefont {Teraji}, \citenamefont {Abe}, \citenamefont
  {Onoda}, \citenamefont {Yamamoto}, \citenamefont {Ohshima},\ and\
  \citenamefont {Isoya}}]{zhang_2013}%
  \BibitemOpen
  \bibfield  {author} {\bibinfo {author} {\bibfnamefont {J.}~\bibnamefont
  {Zhang}}, \bibinfo {author} {\bibfnamefont {J.}~\bibnamefont {Hyun~Shim}},
  \bibinfo {author} {\bibfnamefont {I.}~\bibnamefont {Niemeyer}}, \bibinfo
  {author} {\bibfnamefont {T.}~\bibnamefont {Taniguchi}}, \bibinfo {author}
  {\bibfnamefont {T.}~\bibnamefont {Teraji}}, \bibinfo {author} {\bibfnamefont
  {H.}~\bibnamefont {Abe}}, \bibinfo {author} {\bibfnamefont {S.}~\bibnamefont
  {Onoda}}, \bibinfo {author} {\bibfnamefont {T.}~\bibnamefont {Yamamoto}},
  \bibinfo {author} {\bibfnamefont {T.}~\bibnamefont {Ohshima}}, \ and\
  \bibinfo {author} {\bibfnamefont {D.}~\bibnamefont {Isoya}, \bibfnamefont
  {J.and~Suter}},\ }\bibfield  {title} {\enquote {\bibinfo {title}
  {Experimental implementation of assisted quantum adiabatic passage in a
  single spin},}\ }\href@noop {} {\bibfield  {journal} {\bibinfo  {journal}
  {\PRL},\ }\textbf {\bibinfo {volume} {110}},\ \bibinfo {pages} {240501}
  (\bibinfo {year} {2013})}\BibitemShut {NoStop}%
\bibitem [{\citenamefont {del Campo}(2013)}]{campo_2013}%
  \BibitemOpen
  \bibfield  {author} {\bibinfo {author} {\bibfnamefont {A.}~\bibnamefont {del
  Campo}},\ }\bibfield  {title} {\enquote {\bibinfo {title} {Shortcuts to
  adiabaticity by counterdiabatic driving},}\ }\href@noop {} {\bibfield
  {journal} {\bibinfo  {journal} {\PRL},\ }\textbf {\bibinfo {volume} {111}},\
  \bibinfo {pages} {100502} (\bibinfo {year} {2013})}\BibitemShut {NoStop}%
\bibitem [{\citenamefont {Mostafazadeh}(2001)}]{mostafazadeh_2001}%
  \BibitemOpen
  \bibfield  {author} {\bibinfo {author} {\bibfnamefont {A.}~\bibnamefont
  {Mostafazadeh}},\ }\href@noop {} {\emph {\bibinfo {title} {Dynamical
  invariants, adiabatic approximation and the geometric phase}}}\ (\bibinfo
  {publisher} {Nova},\ \bibinfo {address} {New York, NY, USA},\ \bibinfo {year}
  {2001})\BibitemShut {NoStop}%
\bibitem [{\citenamefont {Muga}\ \emph {et~al.}(2010)\citenamefont {Muga},
  \citenamefont {Chen}, \citenamefont {Ib\'a\~nez}, \citenamefont {Lizuain},\
  and\ \citenamefont {Ruschhaupt}}]{muga_2010}%
  \BibitemOpen
  \bibfield  {author} {\bibinfo {author} {\bibfnamefont {J.~G.}\ \bibnamefont
  {Muga}}, \bibinfo {author} {\bibfnamefont {X.}~\bibnamefont {Chen}}, \bibinfo
  {author} {\bibfnamefont {S.}~\bibnamefont {Ib\'a\~nez}}, \bibinfo {author}
  {\bibfnamefont {I.}~\bibnamefont {Lizuain}}, \ and\ \bibinfo {author}
  {\bibfnamefont {A.}~\bibnamefont {Ruschhaupt}},\ }\bibfield  {title}
  {\enquote {\bibinfo {title} {Transitionless quantum drivings for the harmonic
  oscillator},}\ }\href@noop {} {\bibfield  {journal} {\bibinfo  {journal} {J.
  Phys. B: At. Mol. Opt. Phys.},\ }\textbf {\bibinfo {volume} {43}},\ \bibinfo
  {pages} {085509} (\bibinfo {year} {2010})}\BibitemShut {NoStop}%
\bibitem [{\citenamefont {Torrontegui}\ \emph {et~al.}(2011)\citenamefont
  {Torrontegui}, \citenamefont {Ib\'a\~nez}, \citenamefont {Chen},
  \citenamefont {Ruschhaupt}, \citenamefont {Gu\'ery-Odelin},\ and\
  \citenamefont {Muga}}]{torrontegui_2011}%
  \BibitemOpen
  \bibfield  {author} {\bibinfo {author} {\bibfnamefont {E.}~\bibnamefont
  {Torrontegui}}, \bibinfo {author} {\bibfnamefont {S.}~\bibnamefont
  {Ib\'a\~nez}}, \bibinfo {author} {\bibfnamefont {Xi}~\bibnamefont {Chen}},
  \bibinfo {author} {\bibfnamefont {A.}~\bibnamefont {Ruschhaupt}}, \bibinfo
  {author} {\bibfnamefont {D.}~\bibnamefont {Gu\'ery-Odelin}}, \ and\ \bibinfo
  {author} {\bibfnamefont {J.~G.}\ \bibnamefont {Muga}},\ }\bibfield  {title}
  {\enquote {\bibinfo {title} {Fast atomic transport without vibrational
  heating},}\ }\href@noop {} {\bibfield  {journal} {\bibinfo  {journal}
  {\PRA},\ }\textbf {\bibinfo {volume} {83}},\ \bibinfo {pages} {013415}
  (\bibinfo {year} {2011})}\BibitemShut {NoStop}%
\bibitem [{\citenamefont {Deng}\ \emph {et~al.}(2013)\citenamefont {Deng},
  \citenamefont {Wang}, \citenamefont {Liu}, \citenamefont {H\"anggi},\ and\
  \citenamefont {Gong}}]{deng_2013}%
  \BibitemOpen
  \bibfield  {author} {\bibinfo {author} {\bibfnamefont {J.}~\bibnamefont
  {Deng}}, \bibinfo {author} {\bibfnamefont {Q.}~\bibnamefont {Wang}}, \bibinfo
  {author} {\bibfnamefont {Z.}~\bibnamefont {Liu}}, \bibinfo {author}
  {\bibfnamefont {P.}~\bibnamefont {H\"anggi}}, \ and\ \bibinfo {author}
  {\bibfnamefont {J.}~\bibnamefont {Gong}},\ }\bibfield  {title} {\enquote
  {\bibinfo {title} {Boosting work characteristics and overall heat engine
  performance via accelerated adiabatic control: quantum and classical},}\
  }\href@noop {} {\bibfield  {journal} {\bibinfo  {journal} {\PRE},\ }\textbf
  {\bibinfo {volume} {88}},\ \bibinfo {pages} {06222} (\bibinfo {year}
  {2013})}\BibitemShut {NoStop}%
\bibitem [{\citenamefont {Jarzynski}(2013)}]{jarzynski_2013}%
  \BibitemOpen
  \bibfield  {author} {\bibinfo {author} {\bibfnamefont {C.}~\bibnamefont
  {Jarzynski}},\ }\bibfield  {title} {\enquote {\bibinfo {title} {Generating
  shortcuts to adiabaticity in quantum and classical dynamics},}\ }\href@noop
  {} {\bibfield  {journal} {\bibinfo  {journal} {\PRA},\ }\textbf {\bibinfo
  {volume} {88}},\ \bibinfo {pages} {040101(R)} (\bibinfo {year}
  {2013})}\BibitemShut {NoStop}%
\bibitem [{\citenamefont {Gambardella}(1975)}]{gambardella_1975}%
  \BibitemOpen
  \bibfield  {author} {\bibinfo {author} {\bibfnamefont {P.~J.}\ \bibnamefont
  {Gambardella}},\ }\bibfield  {title} {\enquote {\bibinfo {title} {Exact
  results in quantum many-body systems of interacting particles in many
  dimensions with $\overline{SU(1,1)}$ as the dynamical group},}\ }\href@noop
  {} {\bibfield  {journal} {\bibinfo  {journal} {J. Math. Phys.},\ }\textbf
  {\bibinfo {volume} {16}},\ \bibinfo {pages} {1172} (\bibinfo {year}
  {1975})}\BibitemShut {NoStop}%
\bibitem [{\citenamefont {Di~Martino}\ \emph {et~al.}(2013)\citenamefont
  {Di~Martino}, \citenamefont {Anz\'{a}}, \citenamefont {Facchi}, \citenamefont
  {Kossakowski}, \citenamefont {Marmo}, \citenamefont {Messina}, \citenamefont
  {Militello},\ and\ \citenamefont {Pascazio}}]{dimartino_2013}%
  \BibitemOpen
  \bibfield  {author} {\bibinfo {author} {\bibfnamefont {S.}~\bibnamefont
  {Di~Martino}}, \bibinfo {author} {\bibfnamefont {F.}~\bibnamefont
  {Anz\'{a}}}, \bibinfo {author} {\bibfnamefont {P.}~\bibnamefont {Facchi}},
  \bibinfo {author} {\bibfnamefont {A.}~\bibnamefont {Kossakowski}}, \bibinfo
  {author} {\bibfnamefont {G.}~\bibnamefont {Marmo}}, \bibinfo {author}
  {\bibfnamefont {A.}~\bibnamefont {Messina}}, \bibinfo {author} {\bibfnamefont
  {B.}~\bibnamefont {Militello}}, \ and\ \bibinfo {author} {\bibfnamefont
  {S.}~\bibnamefont {Pascazio}},\ }\bibfield  {title} {\enquote {\bibinfo
  {title} {A quantum particle in a box with moving walls},}\ }\href@noop {}
  {\bibfield  {journal} {\bibinfo  {journal} {J. Phys. A: Math. Theor.},\
  }\textbf {\bibinfo {volume} {46}},\ \bibinfo {pages} {365301} (\bibinfo
  {year} {2013})}\BibitemShut {NoStop}%
\bibitem [{Note1()}]{Note1}%
  \BibitemOpen
  \bibinfo {note} {Hamiltonians that include products of space and momentum
  operator, $q$ and $p$, are \protect \textit {non-local}, whereas \protect
  \textit {local} Hamiltonians contain only terms that depend on at most sums
  of $q$ and $p$.}\BibitemShut {Stop}%
\bibitem [{\citenamefont {Goldstein}(1959)}]{goldstein_1959}%
  \BibitemOpen
  \bibfield  {author} {\bibinfo {author} {\bibfnamefont {H.}~\bibnamefont
  {Goldstein}},\ }\href@noop {} {\emph {\bibinfo {title} {Classical
  Mechanics}}}\ (\bibinfo  {publisher} {Addison-Wesley Publishing Company,
  inc.},\ \bibinfo {address} {Reading, MA, USA},\ \bibinfo {year}
  {1959})\BibitemShut {NoStop}%
\bibitem [{\citenamefont {Duru}(1989)}]{duru_1989}%
  \BibitemOpen
  \bibfield  {author} {\bibinfo {author} {\bibfnamefont {I.~H.}\ \bibnamefont
  {Duru}},\ }\bibfield  {title} {\enquote {\bibinfo {title} {Quantum treatment
  of a class of time-dependent potentials},}\ }\href@noop {} {\bibfield
  {journal} {\bibinfo  {journal} {J. Phys. A: Math. Gen.},\ }\textbf {\bibinfo
  {volume} {22}},\ \bibinfo {pages} {4827} (\bibinfo {year}
  {1989})}\BibitemShut {NoStop}%
\bibitem [{\citenamefont {Berry}\ and\ \citenamefont
  {Klein}(1984)}]{berry_klein_84}%
  \BibitemOpen
  \bibfield  {author} {\bibinfo {author} {\bibfnamefont {M.~V.}\ \bibnamefont
  {Berry}}\ and\ \bibinfo {author} {\bibfnamefont {G.}~\bibnamefont {Klein}},\
  }\bibfield  {title} {\enquote {\bibinfo {title} {Newtonian trajectories and
  quantum waves in expaning force fields},}\ }\href@noop {} {\bibfield
  {journal} {\bibinfo  {journal} {J. Phys. A: Math. Gen.},\ }\textbf {\bibinfo
  {volume} {17}},\ \bibinfo {pages} {1805} (\bibinfo {year}
  {1984})}\BibitemShut {NoStop}%
\bibitem [{\citenamefont {Lewis}(1968)}]{lewis_1968}%
  \BibitemOpen
  \bibfield  {author} {\bibinfo {author} {\bibfnamefont {H.~R.}\ \bibnamefont
  {Lewis}},\ }\bibfield  {title} {\enquote {\bibinfo {title} {Class of exact
  invariants for classical and quantum time-dependent harmonic oscillators},}\
  }\href@noop {} {\bibfield  {journal} {\bibinfo  {journal} {J. Math. Phys.},\
  }\textbf {\bibinfo {volume} {9}},\ \bibinfo {pages} {1976} (\bibinfo {year}
  {1968})}\BibitemShut {NoStop}%
\bibitem [{\citenamefont {Lewis}\ and\ \citenamefont
  {Riesenfeld}(1969)}]{lewis_1969a}%
  \BibitemOpen
  \bibfield  {author} {\bibinfo {author} {\bibfnamefont {H.~R.}\ \bibnamefont
  {Lewis}}\ and\ \bibinfo {author} {\bibfnamefont {W.~B.}\ \bibnamefont
  {Riesenfeld}},\ }\bibfield  {title} {\enquote {\bibinfo {title} {An exact
  quantum theory of the time-dependent harmonic oscillator and of a charged
  particle in a time-dependent electromagnetic field},}\ }\href@noop {}
  {\bibfield  {journal} {\bibinfo  {journal} {J. Math. Phys.},\ }\textbf
  {\bibinfo {volume} {10}},\ \bibinfo {pages} {1458} (\bibinfo {year}
  {1969})}\BibitemShut {NoStop}%
\bibitem [{\citenamefont {G\"unther}\ and\ \citenamefont
  {Leach}(1977)}]{gunther_1977}%
  \BibitemOpen
  \bibfield  {author} {\bibinfo {author} {\bibfnamefont {N.~J.}\ \bibnamefont
  {G\"unther}}\ and\ \bibinfo {author} {\bibfnamefont {P.~G.~L.}\ \bibnamefont
  {Leach}},\ }\bibfield  {title} {\enquote {\bibinfo {title} {Generalized
  invariants for the time-dependent harmonic oscillator},}\ }\href@noop {}
  {\bibfield  {journal} {\bibinfo  {journal} {J. Math. Phys.},\ }\textbf
  {\bibinfo {volume} {18}},\ \bibinfo {pages} {572} (\bibinfo {year}
  {1977})}\BibitemShut {NoStop}%
\bibitem [{\citenamefont {Leach}(1977){\natexlab{a}}}]{leach_1977}%
  \BibitemOpen
  \bibfield  {author} {\bibinfo {author} {\bibfnamefont {P.~G.~L.}\
  \bibnamefont {Leach}},\ }\bibfield  {title} {\enquote {\bibinfo {title}
  {Invariants and wavefunctions for some time-dependent harmonic
  oscillator-type {H}amiltonians},}\ }\href@noop {} {\bibfield  {journal}
  {\bibinfo  {journal} {J. Math. Phys.},\ }\textbf {\bibinfo {volume} {18}},\
  \bibinfo {pages} {1902} (\bibinfo {year} {1977}{\natexlab{a}})}\BibitemShut
  {NoStop}%
\bibitem [{\citenamefont {Leach}(1977){\natexlab{b}}}]{leach_1977b}%
  \BibitemOpen
  \bibfield  {author} {\bibinfo {author} {\bibfnamefont {P.~G.~L.}\
  \bibnamefont {Leach}},\ }\bibfield  {title} {\enquote {\bibinfo {title} {On a
  direct method for the determination of an exact invariant for the
  time-dependent harmonic oscillator},}\ }\href@noop {} {\bibfield  {journal}
  {\bibinfo  {journal} {J. Austral. Math. Soc. Ser. B},\ }\textbf {\bibinfo
  {volume} {20}},\ \bibinfo {pages} {97} (\bibinfo {year}
  {1977}{\natexlab{b}})}\BibitemShut {NoStop}%
\bibitem [{\citenamefont {Leach}(1977){\natexlab{c}}}]{leach_1977a}%
  \BibitemOpen
  \bibfield  {author} {\bibinfo {author} {\bibfnamefont {P.~G.~L.}\
  \bibnamefont {Leach}},\ }\bibfield  {title} {\enquote {\bibinfo {title} {On
  the theory of time-dependent linear canonical transformations as applied to
  {H}amiltonians of the harmonic oscillator type},}\ }\href@noop {} {\bibfield
  {journal} {\bibinfo  {journal} {J. Math. Phys.},\ }\textbf {\bibinfo {volume}
  {18}},\ \bibinfo {pages} {1608} (\bibinfo {year}
  {1977}{\natexlab{c}})}\BibitemShut {NoStop}%
\bibitem [{\citenamefont {Lohe}(2009)}]{lohe_2009}%
  \BibitemOpen
  \bibfield  {author} {\bibinfo {author} {\bibfnamefont {M.~A.}\ \bibnamefont
  {Lohe}},\ }\bibfield  {title} {\enquote {\bibinfo {title} {Exact time
  dependence of solutions to the time-dependent schr\"odinger equation},}\
  }\href@noop {} {\bibfield  {journal} {\bibinfo  {journal} {J. Phys. A: Math.
  Theor.},\ }\textbf {\bibinfo {volume} {42}},\ \bibinfo {pages} {035307}
  (\bibinfo {year} {2009})}\BibitemShut {NoStop}%
\bibitem [{\citenamefont {Yurovsky}\ \emph {et~al.}(2008)\citenamefont
  {Yurovsky}, \citenamefont {Olshanii},\ and\ \citenamefont
  {Weiss}}]{yurovsky_2008}%
  \BibitemOpen
  \bibfield  {author} {\bibinfo {author} {\bibfnamefont {V.~A.}\ \bibnamefont
  {Yurovsky}}, \bibinfo {author} {\bibfnamefont {M.}~\bibnamefont {Olshanii}},
  \ and\ \bibinfo {author} {\bibfnamefont {D.~S.}\ \bibnamefont {Weiss}},\
  }\href@noop {} {\emph {\bibinfo {title} {Collisions, correlations, and
  integrability in atom waveguides}}},\ edited by\ \bibinfo {editor}
  {\bibfnamefont {Paul R.~Berman}\ \bibnamefont {Ennio~Arimondo}}\ and\
  \bibinfo {editor} {\bibfnamefont {Chun~C.}\ \bibnamefont {Lin}},\ \bibinfo
  {series} {Adv. At. Mol. Opt. Phys.}, Vol.~\bibinfo {volume} {55}\ (\bibinfo
  {publisher} {Academic Press},\ \bibinfo {year} {2008})\ p.~\bibinfo {pages}
  {61}\BibitemShut {NoStop}%
\bibitem [{\citenamefont {Huang}(1996)}]{huang_1996}%
  \BibitemOpen
  \bibfield  {author} {\bibinfo {author} {\bibfnamefont {K.}~\bibnamefont
  {Huang}},\ }\bibfield  {title} {\enquote {\bibinfo {title} {Bose condensate
  in external potential: {A Thomas-Fermi} approach},}\ }\href@noop {}
  {\bibfield  {journal} {\bibinfo  {journal} {arXiv:cond-mat/9609033}}
  (\bibinfo {year} {1996})}\BibitemShut {NoStop}%
\bibitem [{\citenamefont {Perelomov}(1978)}]{perelomov_1978}%
  \BibitemOpen
  \bibfield  {author} {\bibinfo {author} {\bibfnamefont {A.~M.}\ \bibnamefont
  {Perelomov}},\ }\bibfield  {title} {\enquote {\bibinfo {title} {The simple
  relation between certain dynamical systems},}\ }\href@noop {} {\bibfield
  {journal} {\bibinfo  {journal} {Commun. Math. Phys.},\ }\textbf {\bibinfo
  {volume} {63}},\ \bibinfo {pages} {9} (\bibinfo {year} {1978})}\BibitemShut
  {NoStop}%
\bibitem [{\citenamefont {Timmermans}\ \emph {et~al.}(1999)\citenamefont
  {Timmermans}, \citenamefont {Tommasini}, \citenamefont {Hussein},\ and\
  \citenamefont {Kerman}}]{timmermans_1999}%
  \BibitemOpen
  \bibfield  {author} {\bibinfo {author} {\bibfnamefont {E.}~\bibnamefont
  {Timmermans}}, \bibinfo {author} {\bibfnamefont {P.}~\bibnamefont
  {Tommasini}}, \bibinfo {author} {\bibfnamefont {M.}~\bibnamefont {Hussein}},
  \ and\ \bibinfo {author} {\bibfnamefont {A.}~\bibnamefont {Kerman}},\
  }\bibfield  {title} {\enquote {\bibinfo {title} {Feshbach resonances in
  atomic {B}ose-{E}instein condensates},}\ }\href@noop {} {\bibfield  {journal}
  {\bibinfo  {journal} {Phys. Rep.},\ }\textbf {\bibinfo {volume} {315}},\
  \bibinfo {pages} {199} (\bibinfo {year} {1999})}\BibitemShut {NoStop}%
\bibitem [{\citenamefont {Staliunas}\ \emph {et~al.}(2004)\citenamefont
  {Staliunas}, \citenamefont {Longhi},\ and\ \citenamefont
  {de~Valc\'arcel}}]{staliunas_2004}%
  \BibitemOpen
  \bibfield  {author} {\bibinfo {author} {\bibfnamefont {K.}~\bibnamefont
  {Staliunas}}, \bibinfo {author} {\bibfnamefont {S.}~\bibnamefont {Longhi}}, \
  and\ \bibinfo {author} {\bibfnamefont {G.~J.}\ \bibnamefont
  {de~Valc\'arcel}},\ }\bibfield  {title} {\enquote {\bibinfo {title} {Faraday
  patterns in low-dimensional {Bose-Einstein} condensates},}\ }\href@noop {}
  {\bibfield  {journal} {\bibinfo  {journal} {Phys. Rev. A},\ }\textbf
  {\bibinfo {volume} {70}},\ \bibinfo {pages} {011601} (\bibinfo {year}
  {2004})}\BibitemShut {NoStop}%
\bibitem [{\citenamefont {del Campo}(2011){\natexlab{b}}}]{campo_epl_2011}%
  \BibitemOpen
  \bibfield  {author} {\bibinfo {author} {\bibfnamefont {A.}~\bibnamefont {del
  Campo}},\ }\bibfield  {title} {\enquote {\bibinfo {title} {Fast frictionless
  dynamics as a toolbox for low-dimensional {Bose-Einstein} condensates},}\
  }\href@noop {} {\bibfield  {journal} {\bibinfo  {journal} {\EPL},\ }\textbf
  {\bibinfo {volume} {96}},\ \bibinfo {pages} {60005} (\bibinfo {year}
  {2011}{\natexlab{b}})}\BibitemShut {NoStop}%
\bibitem [{\citenamefont {Moshinsky}\ and\ \citenamefont
  {Smirnov}(1996)}]{moshinsky_96}%
  \BibitemOpen
  \bibfield  {author} {\bibinfo {author} {\bibfnamefont {M.}~\bibnamefont
  {Moshinsky}}\ and\ \bibinfo {author} {\bibfnamefont {Y.~F.}\ \bibnamefont
  {Smirnov}},\ }\href@noop {} {\emph {\bibinfo {title} {The Harmonic Oscillator
  in Modern Physics}}}\ (\bibinfo  {publisher} {Harwood Academic Publishers},\
  \bibinfo {address} {Amsterdam, Netherlands},\ \bibinfo {year}
  {1996})\BibitemShut {NoStop}%
\bibitem [{\citenamefont {Henderson}\ \emph {et~al.}(2009)\citenamefont
  {Henderson}, \citenamefont {Ryu}, \citenamefont {MacCormick},\ and\
  \citenamefont {Boshier}}]{henderson_2009}%
  \BibitemOpen
  \bibfield  {author} {\bibinfo {author} {\bibfnamefont {K.}~\bibnamefont
  {Henderson}}, \bibinfo {author} {\bibfnamefont {C.}~\bibnamefont {Ryu}},
  \bibinfo {author} {\bibfnamefont {C.}~\bibnamefont {MacCormick}}, \ and\
  \bibinfo {author} {\bibfnamefont {M.~G.}\ \bibnamefont {Boshier}},\
  }\bibfield  {title} {\enquote {\bibinfo {title} {Experimental demonstration
  of painting arbitrary and dynamic potentials for {B}ose-{E}instein
  condensates},}\ }\href@noop {} {\bibfield  {journal} {\bibinfo  {journal}
  {\NJP},\ }\textbf {\bibinfo {volume} {11}},\ \bibinfo {pages} {043030}
  (\bibinfo {year} {2009})}\BibitemShut {NoStop}%
\bibitem [{\citenamefont {Fl\"ugge}(1971)}]{flugge_1971}%
  \BibitemOpen
  \bibfield  {author} {\bibinfo {author} {\bibfnamefont {S.}~\bibnamefont
  {Fl\"ugge}},\ }\href@noop {} {\emph {\bibinfo {title} {Practical quantum
  mechanics}}},\ Vol.~\bibinfo {volume} {I}\ (\bibinfo  {publisher}
  {Springer},\ \bibinfo {address} {Heidelberg, Germany},\ \bibinfo {year}
  {1971})\BibitemShut {NoStop}%
\bibitem [{\citenamefont {Cooper}\ \emph {et~al.}(1995)\citenamefont {Cooper},
  \citenamefont {Khare},\ and\ \citenamefont {Sukhatme}}]{cooper_1995}%
  \BibitemOpen
  \bibfield  {author} {\bibinfo {author} {\bibfnamefont {F.}~\bibnamefont
  {Cooper}}, \bibinfo {author} {\bibfnamefont {A.}~\bibnamefont {Khare}}, \
  and\ \bibinfo {author} {\bibfnamefont {U.}~\bibnamefont {Sukhatme}},\
  }\bibfield  {title} {\enquote {\bibinfo {title} {Supersymmetry and quantum
  mechanics},}\ }\href@noop {} {\bibfield  {journal} {\bibinfo  {journal}
  {Phys. Rep.},\ }\textbf {\bibinfo {volume} {251}},\ \bibinfo {pages} {267}
  (\bibinfo {year} {1995})}\BibitemShut {NoStop}%
\bibitem [{\citenamefont {Castin}\ and\ \citenamefont
  {Dum}(1996)}]{castin_1996}%
  \BibitemOpen
  \bibfield  {author} {\bibinfo {author} {\bibfnamefont {Y.}~\bibnamefont
  {Castin}}\ and\ \bibinfo {author} {\bibfnamefont {R.}~\bibnamefont {Dum}},\
  }\bibfield  {title} {\enquote {\bibinfo {title} {Bose-{E}instein condensates
  in time dependent traps},}\ }\href@noop {} {\bibfield  {journal} {\bibinfo
  {journal} {\PRL},\ }\textbf {\bibinfo {volume} {77}},\ \bibinfo {pages}
  {5315} (\bibinfo {year} {1996})}\BibitemShut {NoStop}%
\bibitem [{\citenamefont {Kagan}\ \emph {et~al.}(1997)\citenamefont {Kagan},
  \citenamefont {Surkov},\ and\ \citenamefont {Shlyapnikov}}]{kagan_1996}%
  \BibitemOpen
  \bibfield  {author} {\bibinfo {author} {\bibfnamefont {Yu.}\ \bibnamefont
  {Kagan}}, \bibinfo {author} {\bibfnamefont {E.~L.}\ \bibnamefont {Surkov}}, \
  and\ \bibinfo {author} {\bibfnamefont {G.~V.}\ \bibnamefont {Shlyapnikov}},\
  }\bibfield  {title} {\enquote {\bibinfo {title} {Evolution of a {B}ose gas in
  anisotropic time-dependent traps},}\ }\href@noop {} {\bibfield  {journal}
  {\bibinfo  {journal} {Phys. Rev. A},\ }\textbf {\bibinfo {volume} {55}},\
  \bibinfo {pages} {R18} (\bibinfo {year} {1997})}\BibitemShut {NoStop}%
\bibitem [{\citenamefont {Egusquiza}\ \emph {et~al.}(2011)\citenamefont
  {Egusquiza}, \citenamefont {Modugno},\ and\ \citenamefont
  {Valle~Basagoiti}}]{egusquiza_2011}%
  \BibitemOpen
  \bibfield  {author} {\bibinfo {author} {\bibfnamefont {I.~L.}\ \bibnamefont
  {Egusquiza}}, \bibinfo {author} {\bibfnamefont {M.}~\bibnamefont {Modugno}},
  \ and\ \bibinfo {author} {\bibfnamefont {M.~A.}\ \bibnamefont
  {Valle~Basagoiti}},\ }\bibfield  {title} {\enquote {\bibinfo {title}
  {Multiple-scale approach for the expansion scaling of superfluid quantum
  gases},}\ }\href@noop {} {\bibfield  {journal} {\bibinfo  {journal} {Phys.
  Rev. A},\ }\textbf {\bibinfo {volume} {84}},\ \bibinfo {pages} {043629}
  (\bibinfo {year} {2011})}\BibitemShut {NoStop}%
\bibitem [{\citenamefont {Kundu}(2009)}]{kundu_2009}%
  \BibitemOpen
  \bibfield  {author} {\bibinfo {author} {\bibfnamefont {A.}~\bibnamefont
  {Kundu}},\ }\bibfield  {title} {\enquote {\bibinfo {title} {Integrable
  nonautonomous nonlinear {S}chr\"odinger equations are equivalent to the
  standard autonomous equation},}\ }\href@noop {} {\bibfield  {journal}
  {\bibinfo  {journal} {\PRE},\ }\textbf {\bibinfo {volume} {79}},\ \bibinfo
  {pages} {015601(R)} (\bibinfo {year} {2009})}\BibitemShut {NoStop}%
\bibitem [{\citenamefont {Fadeev}\ and\ \citenamefont
  {Takhtajan}(2007)}]{fadeev_2007}%
  \BibitemOpen
  \bibfield  {author} {\bibinfo {author} {\bibfnamefont {L.~D.}\ \bibnamefont
  {Fadeev}}\ and\ \bibinfo {author} {\bibfnamefont {L.~A.}\ \bibnamefont
  {Takhtajan}},\ }\href@noop {} {\emph {\bibinfo {title} {Hamiltonian Methods
  in the Theory of Solitons}}}\ (\bibinfo  {publisher} {Springer},\ \bibinfo
  {address} {Heidelberg},\ \bibinfo {year} {2007})\BibitemShut {NoStop}%
\bibitem [{\citenamefont {Kolomeisky}\ \emph {et~al.}(2000)\citenamefont
  {Kolomeisky}, \citenamefont {Newman}, \citenamefont {Straley},\ and\
  \citenamefont {Qi}}]{kolomeisky_2000}%
  \BibitemOpen
  \bibfield  {author} {\bibinfo {author} {\bibfnamefont {E.~B.}\ \bibnamefont
  {Kolomeisky}}, \bibinfo {author} {\bibfnamefont {T.~J.}\ \bibnamefont
  {Newman}}, \bibinfo {author} {\bibfnamefont {J.~P.}\ \bibnamefont {Straley}},
  \ and\ \bibinfo {author} {\bibfnamefont {Xiaoya}\ \bibnamefont {Qi}},\
  }\bibfield  {title} {\enquote {\bibinfo {title} {Low-dimensional {Bose}
  liquids: {B}eyond the {Gross-Pitaevskii} approximation},}\ }\href@noop {}
  {\bibfield  {journal} {\bibinfo  {journal} {\PRL},\ }\textbf {\bibinfo
  {volume} {85}},\ \bibinfo {pages} {1146} (\bibinfo {year}
  {2000})}\BibitemShut {NoStop}%
\bibitem [{\citenamefont {Kolomeisky}\ and\ \citenamefont
  {Straley}(1992)}]{kolomeisky_1992}%
  \BibitemOpen
  \bibfield  {author} {\bibinfo {author} {\bibfnamefont {E.~B.}\ \bibnamefont
  {Kolomeisky}}\ and\ \bibinfo {author} {\bibfnamefont {J.~P.}\ \bibnamefont
  {Straley}},\ }\bibfield  {title} {\enquote {\bibinfo {title}
  {Renormalization-group analysis of the ground-state properties of dilute bose
  systems in \textit{d} spatial dimensions},}\ }\href@noop {} {\bibfield
  {journal} {\bibinfo  {journal} {Phys. Rev. B},\ }\textbf {\bibinfo {volume}
  {46}},\ \bibinfo {pages} {11749} (\bibinfo {year} {1992})}\BibitemShut
  {NoStop}%
\bibitem [{\citenamefont {Girardeau}(1960)}]{girardeau_1960}%
  \BibitemOpen
  \bibfield  {author} {\bibinfo {author} {\bibfnamefont {M.~D.}\ \bibnamefont
  {Girardeau}},\ }\bibfield  {title} {\enquote {\bibinfo {title} {Relationship
  between systems of impenetrable bosons and fermions in one dimension},}\
  }\href@noop {} {\bibfield  {journal} {\bibinfo  {journal} {J. Math. Phys.},\
  }\textbf {\bibinfo {volume} {1}},\ \bibinfo {pages} {516} (\bibinfo {year}
  {1960})}\BibitemShut {NoStop}%
\bibitem [{\citenamefont {Girardeau}\ and\ \citenamefont
  {Wright}(2000)}]{girardeau_2000}%
  \BibitemOpen
  \bibfield  {author} {\bibinfo {author} {\bibfnamefont {M.~D.}\ \bibnamefont
  {Girardeau}}\ and\ \bibinfo {author} {\bibfnamefont {E.~M.}\ \bibnamefont
  {Wright}},\ }\bibfield  {title} {\enquote {\bibinfo {title} {Breakdown of
  time-dependent mean-field theory for a one-dimensional condensate of
  impenetrable bosons},}\ }\href@noop {} {\bibfield  {journal} {\bibinfo
  {journal} {\PRL},\ }\textbf {\bibinfo {volume} {84}},\ \bibinfo {pages}
  {5239} (\bibinfo {year} {2000})}\BibitemShut {NoStop}%
\bibitem [{Note2()}]{Note2}%
  \BibitemOpen
  \bibinfo {note} {The exact many-body quantum system describing a
  Tonks-Girardeau gas ($N$ one-dimensional bosons with contact interactions of
  infinite amplitude) is exactly solvable by means of the Bose-Fermi duality as
  discussed by Girardeau \cite {girardeau_1960}. It is remarkable that the
  classical integrability condition in terms of the zero-curvature
  representation for the mean-field Kolomeisky equation describing this system
  has not yet been found.}\BibitemShut {Stop}%
\bibitem [{\citenamefont {Kim}\ and\ \citenamefont {Zubarev}(2003)}]{kim_2003}%
  \BibitemOpen
  \bibfield  {author} {\bibinfo {author} {\bibfnamefont {Y.~E.}\ \bibnamefont
  {Kim}}\ and\ \bibinfo {author} {\bibfnamefont {A.~L.}\ \bibnamefont
  {Zubarev}},\ }\bibfield  {title} {\enquote {\bibinfo {title}
  {Density-functional theory of bosons in a trap},}\ }\href@noop {} {\bibfield
  {journal} {\bibinfo  {journal} {\PRA},\ }\textbf {\bibinfo {volume} {67}},\
  \bibinfo {pages} {015602} (\bibinfo {year} {2003})}\BibitemShut {NoStop}%
\bibitem [{\citenamefont {Damski}(2004)}]{damski_2004}%
  \BibitemOpen
  \bibfield  {author} {\bibinfo {author} {\bibfnamefont {B.}~\bibnamefont
  {Damski}},\ }\bibfield  {title} {\enquote {\bibinfo {title} {Shock waves in
  ultracold {F}ermi ({T}onks) gases},}\ }\href@noop {} {\bibfield  {journal}
  {\bibinfo  {journal} {J. Phys. B: At. Mol. Opt. Phys.},\ }\textbf {\bibinfo
  {volume} {37}},\ \bibinfo {pages} {L85} (\bibinfo {year} {2004})}\BibitemShut
  {NoStop}%
\bibitem [{\citenamefont {Ozcakmakli}\ and\ \citenamefont
  {Yuce}(2012)}]{ozcakmakli_2012}%
  \BibitemOpen
  \bibfield  {author} {\bibinfo {author} {\bibfnamefont {Z.}~\bibnamefont
  {Ozcakmakli}}\ and\ \bibinfo {author} {\bibfnamefont {C.}~\bibnamefont
  {Yuce}},\ }\bibfield  {title} {\enquote {\bibinfo {title} {Shortcuts to
  adiabaticity for growing condensates},}\ }\href@noop {} {\bibfield  {journal}
  {\bibinfo  {journal} {Physica Scripta},\ }\textbf {\bibinfo {volume} {86}},\
  \bibinfo {pages} {055001} (\bibinfo {year} {2012})}\BibitemShut {NoStop}%
\bibitem [{\citenamefont {{Lamporesi}}\ \emph {et~al.}(2013)\citenamefont
  {{Lamporesi}}, \citenamefont {{Donadello}}, \citenamefont {{Serafini}},
  \citenamefont {{Dalfovo}},\ and\ \citenamefont {{Ferrari}}}]{lamporesi_2013}%
  \BibitemOpen
  \bibfield  {author} {\bibinfo {author} {\bibfnamefont {G.}~\bibnamefont
  {{Lamporesi}}}, \bibinfo {author} {\bibfnamefont {S.}~\bibnamefont
  {{Donadello}}}, \bibinfo {author} {\bibfnamefont {S.}~\bibnamefont
  {{Serafini}}}, \bibinfo {author} {\bibfnamefont {F.}~\bibnamefont
  {{Dalfovo}}}, \ and\ \bibinfo {author} {\bibfnamefont {G.}~\bibnamefont
  {{Ferrari}}},\ }\bibfield  {title} {\enquote {\bibinfo {title} {Spontaneous
  creation of {Kibble-Zurek} solitons in a {Bose-Einstein} condensate},}\
  }\href@noop {} {\bibfield  {journal} {\bibinfo  {journal} {Nature Physics},\
  }\textbf {\bibinfo {volume} {9}},\ \bibinfo {pages} {656} (\bibinfo {year}
  {2013})}\BibitemShut {NoStop}%
\bibitem [{\citenamefont {Weiler}\ \emph {et~al.}(2008)\citenamefont {Weiler},
  \citenamefont {Neely}, \citenamefont {Scherer}, \citenamefont {Bradley},
  \citenamefont {Davis},\ and\ \citenamefont {Anderson}}]{weiler_2008}%
  \BibitemOpen
  \bibfield  {author} {\bibinfo {author} {\bibfnamefont {C.~N.}\ \bibnamefont
  {Weiler}}, \bibinfo {author} {\bibfnamefont {T.~W.}\ \bibnamefont {Neely}},
  \bibinfo {author} {\bibfnamefont {D.~R.}\ \bibnamefont {Scherer}}, \bibinfo
  {author} {\bibfnamefont {A.~S.}\ \bibnamefont {Bradley}}, \bibinfo {author}
  {\bibfnamefont {M.~J.}\ \bibnamefont {Davis}}, \ and\ \bibinfo {author}
  {\bibfnamefont {B.~P.}\ \bibnamefont {Anderson}},\ }\bibfield  {title}
  {\enquote {\bibinfo {title} {Spontaneous vortices in the formation of
  {B}ose-{E}instein condensates},}\ }\href@noop {} {\bibfield  {journal}
  {\bibinfo  {journal} {Nature},\ }\textbf {\bibinfo {volume} {455}},\ \bibinfo
  {pages} {948} (\bibinfo {year} {2008})}\BibitemShut {NoStop}%
\bibitem [{\citenamefont {del Campo}\ \emph {et~al.}(2013)\citenamefont {del
  Campo}, \citenamefont {Kibble},\ and\ \citenamefont {Zurek}}]{delcampo_2013}%
  \BibitemOpen
  \bibfield  {author} {\bibinfo {author} {\bibfnamefont {A.}~\bibnamefont {del
  Campo}}, \bibinfo {author} {\bibfnamefont {T.~W.~B.}\ \bibnamefont {Kibble}},
  \ and\ \bibinfo {author} {\bibfnamefont {W.~H.}\ \bibnamefont {Zurek}},\
  }\bibfield  {title} {\enquote {\bibinfo {title} {Causality and
  non-equilibrium second-order phase transitions in inhomogeneous systems},}\
  }\href@noop {} {\bibfield  {journal} {\bibinfo  {journal} {J. Phys.: Condens.
  Matter},\ }\textbf {\bibinfo {volume} {25}},\ \bibinfo {pages} {404210}
  (\bibinfo {year} {2013})}\BibitemShut {NoStop}%
\bibitem [{\citenamefont {del Campo}\ and\ \citenamefont
  {Zurek}(2013)}]{delcampo_zurek_2013}%
  \BibitemOpen
  \bibfield  {author} {\bibinfo {author} {\bibfnamefont {A.}~\bibnamefont {del
  Campo}}\ and\ \bibinfo {author} {\bibfnamefont {W.~H.}\ \bibnamefont
  {Zurek}},\ }\bibfield  {title} {\enquote {\bibinfo {title} {Universality of
  phase transition dynamics: Topological defects from symmetry breaking},}\
  }\href@noop {} {\bibfield  {journal} {\bibinfo  {journal} {arXiv:1310.1600}}
  (\bibinfo {year} {2013})}\BibitemShut {NoStop}%
\bibitem [{\citenamefont {Gritsev}\ \emph {et~al.}(2010)\citenamefont
  {Gritsev}, \citenamefont {Barmettler},\ and\ \citenamefont
  {Demler}}]{gritsev_2010}%
  \BibitemOpen
  \bibfield  {author} {\bibinfo {author} {\bibfnamefont {V.}~\bibnamefont
  {Gritsev}}, \bibinfo {author} {\bibfnamefont {P.}~\bibnamefont {Barmettler}},
  \ and\ \bibinfo {author} {\bibfnamefont {E.}~\bibnamefont {Demler}},\
  }\bibfield  {title} {\enquote {\bibinfo {title} {Scaling approach to quantum
  non-equilibrium dynamics of many-body systems},}\ }\href@noop {} {\bibfield
  {journal} {\bibinfo  {journal} {\NJP},\ }\textbf {\bibinfo {volume} {12}},\
  \bibinfo {pages} {113005} (\bibinfo {year} {2010})}\BibitemShut {NoStop}%
\bibitem [{\citenamefont {Muga}\ \emph {et~al.}(2009)\citenamefont {Muga},
  \citenamefont {Chen}, \citenamefont {Ruschhaupt},\ and\ \citenamefont
  {Gu\'ery-Odelin}}]{muga_2009}%
  \BibitemOpen
  \bibfield  {author} {\bibinfo {author} {\bibfnamefont {J.~G.}\ \bibnamefont
  {Muga}}, \bibinfo {author} {\bibfnamefont {Xi}~\bibnamefont {Chen}}, \bibinfo
  {author} {\bibfnamefont {A.}~\bibnamefont {Ruschhaupt}}, \ and\ \bibinfo
  {author} {\bibfnamefont {D.}~\bibnamefont {Gu\'ery-Odelin}},\ }\bibfield
  {title} {\enquote {\bibinfo {title} {Frictionless dynamics of {Bose-Einstein}
  condensates under fast trap variations},}\ }\href@noop {} {\bibfield
  {journal} {\bibinfo  {journal} {J. Phys. B: At. Mol. Opt. Phys.},\ }\textbf
  {\bibinfo {volume} {42}},\ \bibinfo {pages} {241001} (\bibinfo {year}
  {2009})}\BibitemShut {NoStop}%
\bibitem [{\citenamefont {Palmero}\ \emph {et~al.}(2013)\citenamefont
  {Palmero}, \citenamefont {Torrontegui}, \citenamefont {Gu\'ery-Odelin},\ and\
  \citenamefont {Muga}}]{palmero_2013}%
  \BibitemOpen
  \bibfield  {author} {\bibinfo {author} {\bibfnamefont {M.}~\bibnamefont
  {Palmero}}, \bibinfo {author} {\bibfnamefont {E.}~\bibnamefont
  {Torrontegui}}, \bibinfo {author} {\bibfnamefont {D.}~\bibnamefont
  {Gu\'ery-Odelin}}, \ and\ \bibinfo {author} {\bibfnamefont {J.~G.}\
  \bibnamefont {Muga}},\ }\bibfield  {title} {\enquote {\bibinfo {title} {Fast
  transport of two ions in an anharmonic trap},}\ }\href@noop {} {\bibfield
  {journal} {\bibinfo  {journal} {Phys. Rev. A},\ }\textbf {\bibinfo {volume}
  {88}},\ \bibinfo {pages} {053423} (\bibinfo {year} {2013})}\BibitemShut
  {NoStop}%
\bibitem [{\citenamefont {Chen}\ \emph
  {et~al.}(2011){\natexlab{b}}\citenamefont {Chen}, \citenamefont
  {Torrontegui}, \citenamefont {Stefanatos}, \citenamefont {Li},\ and\
  \citenamefont {Muga}}]{Chen_2011x}%
  \BibitemOpen
  \bibfield  {author} {\bibinfo {author} {\bibfnamefont {Xi}~\bibnamefont
  {Chen}}, \bibinfo {author} {\bibfnamefont {E.}~\bibnamefont {Torrontegui}},
  \bibinfo {author} {\bibfnamefont {Dionisis}\ \bibnamefont {Stefanatos}},
  \bibinfo {author} {\bibfnamefont {Jr-Shin}\ \bibnamefont {Li}}, \ and\
  \bibinfo {author} {\bibfnamefont {J.~G.}\ \bibnamefont {Muga}},\ }\bibfield
  {title} {\enquote {\bibinfo {title} {Optimal trajectories for efficient
  atomic transport without final excitation},}\ }\href@noop {} {\bibfield
  {journal} {\bibinfo  {journal} {Phys. Rev. A},\ }\textbf {\bibinfo {volume}
  {84}},\ \bibinfo {pages} {043415} (\bibinfo {year}
  {2011}{\natexlab{b}})}\BibitemShut {NoStop}%
\bibitem [{\citenamefont {Stefanatos}(2013)}]{stefanatos_2013}%
  \BibitemOpen
  \bibfield  {author} {\bibinfo {author} {\bibfnamefont {D.}~\bibnamefont
  {Stefanatos}},\ }\bibfield  {title} {\enquote {\bibinfo {title} {Optimal
  shortcuts to adiabaticity for a quantum piston},}\ }\href@noop {} {\bibfield
  {journal} {\bibinfo  {journal} {Automatica},\ }\textbf {\bibinfo {volume}
  {49}},\ \bibinfo {pages} {3079} (\bibinfo {year} {2013})}\BibitemShut
  {NoStop}%
\bibitem [{\citenamefont {Choi}\ \emph {et~al.}(2013)\citenamefont {Choi},
  \citenamefont {Onofrio},\ and\ \citenamefont {Sundaram}}]{choi_2013}%
  \BibitemOpen
  \bibfield  {author} {\bibinfo {author} {\bibfnamefont {S.}~\bibnamefont
  {Choi}}, \bibinfo {author} {\bibfnamefont {R.}~\bibnamefont {Onofrio}}, \
  and\ \bibinfo {author} {\bibfnamefont {B.}~\bibnamefont {Sundaram}},\
  }\bibfield  {title} {\enquote {\bibinfo {title} {Ehrenfest dynamics and
  frictionless cooling methods},}\ }\href@noop {} {\bibfield  {journal}
  {\bibinfo  {journal} {Phys. Rev. A},\ }\textbf {\bibinfo {volume} {88}},\
  \bibinfo {pages} {053401} (\bibinfo {year} {2013})}\BibitemShut {NoStop}%
\bibitem [{\citenamefont {Salamon}\ \emph
  {et~al.}(2009){\natexlab{a}}\citenamefont {Salamon}, \citenamefont
  {Hoffmann}, \citenamefont {Rezek},\ and\ \citenamefont
  {Kosloff}}]{salamon_2008}%
  \BibitemOpen
  \bibfield  {author} {\bibinfo {author} {\bibfnamefont {P.}~\bibnamefont
  {Salamon}}, \bibinfo {author} {\bibfnamefont {K.~H.}\ \bibnamefont
  {Hoffmann}}, \bibinfo {author} {\bibfnamefont {Y.}~\bibnamefont {Rezek}}, \
  and\ \bibinfo {author} {\bibfnamefont {R.}~\bibnamefont {Kosloff}},\
  }\bibfield  {title} {\enquote {\bibinfo {title} {Maximum work in minimum time
  from a conservative quantum system},}\ }\href@noop {} {\bibfield  {journal}
  {\bibinfo  {journal} {Phys. Chem. Chem. Phys.},\ }\textbf {\bibinfo {volume}
  {11}},\ \bibinfo {pages} {1027} (\bibinfo {year}
  {2009}{\natexlab{a}})}\BibitemShut {NoStop}%
\bibitem [{\citenamefont {Salamon}\ \emph
  {et~al.}(2009){\natexlab{b}}\citenamefont {Salamon}, \citenamefont
  {Hoffmann}, \citenamefont {Rezek},\ and\ \citenamefont
  {Kosloff}}]{salamon_2009}%
  \BibitemOpen
  \bibfield  {author} {\bibinfo {author} {\bibfnamefont {P.}~\bibnamefont
  {Salamon}}, \bibinfo {author} {\bibfnamefont {K.~H.}\ \bibnamefont
  {Hoffmann}}, \bibinfo {author} {\bibfnamefont {Y.}~\bibnamefont {Rezek}}, \
  and\ \bibinfo {author} {\bibfnamefont {R.}~\bibnamefont {Kosloff}},\
  }\bibfield  {title} {\enquote {\bibinfo {title} {Maximum work in minimum time
  from a conservative quantum system},}\ }\href@noop {} {\bibfield  {journal}
  {\bibinfo  {journal} {Phys. Chem. Chem. Phys.},\ }\textbf {\bibinfo {volume}
  {11}},\ \bibinfo {pages} {1027} (\bibinfo {year}
  {2009}{\natexlab{b}})}\BibitemShut {NoStop}%
\bibitem [{\citenamefont {Hoffmann}\ \emph {et~al.}(2011)\citenamefont
  {Hoffmann}, \citenamefont {Salamon}, \citenamefont {Rezek},\ and\
  \citenamefont {Kosloff}}]{hoffmann_2011}%
  \BibitemOpen
  \bibfield  {author} {\bibinfo {author} {\bibfnamefont {K.~H.}\ \bibnamefont
  {Hoffmann}}, \bibinfo {author} {\bibfnamefont {P.}~\bibnamefont {Salamon}},
  \bibinfo {author} {\bibfnamefont {Y.}~\bibnamefont {Rezek}}, \ and\ \bibinfo
  {author} {\bibfnamefont {R.}~\bibnamefont {Kosloff}},\ }\bibfield  {title}
  {\enquote {\bibinfo {title} {Time-optimal controls for frictionless cooling
  in harmonic traps},}\ }\href@noop {} {\bibfield  {journal} {\bibinfo
  {journal} {EPL (Europhysics Letters)},\ }\textbf {\bibinfo {volume} {96}},\
  \bibinfo {pages} {60015} (\bibinfo {year} {2011})}\BibitemShut {NoStop}%
\bibitem [{\citenamefont {{Choi}}\ \emph {et~al.}(2011)\citenamefont {{Choi}},
  \citenamefont {{Onofrio}},\ and\ \citenamefont {{Sundaram}}}]{choi_2011}%
  \BibitemOpen
  \bibfield  {author} {\bibinfo {author} {\bibfnamefont {S.}~\bibnamefont
  {{Choi}}}, \bibinfo {author} {\bibfnamefont {R.}~\bibnamefont {{Onofrio}}}, \
  and\ \bibinfo {author} {\bibfnamefont {B.}~\bibnamefont {{Sundaram}}},\
  }\bibfield  {title} {\enquote {\bibinfo {title} {Optimized sympathetic
  cooling of atomic mixtures via fast adiabatic strategies},}\ }\href@noop {}
  {\bibfield  {journal} {\bibinfo  {journal} {\PRA},\ }\textbf {\bibinfo
  {volume} {84}},\ \bibinfo {pages} {051601} (\bibinfo {year}
  {2011})}\BibitemShut {NoStop}%
\bibitem [{\citenamefont {Salamon}\ \emph {et~al.}(2011)\citenamefont
  {Salamon}, \citenamefont {Hoffmann},\ and\ \citenamefont
  {Tsirlin}}]{salamon_2012}%
  \BibitemOpen
  \bibfield  {author} {\bibinfo {author} {\bibfnamefont {P.}~\bibnamefont
  {Salamon}}, \bibinfo {author} {\bibfnamefont {K.~H.}\ \bibnamefont
  {Hoffmann}}, \ and\ \bibinfo {author} {\bibfnamefont {A.}~\bibnamefont
  {Tsirlin}},\ }\bibfield  {title} {\enquote {\bibinfo {title} {Optimal control
  in a quantum cooling problem},}\ }\href@noop {} {\bibfield  {journal}
  {\bibinfo  {journal} {Appl. Math. Lett.},\ }\textbf {\bibinfo {volume}
  {25}},\ \bibinfo {pages} {1263} (\bibinfo {year} {2011})}\BibitemShut
  {NoStop}%
\bibitem [{\citenamefont {Yuce}(2012)}]{yuce_2012}%
  \BibitemOpen
  \bibfield  {author} {\bibinfo {author} {\bibfnamefont {C.}~\bibnamefont
  {Yuce}},\ }\bibfield  {title} {\enquote {\bibinfo {title} {Fast frictionless
  expansion of an optical lattice},}\ }\href@noop {} {\bibfield  {journal}
  {\bibinfo  {journal} {Phys. Lett. A},\ }\textbf {\bibinfo {volume} {376}},\
  \bibinfo {pages} {1717} (\bibinfo {year} {2012})}\BibitemShut {NoStop}%
\bibitem [{\citenamefont {{Schaff}}\ \emph {et~al.}(2010)\citenamefont
  {{Schaff}}, \citenamefont {{Song}}, \citenamefont {{Vignolo}},\ and\
  \citenamefont {{Labeyrie}}}]{schaff_2010}%
  \BibitemOpen
  \bibfield  {author} {\bibinfo {author} {\bibfnamefont {J.-F.}\ \bibnamefont
  {{Schaff}}}, \bibinfo {author} {\bibfnamefont {X.-L.}\ \bibnamefont
  {{Song}}}, \bibinfo {author} {\bibfnamefont {P.}~\bibnamefont {{Vignolo}}}, \
  and\ \bibinfo {author} {\bibfnamefont {G.}~\bibnamefont {{Labeyrie}}},\
  }\bibfield  {title} {\enquote {\bibinfo {title} {{Fast optimal transition
  between two equilibrium states}},}\ }\href@noop {} {\bibfield  {journal}
  {\bibinfo  {journal} {\PRA},\ }\textbf {\bibinfo {volume} {82}},\ \bibinfo
  {pages} {033430} (\bibinfo {year} {2010})}\BibitemShut {NoStop}%
\bibitem [{\citenamefont {{del Campo}}\ \emph {et~al.}(2013)\citenamefont {{del
  Campo}}, \citenamefont {{Goold}},\ and\ \citenamefont
  {{Paternostro}}}]{campo_2013b}%
  \BibitemOpen
  \bibfield  {author} {\bibinfo {author} {\bibfnamefont {A.}~\bibnamefont {{del
  Campo}}}, \bibinfo {author} {\bibfnamefont {J.}~\bibnamefont {{Goold}}}, \
  and\ \bibinfo {author} {\bibfnamefont {M.}~\bibnamefont {{Paternostro}}},\
  }\bibfield  {title} {\enquote {\bibinfo {title} {More bang for your buck:
  Towards super-adiabatic quantum engines},}\ }\href@noop {} {\bibfield
  {journal} {\bibinfo  {journal} {arXiv:1305.3223}} (\bibinfo {year}
  {2013})}\BibitemShut {NoStop}%
\bibitem [{\citenamefont {{Schaff}}\ \emph {et~al.}(2011)\citenamefont
  {{Schaff}}, \citenamefont {{Song}}, \citenamefont {{Capuzzi}}, \citenamefont
  {{Vignolo}},\ and\ \citenamefont {{Labeyrie}}}]{schaff_2011}%
  \BibitemOpen
  \bibfield  {author} {\bibinfo {author} {\bibfnamefont {J.-F.}\ \bibnamefont
  {{Schaff}}}, \bibinfo {author} {\bibfnamefont {X.-L.}\ \bibnamefont
  {{Song}}}, \bibinfo {author} {\bibfnamefont {P.}~\bibnamefont {{Capuzzi}}},
  \bibinfo {author} {\bibfnamefont {P.}~\bibnamefont {{Vignolo}}}, \ and\
  \bibinfo {author} {\bibfnamefont {G.}~\bibnamefont {{Labeyrie}}},\ }\bibfield
   {title} {\enquote {\bibinfo {title} {Shortcut to adiabaticity for an
  interacting {B}ose-{E}instein condensate},}\ }\href@noop {} {\bibfield
  {journal} {\bibinfo  {journal} {EPL (Europhysics Letters)},\ }\textbf
  {\bibinfo {volume} {93}},\ \bibinfo {pages} {23001} (\bibinfo {year}
  {2011})}\BibitemShut {NoStop}%
\bibitem [{\citenamefont {Negretti}\ \emph {et~al.}(2013)\citenamefont
  {Negretti}, \citenamefont {Benseny}, \citenamefont {Mompart},\ and\
  \citenamefont {Calarco}}]{negretti_2013}%
  \BibitemOpen
  \bibfield  {author} {\bibinfo {author} {\bibfnamefont {A.}~\bibnamefont
  {Negretti}}, \bibinfo {author} {\bibfnamefont {A.}~\bibnamefont {Benseny}},
  \bibinfo {author} {\bibfnamefont {J.}~\bibnamefont {Mompart}}, \ and\
  \bibinfo {author} {\bibfnamefont {T.}~\bibnamefont {Calarco}},\ }\bibfield
  {title} {\enquote {\bibinfo {title} {Speeding up the spatial adiabatic
  passage of matter waves in optical microtraps by optimal control},}\
  }\href@noop {} {\bibfield  {journal} {\bibinfo  {journal} {Quantum
  Information Processing},\ }\textbf {\bibinfo {volume} {12}},\ \bibinfo
  {pages} {1439} (\bibinfo {year} {2013})}\BibitemShut {NoStop}%
\bibitem [{\citenamefont {Bowler}\ \emph {et~al.}(2012)\citenamefont {Bowler},
  \citenamefont {Gaebler}, \citenamefont {Lin}, \citenamefont {Tan},
  \citenamefont {Hanneke}, \citenamefont {Jost}, \citenamefont {Home},
  \citenamefont {Leibfried},\ and\ \citenamefont {Wineland}}]{bowler_2012}%
  \BibitemOpen
  \bibfield  {author} {\bibinfo {author} {\bibfnamefont {R.}~\bibnamefont
  {Bowler}}, \bibinfo {author} {\bibfnamefont {J.}~\bibnamefont {Gaebler}},
  \bibinfo {author} {\bibfnamefont {Y.}~\bibnamefont {Lin}}, \bibinfo {author}
  {\bibfnamefont {T.~R.}\ \bibnamefont {Tan}}, \bibinfo {author} {\bibfnamefont
  {D.}~\bibnamefont {Hanneke}}, \bibinfo {author} {\bibfnamefont {J.~D.}\
  \bibnamefont {Jost}}, \bibinfo {author} {\bibfnamefont {J.~P.}\ \bibnamefont
  {Home}}, \bibinfo {author} {\bibfnamefont {D.}~\bibnamefont {Leibfried}}, \
  and\ \bibinfo {author} {\bibfnamefont {D.~J.}\ \bibnamefont {Wineland}},\
  }\bibfield  {title} {\enquote {\bibinfo {title} {Coherent diabatic ion
  transport and separation in a multizone trap array},}\ }\href@noop {}
  {\bibfield  {journal} {\bibinfo  {journal} {Phys. Rev. Lett.},\ }\textbf
  {\bibinfo {volume} {109}},\ \bibinfo {pages} {080502} (\bibinfo {year}
  {2012})}\BibitemShut {NoStop}%
\bibitem [{\citenamefont {Walther}\ \emph {et~al.}(2012)\citenamefont
  {Walther}, \citenamefont {Ziesel}, \citenamefont {Ruster}, \citenamefont
  {Dawkins}, \citenamefont {Ott}, \citenamefont {Hettrich}, \citenamefont
  {Singer}, \citenamefont {Schmidt-Kaler},\ and\ \citenamefont
  {Poschinger}}]{walther_2012}%
  \BibitemOpen
  \bibfield  {author} {\bibinfo {author} {\bibfnamefont {A.}~\bibnamefont
  {Walther}}, \bibinfo {author} {\bibfnamefont {F.}~\bibnamefont {Ziesel}},
  \bibinfo {author} {\bibfnamefont {T.}~\bibnamefont {Ruster}}, \bibinfo
  {author} {\bibfnamefont {S.~T.}\ \bibnamefont {Dawkins}}, \bibinfo {author}
  {\bibfnamefont {K.}~\bibnamefont {Ott}}, \bibinfo {author} {\bibfnamefont
  {M.}~\bibnamefont {Hettrich}}, \bibinfo {author} {\bibfnamefont
  {K.}~\bibnamefont {Singer}}, \bibinfo {author} {\bibfnamefont
  {F.}~\bibnamefont {Schmidt-Kaler}}, \ and\ \bibinfo {author} {\bibfnamefont
  {U.}~\bibnamefont {Poschinger}},\ }\bibfield  {title} {\enquote {\bibinfo
  {title} {Controlling fast transport of cold trapped ions},}\ }\href@noop {}
  {\bibfield  {journal} {\bibinfo  {journal} {Phys. Rev. Lett.},\ }\textbf
  {\bibinfo {volume} {109}},\ \bibinfo {pages} {080501} (\bibinfo {year}
  {2012})}\BibitemShut {NoStop}%
\bibitem [{\citenamefont {{Bakr}}\ \emph {et~al.}(2009)\citenamefont {{Bakr}},
  \citenamefont {{Gillen}}, \citenamefont {{Peng}}, \citenamefont
  {{F\"olling}},\ and\ \citenamefont {{Greiner}}}]{bakr_2009}%
  \BibitemOpen
  \bibfield  {author} {\bibinfo {author} {\bibfnamefont {W.~S.}\ \bibnamefont
  {{Bakr}}}, \bibinfo {author} {\bibfnamefont {J.~I.}\ \bibnamefont
  {{Gillen}}}, \bibinfo {author} {\bibfnamefont {A.}~\bibnamefont {{Peng}}},
  \bibinfo {author} {\bibfnamefont {S.}~\bibnamefont {{F\"olling}}}, \ and\
  \bibinfo {author} {\bibfnamefont {M.}~\bibnamefont {{Greiner}}},\ }\bibfield
  {title} {\enquote {\bibinfo {title} {A quantum gas microscope for detecting
  single atoms in a {H}ubbard-regime optical lattice},}\ }\href@noop {}
  {\bibfield  {journal} {\bibinfo  {journal} {Nature},\ }\textbf {\bibinfo
  {volume} {462}},\ \bibinfo {pages} {74} (\bibinfo {year} {2009})}\BibitemShut
  {NoStop}%
\bibitem [{\citenamefont {Endres}\ \emph {et~al.}(2011)\citenamefont {Endres},
  \citenamefont {Cheneau}, \citenamefont {Fukuhara}, \citenamefont
  {Weitenberg}, \citenamefont {Schau�}, \citenamefont {Gross}, \citenamefont
  {Mazza}, \citenamefont {Ba\~{n}uls}, \citenamefont {Pollet}, \citenamefont
  {Bloch},\ and\ \citenamefont {Kuhr}}]{endres_2011}%
  \BibitemOpen
  \bibfield  {author} {\bibinfo {author} {\bibfnamefont {M.}~\bibnamefont
  {Endres}}, \bibinfo {author} {\bibfnamefont {M.}~\bibnamefont {Cheneau}},
  \bibinfo {author} {\bibfnamefont {T.}~\bibnamefont {Fukuhara}}, \bibinfo
  {author} {\bibfnamefont {C.}~\bibnamefont {Weitenberg}}, \bibinfo {author}
  {\bibfnamefont {P.}~\bibnamefont {Schau�}}, \bibinfo {author}
  {\bibfnamefont {C.}~\bibnamefont {Gross}}, \bibinfo {author} {\bibfnamefont
  {L.}~\bibnamefont {Mazza}}, \bibinfo {author} {\bibfnamefont {M.~C.}\
  \bibnamefont {Ba\~{n}uls}}, \bibinfo {author} {\bibfnamefont
  {L.}~\bibnamefont {Pollet}}, \bibinfo {author} {\bibfnamefont
  {I.}~\bibnamefont {Bloch}}, \ and\ \bibinfo {author} {\bibfnamefont
  {S.}~\bibnamefont {Kuhr}},\ }\bibfield  {title} {\enquote {\bibinfo {title}
  {Observation of correlated particle-hole pairs and string order in
  low-dimensional {M}ott insulators},}\ }\href@noop {} {\bibfield  {journal}
  {\bibinfo  {journal} {Science},\ }\textbf {\bibinfo {volume} {334}},\
  \bibinfo {pages} {200} (\bibinfo {year} {2011})}\BibitemShut {NoStop}%
\end{thebibliography}%
\end{document}